\documentclass[10pt,draftclsnofoot,onecolumn]{IEEEtran}
\usepackage[utf8]{inputenc}
\usepackage{amsmath}
\usepackage{amssymb}
\usepackage{upgreek}
\usepackage{tipa}
\usepackage{dsfont}
\usepackage{enumerate}
\usepackage{subfig}
\usepackage{float}
\usepackage{multicol}
\usepackage[usenames,dvipsnames]{pstricks}
\usepackage{epsfig}
\usepackage{pst-grad}
\usepackage{pst-plot}
\usepackage[noadjust]{cite}
\usepackage{caption}
\usepackage{graphicx}
\usepackage{epstopdf}
\usepackage{psfrag}
\usepackage{color}
\usepackage{amsfonts}
\usepackage{amsmath,mathtools}
\usepackage{booktabs}
\usepackage{stackengine}
\usepackage{array}
\usepackage{balance}
\usepackage{comment} 
\usepackage{multirow}
\usepackage{url}
\makeatletter
\newcommand{\vast}{\bBigg@{4}}
\newcommand{\Vast}{\bBigg@{5}}

\begin{document}

\title{Compound TCP with Random Early Detection (RED): stability, bifurcation and performance analyses}

\author{\IEEEauthorblockN{Sreelakshmi Manjunath and Gaurav Raina} \\
\IEEEauthorblockA{Department of Electrical Engineering, Indian Institute of Technology Madras, Chennai 600 036, India\\
Email: $\lbrace \text{sreelakshmi, gaurav} \rbrace$@ee.iitm.ac.in}
\thanks{This is an extension of our preliminary work that appeared in Proceedings of the 27th IEEE Chinese Control and Decision Conference (CCDC),
2015. DOI: 10.1109/CCDC.2015.7162875}
}

% Grants or other notes about the article that should go on the front
% page should be placed within the \thanks{} command in the title
% (and the %-sign in front of \thanks{} should be deleted)
%
% General acknowledgments should be placed at the end of the article.

% \subtitle{Do you have a subtitle?\\ If so, write it here}

%\titlerunning{Short form of title}        % if too long for running head

% The correct dates will be entered by the editor

\maketitle

\begin{abstract}
The persistent problem of increased queueing delays in the Internet motivates the study of currently implemented transport protocols and active queue management (AQM) policies. We study Compound TCP (default protocol in Windows) with Random Early Detection (RED).  RED is an early queue policy that is implemented in today's routers, but is not deployed in live operating networks. RED uses an exponentially weighted moving average of the queue size to make packet-dropping decisions, aiming to control the queue size\textemdash and hence the queueing delay. One must study RED with current protocols in order to explore its viability in the context of the issue of increased queueing delays.

We first derive a non-linear time-delayed model for the Compound TCP-RED system. As time-delayed systems are prone to instability, we begin by analysing the local stability of this model. We derive a sufficient condition for its local stability, and examine the impact of the (i) round-trip time (RTT) of the TCP flows, (ii) queue averaging parameter and (iii) packet-dropping thresholds, on system stability. Further, we establish that the system undergoes a Hopf bifurcation as any of the above three parameters is varied. This suggests the emergence of limit cycle oscillations in the queue size, which may lead to synchronisation of TCP flows and loss of link utilisation. Next, we study a regime where averaging over the queue size is not performed, and packet-dropping decisions are based on the instantaneous queue size. For a fluid model befitting this regime, we derive the necessary and sufficient condition for local stability. We observe that the system loses stability as either the RTT of the TCP flows or the packet-dropping threshold is increased, in this regime as well. A comparison of the stability results for Compound TCP-RED in the two regimes\textemdash with and without queue size averaging\textemdash
reveals that such averaging may not be beneficial to system stability. Packet-level simulations show that the queue size indeed exhibits limit cycle oscillations as system parameters are varied. We then outline a simple threshold-based queue policy, that could ensure stable and low-latency operation. Through a simulation-based performance evaluation, we also show that the threshold policy outperforms RED in terms of queueing delay, flow completion time and packet loss, in a variety of network settings. Our study highlights that the threshold-based policy could mitigate the issue of increased queueing delays in the Internet.
\end{abstract}

\section{Introduction}
Packet-switched networks use buffers, placed in every router, to store packets that arrive when the link is busy. Buffers play an important role in network performance as they impact packet loss and throughput, and also absorb bursty traffic. During the formative years of the Internet, engineers suggested that these buffers be capable of holding at least a bandwidth-delay product worth of packets~\cite{villamizar1994high}. This eventually lead to excessive provisioning and proliferation of large buffers across the Internet. Further, these routers employ the Drop-Tail queue policy, which means that packets get dropped only if they arrive to find a full buffer. Transmission Control Protocol (TCP), a fundamental protocol of the Internet, increases the window size (number of packets that can be sent in one round-trip time), until the first packet loss is detected. In Drop-Tail queues, TCP ends up sending enough packets to occupy the entire buffer, as it encounters the first packet loss only when the buffer is full. Such filling of buffers leads to an increase in queueing delays across the network.

The problem of persistently full large buffers\textemdash commonly known as \emph{bufferbloat}\textemdash was first highlighted in~\cite{nagle1987packet}. Subsequently, it was recognised that the deployment of effective Active Queue Management (AQM) strategies at routers could mitigate this problem~\cite{braden1998recommendations}. AQMs aim to drop packets \emph{before} buffers fill up. 
This serves as implicit feedback to TCP sources regarding the onset of congestion, thus prompting them to reduce their respective sending windows. The first major AQM, Random Early Detection (RED) was proposed in~\cite{floyd1993random}. This was followed by numerous other proposals; for example Random Exponential Marking (REM)~\cite{athuraliya2001rem}, Proportional Integral queue policy~\cite{hollot2002analysis}, Exponential RED~\cite{liu2005exponential}, PIE~\cite{pan2013pie} among others. However, due to a lack of consensus on the optimal queue policy, the simple Drop-Tail policy\textemdash which waits for the buffer to completely fill up before dropping a packet\textemdash continues to be widely deployed. Thus, bufferbloat persists even today, and has been identified as the primary cause of the increase in queueing delays in the Internet~\cite{gettys2012bufferbloat,cerf2014bufferbloat}. 
% With technological advancements such as 5G and Internet of Things, it is crucial to find a plausible solution to the issue of bufferbloat~\cite{pretz2001future}. 
Recent reports re-emphasise that use of effective queue management strategies could be one such solution~\cite{nichols2012controlling}.
% Re-emphasising the need for an effective queue management strategy, the authors of~\cite{Nichols_12} proposed an AQM strategy called CoDel. Nevertheless, none of these AQMs are deployed so far.

Most routers today have the Random Early Detection policy implemented in them, though not deployed in live networks. RED works as follows: when a packet arrives at the link, RED computes an exponentially weighted moving average of the queue size, and then drops the packet with a probability that is a function of the average queue size. By doing so, RED aims to drop packets when the average queue size reaches a pre-defined threshold, and notify end systems of incipient congestion~\cite{floyd1993random}. This was expected to avoid synchronisation of TCP flows, and also to tackle burstiness. However, some researchers observed that RED could lead to oscillatory behaviour~\cite{Bonald2000,christiansen2000tuning}, and concluded that RED deployment may not be straight forward~\cite{may1999reasons}. This was primarily attributed to the lack of guidelines for the choice of parameters~\cite{christiansen2000tuning, Bonald2000}. In the following years, RED drew considerable research attention, giving rise to many variants\textemdash e.g. Gentle RED, Adaptive RED, Weighted RED~\cite{hamadneh2011weighted}, Robust RED~\cite{zhang2010rred}, Subsidized RED \cite{wang2005subsidized} etc. However, none of these have been implemented. Consequently, among the aforementioned AQMs, RED is the closest to be brought into effective action, and continues to generate research interest; for example see \cite{siregar2017implementation, abdulkareem2015efred}. Before deploying the RED queue policy, it is imperative to study its performance in the context of protocols that are currently deployed in the end systems.

The Internet today caters to multiple users, on various platforms, communicating using a wide variety of protocols. Consequently, the queue management strategy at the router would encounter data packets from numerous traffic sources. Performance evaluation of AQM policies should ideally consider all varieties of Internet traffic. However, this could be difficult, as the heterogeneity involved would lead to a complex set-up or intractable models. One possible approach would be to consider one transport protocol which forms a dominant component of the Internet traffic, obtain some insight for this setup, and then verify if the insight holds for a wider traffic mix. It has been reported that TCP traffic largely dominates the Internet~\cite{thompson1997wide,williamson2001internet}, amounting to about $70$\% of the total traffic~\cite{aouini2016towards}.
End-systems today deploy different flavours of TCP; for example, Compound~\cite{tan2006compound} is implemented in Windows, and CUBIC~\cite{ha2008cubic} is the default in Linux. A recent study conducted using a TCP algorithm identification tool reports that about 25\% of the end systems considered used Compound TCP~\cite{yang2014tcp}. Further, it has been reported that Compound TCP will play a significant role in future networks supporting Internet of Things applications~\cite{pokhrel2018modeling}. Therefore, a detailed study of the RED queue policy in conjunction with Compound TCP is valuable. To that end, we begin by studying a system of Compound TCP flows operating over a router with the RED policy, and then verify if the insight obtained would hold for a wider variety of traffic scenarios.  

One of the primary goals of the RED policy is to regulate the queue size, while minimising packet losses~\cite{floyd1993random}. This trade-off between queue size and packet loss is to be maintained, despite changes in other parameters, such as the feedback delay. In order to understand these trade-offs better one may study mathematical models of the underlying system. Mathematical models that can capture the interaction of the additive increase multiplicative decrease (AIMD) behaviour of Compound TCP with the feedback provided by the queue policy are essentially non-linear~\cite{tan2006compound,raja2012delay}. In addition to this, feedback in TCP-AQM systems is non-instantaneous owing to queueing and propagation delays. As a result, Compound TCP-RED is a non-linear time-delayed system. Such time-delayed systems are prone to loss of stability, as either the feedback delay or any of the system parameters varies~\cite{Franklin1995}. In this case, loss of stability may imply that RED would fail to regulate the queue size as intended. Arguably, performance evaluation of these systems would have to incorporate the study of system stability. Control-theoretic analysis enables one to study a suitable mathematical model, and understand the stability, of the underlying system. In particular, local stability analysis is often used to define bounds on the system parameters for stable operation~\cite{srikant2012mathematics}. Such analysis of RED, along with TCP Reno, was conducted in~\cite{misra2000fluid,hollot2001control,low2002dynamics, cicco2016control}. A control-theoretic analysis of Compound TCP-RED is still in order.

It is well known, in dynamical systems' literature, that non-linear systems often exhibit limit cycles, as they lose stability~\cite{guckenheimer2013nonlinear}. In case of TCP-AQM systems, such limit cycles in the system dynamics could manifest as limit cycle oscillations in the queue size. Such queue size oscillations are known to cause synchronisation of TCP flows and loss of throughput~\cite{raina2005part,zhang1990oscillating}. Therefore, loss of stability could hamper network performance. This makes it useful to study the transition of the system into the unstable regime, in addition to studying system stability itself. In our study of Compound TCP-RED, we use bifurcation-theoretic tools to establish that the system transits into instability via a Hopf bifurcation\textemdash which guarantees the emergence of limit cycle oscillations in the system dynamics. We aim to answer the following questions through our study:
\begin{enumerate}
 \item [(i)] what is the impact of network parameters, such as round-trip time (RTT) of the TCP flows, on the stability of Compound TCP-RED?
 \item [(ii)] is queue size averaging beneficial to system stability?
 \item [(iii)] what is the influence of packet-dropping thresholds on stability?
\end{enumerate}

We first outline a non-linear fluid model for Compound TCP-RED (Section~\ref{sec:models}). We conduct a local stability analysis of this model and find a sufficient condition for local stability. We then explicitly establish that the system transits into instability via a Hopf bifurcation, as system parameters are varied (Section~\ref{sec:CompoundTCP-RED}). Numerically constructed stability charts reveal that Compound TCP-RED may become unstable as the RTT of the TCP flows increases. It is also seen that system stability is sensitive to the choice of the queue averaging parameter. We then study a regime where queue size averaging is not performed, and the packet-drop probability is a function of the instantaneous queue size (Section~\ref{sec:CompoundTCP-RED_no_averaging}). For a fluid model for Compound TCP-RED operating in this regime, we derive the necessary and sufficient condition for local stability. We prove that the system undergoes a Hopf bifurcation in this regime as well. However, our analysis indicates that, Compound TCP-RED remains stable for comparatively larger RTTs in this regime. Local stability results suggest that smaller packet-dropping thresholds could be favourable to system stability.

Packet-level simulations conducted on the Network Simulator NS2~\cite{ns2} are presented to corroborate the analytical insights (Section~\ref{sec:sims}). It is indeed observed that (i) large RTTs are detrimental to system stability, (ii) queue size averaging may not be beneficial as the system remains stable for larger RTTs in the absence of averaging, and (iii) smaller thresholds for dropping packets aid stability. It is observed that, when the system loses stability, TCP flows synchronise and link utilisation drops, leading to deteriorating network performance. These observations are seen to hold even when the traffic setting comprises CUBIC (default TCP in Linux), UDP and HTTP flows. This suggests that the analytical insights may be applicable to a more general setting with a wider variety of traffic as well.

In essence, our theoretical study of the Compound TCP-RED system suggests that the system is susceptible to instabilities as either the RTT of the TCP flows or the packet-dropping threshold, is varied. Additionally, it alerts us to the fact that queue size averaging performed by the RED queue policy may not be beneficial to system stability. The packet-level simulations that we conduct enable us to not only validate the analytical insight regarding system stability, but also to observe phenomena such as flow synchronisation and drop in link utilisation, which occur when system stability is lost. This strengthens our premise regarding the centrality of stability analysis to performance evaluation.

Based on our observations of the Compound TCP-RED system, we outline a simple threshold-based queue policy (Section~\ref{sec:threshold_policy}). We discuss local stability results for a system of Compound TCP flows feeding into a router with the threshold-based queue policy. It is observed that, with the threshold-based policy, local stability does not depend on the round-trip time. Additionally, the stability conditions suggest that the TCP and AQM parameters must be co-designed to ensure stable operation. We then conduct a simulation-based performance evaluation of RED and the threshold-based queue policy, for various network settings and traffic mixes. It is seen that the threshold policy consistently outperforms RED in terms of queueing delay, packet loss and flow completion time. We also discuss some stability results for the threshold policy in conjunction with TCP Reno (an early proposal for loss-based protocol).
% and FAST TCP (a delay-based protocol proposed for high-speed networks). 
It is seen that the threshold-based policy could ensure stable and low-latency operation with this transport protocol as well. We summarise the contributions of our work in Section~\ref{sec:conclude}.

\section{Models}
\label{sec:models}
We now describe the general fluid model for TCP, and then describe the algorithm and model for window update of Compound TCP. This is followed by a description of the models for queue dynamics and the RED queue policy. A non-linear fluid model for the coupled Compound TCP-RED system is also outlined. 

\subsection{Transmission Control Protocol}
\label{sec:CompoundTCP}
Transmission control protocol is a window-based flow control algorithm embedded in the end-systems, to ensure reliable transmission of data packets. TCP uses a sliding window mechanism, that enables it to send a set of packets within a stipulated time frame, instead of waiting for each packet to be acknowledged before sending the next. The size of this sending window is increased or reduced based on the information regarding network congestion. There are different variants of TCP, and these variants primarily differ in the form of feedback they use to infer congestion.
% in the network. 

Some flavours of TCP use either packet loss or an estimate of queueing delay as congestion feedback, while others use a combination of the two. Tahoe TCP~\cite{jacobson1988congestion} and TCP Reno~\cite{jacobson1990modified} are two of the earliest proposals for loss-based TCP. 
% These TCP flavours adopt the AIMD rule, wherein the window size is increased by one packet over a RTT when all the packets are acknowledged, and halved if a packet loss is detected. 
While, Vegas TCP, arguably, laid the foundations for delay-based TCP flavours~\cite{brakmo1995tcp}. FAST TCP, which can be viewed as a high-speed version of Vegas, is a new proposal for delay-based TCP~\cite{Wei2006}. Recognising that the some issues such as efficiency, RTT fairness and TCP fairness can not be simultaneously mitigated either by loss- or delay-based protocols, the authors of~\cite{tan2006compound} proposed Compound TCP, which is a synergy of the two. It is currently the default protocol in the Windows OS. Other variants of loss- and delay-based protocols are TCP Illinois~\cite{liu2008tcp} and TCP-Africa~\cite{king2005tcp}.

We recapitulate the development of the fluid model for TCP Reno presented in~\cite{misra2000fluid,raina2005buffer}. Following which we present a general fluid model that captures the sending window dynamics of a class of delay- and loss-based protocols. Let $W(t)$ represent the sending window size (number of packets sent in one round-trip time) of a \emph{single} TCP flow of round-trip time $\tau$, which is increased by $1$ packet per RTT and reduced to $W/2$ every time a packet loss is detected. If the rate at which packets are sent is approximated as $W(t)/\tau$, then acknowledgements are received at a rate of $W(t-\tau)$ at time $t$. Let $p(t)$ be the packet-loss probability at time $t$. Suppose there are $N$ end users, and let $W^N(t)$ represent the \emph{sum} of all the window sizes. Then, $W^N(t)$ follows
\begin{align*}
 W^N(t+\delta)-W^N(t) \approx \frac{\delta \, N}{\tau} - \delta \frac{W^N(t)}{2N} \bigg(\frac{W^N(t-\tau)}{\tau}p(t-\tau)\bigg).
\end{align*}
Upon dividing both sides of the above equation by $N$, we arrive at the following continuous time approximation for the update of the \emph{average} window size $w(t)=W^N(t)/N$
\begin{align*}
 \dot{w}(t) = \frac{1}{\tau}-\frac{w(t)}{2}\bigg(\frac{w(t-\tau)}{\tau}p(t-\tau)\bigg),
\end{align*}
where $\dot{w}(t) = dw(t)/dt$ and $\tau$ can be regarded as the average round-trip time of the TCP flows. Observe that the number of users $N$ doesn't feature in the above delay-differential equation for window update. We shall remark on this assertion, as well as a few model assumptions, towards the end of this section, once the closed-loop TCP-RED model is outlined.

Now consider a TCP variant that increases the sending window by $i(w)$ per acknowledgement received and reduces it by $d(w)$ per packet drop. Further, if a packet is lost with a probability of $p(t)$, it is acknowledged with a probability $1-p(t)$. The average window size of the TCP flows is then updated as per the following delay-differential equation~\cite{raina2005part}
\begin{align}
 \dot{w}(t) = \bigg(i\big(w(t)\big)\big(1-p(t)\big)-d\big(w(t)\big)p(t-\tau)\bigg)\frac{w(t-\tau)}{\tau}.\label{eq:gen_TCP}
\end{align}
The functions $i(w)$ and $d(w)$ are specific to each TCP variant. Below we discuss the algorithm for Compound TCP, and derive the corresponding window update functions. 
While studying Compound TCP would be desirable, as it is implemented in today's Internet, it is natural to extend the study to other TCP variants which (i) use queueing delay and packet loss as feedback like Compound TCP, and (ii) can be modelled using the above equation~\eqref{eq:gen_TCP} for window update. To that end, we list the window update functions for some TCP variants in Table~\ref{tab:TCP_functional_forms}. One may use these functional forms to repeat the analysis, we conduct for Compound TCP in the later sections, for these TCP variants.

\subsubsection{Compound TCP}
Compound TCP is a congestion control protocol proposed for high-speed and long-delay networks~\cite{tan2006compound}. It is currently the default protocol in the Windows OS. Compound TCP incorporates a scalable delay-based component into the additive increase multiplicative decrease algorithm of TCP Reno, which is a loss-based protocol. The delay-based component aims to increase the window size when the network is sensed to be under-utilised, and reduce it when congestion is detected. 

The congestion control algorithm of Compound TCP employs two variables $cwnd$: the congestion window (controls the loss-based component), and $dwnd$: the delay window (controls the delay-based component). The sending window is calculated as~\cite{tan2006compound}
\begin{align*}
 win = \text{min}(cwnd+dwnd,awnd),
\end{align*}
where $awnd$ is the advertised window size from the receiver.
In its congestion avoidance phase, $cwnd$ is increased by $1$ packet per RTT, and halved when packet loss is detected. In one RTT, Compound TCP sends $cwnd + dwnd$ packets. Therefore, upon receiving an acknowledgement, the congestion window is updated as
\begin{align*}
 cwnd = cwnd + 1/(cwnd + dwnd).
\end{align*}

The algorithm for the delay-based component is designed as follows
\[
 dwnd(t+1) = \begin{cases}
              dwnd(t) + (\alpha\, win(t)^k - 1)^{+}, & \text{if } diff < \tilde{\gamma},\\
              (dwnd(t) - \zeta\, diff)^{+}, & \text{if } diff \geq \tilde{\gamma},\\
              (win(t)\, (1-\beta) - cwnd/2)^{+}, & \text{if loss},
             \end{cases}
\]
where $\tilde{\gamma}$ is a threshold for congestion detection, and $\zeta$ defines how rapidly Compound TCP reduces the window size when early congestion is detected. The parameters $\alpha, k,$ and $\beta$ govern the increase and decrease of the window size in Compound TCP.
The default values of these parameters are: $\alpha=0.125,k=0.75,\beta=0.5,\tilde{\gamma} = 30,~\zeta = 0.5$. The variable $diff$ is the number of backlogged packets per TCP connection in the bottleneck queue, and is estimated as
\begin{align*}
 diff = \bigg(\frac{win}{baseRTT} - \frac{win}{RTT}\bigg) baseRTT.
\end{align*}
In the above equation, $baseRTT$ is an estimation of the transmission delay of a packet which is updated by observing the smallest round-trip time across flows, and $RTT$ is the actual round-trip time of the flow. 

\begin{table}[t]
\centering
\caption{Functional forms for the window update functions of various TCP variants.}
\begin{tabular}{l | l | l | l}
\hline\label{tab:TCP_functional_forms}
\rule{0pt}{1.5\normalbaselineskip}\hspace{-1mm} TCP variant & Increase: $i(w)$ & Decrease: $d(w)$ & Default values \\ [2ex]
% \hline \rule{0pt}{1.5\normalbaselineskip}
% \hspace{-1mm}TCP Reno & $1/w$& $w/2$ & $-$\\[2ex]
\hline\rule{0pt}{1.5\normalbaselineskip}
\hspace{-1mm}
Compound TCP & $\alpha\, w^{k-1}$ & $\beta\,w$ & $\alpha = 0.125,k=0.75,\beta = 0.5$\\ [2ex]
\hline\rule{0pt}{1.5\normalbaselineskip}
\hspace{-1mm}
TCP Illinois & $\alpha_{max}/w$ & $\beta_{min}\,w$ & $\alpha_{max} = 10, \beta_{min} = 0.125$\\[2ex]
\hline\rule{0pt}{1.5\normalbaselineskip}
\hspace{-1mm}
\multirow{3}{*}{AFRICA TCP} & $a(w)/w$ & $w\, b(w)$ \\[2ex]
\cline{2-3}\rule{0pt}{1.5\normalbaselineskip}
\hspace{-1mm}
&\multicolumn{3}{c}{$a(w) = \frac{0.156\,w^2\, b(w)}{\big(2-b(w)\big)w^{1.2}},\,\,\,\,\,\, b(w) = \frac{-0.4\big(\log(w)-\log(38)\big)}{\big(\log(83000)-\log(38)\big)}+0.5$} \\
\hline \end{tabular}
\end{table} 

Summing the loss window and delay window we get the window update equation as~\cite{raja2012delay}
\begin{align}
 w(t+1) =& \begin{cases}
           w(t) + \alpha\, w(t)^k, & \text{if no loss},\\
           w(t)\,(1-\beta), & \text{if loss}.
          \end{cases}\label{eq:window_evolution}
\end{align}
From~\eqref{eq:window_evolution}, one may derive the increase and decrease functions that govern the evolution of the sending window of a Compound TCP flow as~\cite{raja2012delay}
$$ i(w) =\, \alpha w^{k-1}, \qquad d(w) =\, \beta w,$$
respectively, where $\alpha,\beta$, $k > 0$ are the Compound TCP parameters defined earlier.
 
\subsection{Queue dynamics}
We now outline a model for the evolution of the queue size. Let the $X^N(t) = W^N(t)/\tau$ be an approximation of the rate at which packets arrive at the queue at time $t$, and let the average arrival be $x(t) = X^N(t)/N$. Recall that, the average arrival rate can also be written in terms of the average window size as $x(t) = w(t)/\tau$. If packets are dropped with a probability $p(\cdot)$, then they would be queued with a probability $1-p(\cdot)$. Thus, the total arrival at the queue can be written as $\big(1-p(t)\big)\delta Nx(t)$. Next, if $\tilde{C}$ be the per-flow service capacity, the packets are served at a rate of $C = N\tilde{C}$. Then, in the time interval $(t,t+\delta)$, the queue size $Q^N(t)$ evolves according to~\cite{raina2005buffer}
\begin{align}
 Q^N(t+\delta) \approx \Big[Q^N(t) + \big(1-p(t)\big)\delta Nx(t) - \delta N \tilde{C}\Big]^B_0,\label{eq:total_queue_evolution}
\end{align}
where the notation $[q]^b_0=\min\big(\max(q,0),b\big)$ is used, and $B$ represents the size of the router buffer.

Now, consider a regime where the buffer is sized as per the bandwidth-delay product rule, and the router deploys an AQM policy.  Then, the scaled queue size $q(t) = Q^N(t)/N$ follows the following differential equation
\begin{align}
 \dot{q}(t) = \vast\{\begin{array}{l r}
             \big(1-p(t)\big)\frac{w(t)}{\tau} - \tilde{C} & \text{ if } q(t) > 0,\\
             \\
             \text{max}\big\{ \big(1-p(t)\big)\frac{w(t)}{\tau} - \tilde{C},0\big\}& \text{ if } q(t) = 0.
              \end{array}\label{eq:instantaneous_queue}
\end{align}
The packet-drop probability $p(\cdot)$ is specified by the AQM deployed at the router. In equation~\eqref{eq:instantaneous_queue}, note the use of the notation $w(t)/\tau$ for the arrival rate instead of $x(t)$. This is because, the fluid model for TCP is written in terms of the average window size $w(t)$, and this variable will later be treated as a state variable of the Compound TCP-RED system. 
\subsection{Random Early Detection queue policy}
\label{sec:RED}
RED uses an exponentially weighted moving average scheme to compute the average queue size. The recursive computation is given by
\begin{align*}
\mathtt{ avg \leftarrow\, (1-w_q)\, avg + w_q q}
\end{align*}
where $q$ is the instantaneous queue size, and $0 < \mathtt{w_q} < 1$ is the queue weighting parameter. Observe that, the weight given to each sample of the queue size reduces exponentially as time progresses. Note that $\mathtt{w_q} = 1$ would mean that no averaging is performed over the queue size. 

A fluid model for the update of the average queue size was proposed in~\cite{low2002dynamics}. As per this model, the average queue size $r(t)$ is updated as
\begin{align}
 \dot{r}(t) =& -\gamma \tilde{C}\big(r(t)-q(t)\big),\label{eq:ave_queue_RED}
\end{align}
where $\gamma$ is a RED parameter, and the product $\gamma \tilde{C}$ captures the weight given to the instantaneous queue size $q(t)$. Note that, when the average queue size is updated every $1/\tilde{C}$ time instant, then $\gamma \in (0,1]$. Given the average queue length, the packet-drop probability is given by
\begin{align}
 p(t) = \Vast\{\begin{array}{l l}
                0 & r(t) \leq \underline{b},\\
                \rho r(t) - \rho \underline{b} & \underline{b}<r(t)<\overline{b},\\
                \eta r(t) - (1 - 2\overline{p}) & \overline{b} \leq r(t) < 2\overline{b},\\
                1 & r(t) \geq 2\overline{b},
               \end{array}\label{eq:marking_probability_RED}
\end{align}
where $\underline{b}$, $\overline{b}$ and $\overline{p}$ are RED parameters, namely minimum threshold, maximum threshold and maximum packet-drop probability. Using these, the other parameters are defined as
\begin{align*}
  \rho = \frac{\overline{p}}{\overline{b}-\underline{b}}, \qquad \text{and} \qquad \eta = \frac{1-\overline{p}}{\overline{b}}.
\end{align*}
The suggested values of these parameters are: $\gamma=10^{-4}, \underline{b}=50$ pkts, $\overline{b}=550$ pkts, $\overline{p}=0.1, \rho = 2\times 10^{-4}, \eta= 16.36\times 10^{-4}$.
For our analysis, we assume that the system operates in the region $\underline{b}<r(t)<\overline{b}$, adhering to the suggestion made in~\cite{low2002dynamics}. Thus, the packet-drop probability is the following affine function of the average queue length,
\begin{align}
 p(t) = \rho\big(r(t)-\underline{b}\big)\label{eq:prob_raw_RED}.
\end{align}
Using equation~\eqref{eq:prob_raw_RED} and the dynamics of the average queue size given by equation~\eqref{eq:ave_queue_RED}, the dynamics of the packet-drop probability can be written as
\begin{align}
 \dot{p}(t) =&\, -\gamma \tilde{C} \Big(p(t)+\rho \underline{b}-\rho q(t)\Big).\label{eq:prob_RED}
\end{align}

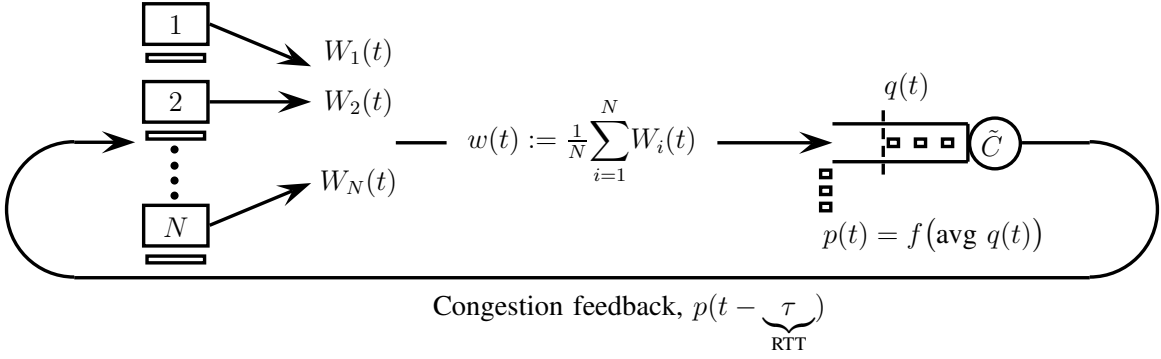
\begin{figure}[tbh]
\newcommand{\myarrowlength}{1.5}
 \newcommand{\myarrowsize}{0.08cm 5.0}
 \newcommand{\mylinewidth}{0.05}
 \begin{center}
\scalebox{0.9} % Change this value to rescale the drawing.
{
\begin{pspicture}(2.5,-3)(10.5,1.7)
\psframe[linewidth=\mylinewidth,dimen=outer](1.0,1.95)(0.0,1.3)
\psframe[linewidth=\mylinewidth,dimen=outer](0.95,1.2)(0.05,1.05)
\psframe[linewidth=\mylinewidth,dimen=outer](1.0,0.8)(0.0,0.15)
\psframe[linewidth=\mylinewidth,dimen=outer](0.95,0.05)(0.05,-0.1)
\psframe[linewidth=\mylinewidth,dimen=outer](1.0,-1.03)(0.0,-1.68)
\psframe[linewidth=\mylinewidth,dimen=outer](0.95,-1.78)(0.05,-1.93)
\psdots[dotsize=0.12](0.48,-0.23)
\psdots[dotsize=0.12](0.48,-0.47)
\psdots[dotsize=0.12](0.48,-0.69)
\psdots[dotsize=0.12](0.48,-0.91)
\rput(0.5,1.615){\large{$1$}}
\rput(0.5,0.475){\large{$2$}}
\rput(0.5,-1.385){\large{$N$}}
% % % 
\psline[linewidth=\mylinewidth,arrowsize=\myarrowsize,arrowlength=\myarrowlength,arrowinset=0.4]{->}(1.0,1.625)(2.5,1)
\rput(3.2,1.2){\large{$W_1(t)$}}
\psline[linewidth=\mylinewidth,arrowsize=\myarrowsize,arrowlength=\myarrowlength,arrowinset=0.4]{->}(1.0,0.475)(2.5,0.475)
\rput(3.2,0.46){\large{$W_2(t)$}}
\psline[linewidth=\mylinewidth,arrowsize=\myarrowsize,arrowlength=\myarrowlength,arrowinset=0.4]{->}(1.0,-1.35)(2.5,-0.725)
\rput(3.2,-0.725){\large{$W_N(t)$}}
% % % 
\psline[linewidth=\mylinewidth](10.2,0.15)(12.2,0.15)
\psline[linewidth=\mylinewidth](12.2,0.15)(12.2,-0.41)
\psline[linewidth=\mylinewidth](10.2,-0.41)(12.2,-0.41)
\psline[linestyle=dashed,linewidth=\mylinewidth](10.95,0.35)(10.95,-0.6)
\psframe[linewidth=\mylinewidth,dimen=outer](11.8,-0.05)(12,-0.21)
\psframe[linewidth=\mylinewidth,dimen=outer](11.4,-0.05)(11.6,-0.21)
\psframe[linewidth=\mylinewidth,dimen=outer](11,-0.05)(11.2,-0.21)
\pscircle[linewidth=\mylinewidth,dimen=outer](12.6,-0.14){0.4}
\rput(12.55,-0.13){\large{$\tilde{C}$}}
\rput(11.3,0.7){\large{$q(t)$}}
% % % 
\psline[linewidth=\mylinewidth](3.75,-0.13)(4.5,-0.13)
\rput(6.5,-0.1){\large{$w(t) := \frac{1}{N}\displaystyle{\sum_{i=1}^{N}}W_i(t)$}}
\psline[linewidth=\mylinewidth,arrowsize=\myarrowsize,arrowlength=\myarrowlength,arrowinset=0.4]{->}(8.5,-0.13)(10.2,-0.13)
% % % 
\psframe[linewidth=\mylinewidth,dimen=outer](10.2,-0.51)(10,-0.67)
\psframe[linewidth=\mylinewidth,dimen=outer](10.2,-0.76)(10,-0.92)
\psframe[linewidth=\mylinewidth,dimen=outer](10.2,-1)(10,-1.16)
\rput(11.7,-1.5){\large{$p(t)=f\big(\text{avg}\,\, q(t)\big)$}}
% % % 
\psline[linewidth=\mylinewidth](13,-0.13)(14,-0.13)
\psarc[linewidth=\mylinewidth](14,-1.115){1}{-90.0}{90.0}
\psline[linewidth=\mylinewidth](14,-2.125)(-1,-2.125)
\psarc[linewidth=\mylinewidth](-1,-1.115){1}{90.0}{-90.0}
\psline[linewidth=\mylinewidth,arrowsize=\myarrowsize,arrowlength=\myarrowlength,arrowinset=0.4]{->}(-1,-0.1225)(-0.1,-0.1225)
\rput(7.2,-2.8){\large{Congestion feedback, $p(t-\underbrace{\tau}_{\text{RTT}})$}}\end{pspicture} 
}
\end{center}
\caption{Schematic diagram explaining the network scenario considered for the construction of the non-linear time-delayed fluid model for Compound TCP-RED~\eqref{eq:Compound_RED}. There are $N$ Compound TCP sources each sending $W_i(t)$ packets per round-trip time $\tau$. The average of these quantities yields the average window size $w(t)$. The router deploys the RED queue management policy and serves the packets at a rate $\tilde{C}$. The packet-drop probability $p(t)$ is computed as a function of the average queue size. The TCP sources update their sending windows based on whether a packet is dropped or not, thereby using packet-drop probability $p(t-\tau)$ as a feedback signal. This feedback is considered to be time delayed owing to the RTT of the TCP flows.}
\label{fig:schematic}
\end{figure}

Combining \eqref{eq:gen_TCP}, \eqref{eq:instantaneous_queue} and \eqref{eq:prob_RED}, we get the following third order, non-linear, time delayed model for a large number of long-lived Compound TCP flows feeding into a router with the RED policy for queue management
% \begin{align}
%  \dot{w}(t) =&\, \Big(i\big(w(t)\big) - d\big(w(t)\big) p(t-\tau)\Big)\frac{w(t-\tau)}{\tau},\notag\\
%  \dot{q}(t) =&\, \big(1-p(t)\big)\frac{w(t)}{\tau}-C,\notag\\
%  \dot{p}(t) =&\, -\gamma C \Big(p(t)+\rho \underline{b}-\rho q(t)\Big).\label{eq:3rd_order}
% \end{align}
%%%%%%%%%%%%%%%%%%%%%%%%%%%%%%%%%%%%%%%%%%%%%% 1-p term %%%%%%%%%%%%%%%%%%%%%%%%%%%%%%%%%%%%%%%%%%%%%%
% \begin{small}
\begin{align}
 \dot{w}(t) =&\, \Big(i\big(w(t)\big)\big(1-p(t-\tau)\big) - d\big(w(t)\big) p(t-\tau)\Big)\frac{w(t-\tau)}{\tau},\notag\\
 \dot{q}(t) =&\, \big(1-p(t)\big)\frac{w(t)}{\tau}-\tilde{C},\notag\\
 \dot{p}(t) =&\, -\gamma \tilde{C} \Big(p(t)+\rho \underline{b}-\rho q(t)\Big).\label{eq:Compound_RED}
\end{align}
% \end{small}
A schematic diagram showing the network scenario considered and the quantities modelled is presented in Figure~\ref{fig:schematic}.
Following are some remarks about the above fluid model for the closed-loop Compound TCP-RED system~\eqref{eq:Compound_RED}:
\begin{enumerate}
 \item [(i)] The fluid model for TCP given by equation~\eqref{eq:gen_TCP} captures the behaviour in the congestion avoidance phase, and does not capture the slow-start phase of TCP. This is a common assumption in the development of various other models for TCP~\cite{poojary2017asymptotic,mathis1997macroscopic}. A few models which capture the slow-start phase, are in the form of partial differential equations, for example see~\cite{baccelli2002mean}.
 \item [(ii)] It is important to note that the TCP fluid model which we motivate and study does not have the number of users as a parameter in the model. This is a departure from previous model developments of TCP; for example see \cite{low2002dynamics,kunniyur2001analysis,hollot2001designing}, where the number of users is present as a model parameter, and also influences the stability results. The TCP model we consider is a fluid model, which rests on the assumption that there are many flows in the system, and that the network is operating in a high bandwidth-delay environment. Later in the paper, using packet-level simulations, we highlight that our model assumptions are justified. 
%  Also, the model considers only long-lived flows and assumes that the system is operating in the large bandwidth-delay product regime. This assumption is benign, as the bandwidth-delay product does get large with increasing bandwidth in current high-speed networks. 
%  \item [(ii)] The TCP fluid model we study does not incorporate the number of users $N$, which is a significant departure from~\cite{hollot2001control, low2002dynamics}. If the TCP-AQM model were to feature the parameter $N$, this parameter would naturally play a role in system stability or the loss of it. By intuition, this may not be representative of the operation of transport protocols and queue policies across the Internet. Further, it is highlighted in~\cite{misra2000fluid}, that the TCP model we study is valid when the number of users is large, and gets more accurate as $N$ increases. Also, the model considers only long-lived flows and assumes that the system is operating in the large bandwidth-delay product regime. This assumption is benign, as the bandwidth-delay product does get large with increasing bandwidth in current high-speed networks.  
%  \item [(iii)] The window update functions for Compound TCP, $i(w)$ and $d(w)$, are derived for the regime where the number of backlogged packets per connection is less than the defined threshold, \emph{i.e.} $diff < \tilde{\gamma}$. We will show, in Section~\ref{sec:sims}, that this usually true in the congestion avoidance phase of Compound for the default value of $\tilde{\gamma}$.
 \item [(iii)] In equation~\eqref{eq:gen_TCP}, $\tau$ represents the feedback delay, which is the average round-trip time of the TCP flows. Round-trip time is the end-to-end delay experienced by a data packet, which comprises of the queueing delay and the propagation delay. The queueing delay depends on the queue size, which is practically limited by the router buffer and is aimed to be controlled by the queue policy. Moreover, in a low-latency regime which is interesting from a networking perspective, the queueing delay would form a negligible component of the end-to-end delay. Thus, one may assume that the propagation delay is much larger than the queueing delay, and the round-trip time can then be approximated by the propagation delay alone. Such an assumption also aids analytical tractability, as it eliminates the possibility of a state-dependent delay.
%  which would naturally make the analysis much harder.
\end{enumerate}

\section{Compound TCP with RED policy}
\label{sec:CompoundTCP-RED}
The Compound TCP-RED system~\eqref{eq:Compound_RED} is a non-linear, time-delayed model. For such non-linear systems, it is natural to start with analysing the system stability in a local neighbourhood of the equilibrium. Therefore, we consider a linear approximation of the above model, and derive some conditions for local stability of the system about its equilibrium.
% point. 

Let $(w^\ast,q^\ast,p^\ast)$ denote a non-trivial equilibrium of system~\eqref{eq:Compound_RED}. Then, the equilibrium satisfies
\begin{align}
i(w^\ast)(1-p^\ast)=& \,d(w^\ast)p^\ast, & q^\ast =& \,\frac{p^\ast}{\rho}+\underline{b}, & w^\ast(1-p^\ast) =& \,\tilde{C}\tau.\label{eq:Compound_RED_equilibrium}
\end{align}
Consider the perturbations $u_1(t)=w(t)-w^\ast$, $u_2(t) = q(t)-q^\ast$ and $u_3(t) = p(t) - p^\ast$. Linearising system~\eqref{eq:Compound_RED} about the non-trivial equilibrium yields
%  \begin{align}
%   \dot{u}_1(t) =&\, \big(i'(w^\ast) - d'(w^\ast)p^\ast\big)\frac{w^\ast}{\tau}u_{1}(t) - d(w^\ast)\frac{w^\ast}{\tau}u_3(t-\tau),\notag\\
%   \dot{u}_2(t) =&\,  \frac{(1-p^\ast)}{\tau}u_1(t)-\frac{w^\ast}{\tau}u_3(t),\notag\\
%   \dot{u}_3(t) =&\,\rho\gamma C u_2(t) -\gamma C u_3(t),\label{eq:3rd_order_lin}
%  \end{align}
 %%%%%%%%%%%%%%%%%%%%%%%%%%%%%%%%%%%%%%%%%%%%%% 1-p term %%%%%%%%%%%%%%%%%%%%%%%%%%%%%%%%%%%%%%%%%%%%%%
  \begin{align}
  \dot{u}_1(t) =&\, \big(i'(w^\ast)(1-p^\ast) - d'(w^\ast)p^\ast\big)\frac{w^\ast}{\tau}u_{1}(t) - \big(i(w^\ast) + d(w^\ast)\big)\frac{w^\ast}{\tau}u_3(t-\tau),\notag\\
  \dot{u}_2(t) =&\,  (1-p^\ast)\frac{1}{\tau}u_1(t)-\frac{w^\ast}{\tau}u_3(t),\notag\\
  \dot{u}_3(t) =&\,\rho\gamma \tilde{C} u_2(t) -\gamma \tilde{C} u_3(t),\label{eq:Compound_RED_lin}
 \end{align}
 where $$i'(w^\ast)=\frac{\mathrm{d}}{\mathrm{d}w}i(w)\bigg|_{w=w^\ast},\,\,\,\,\,\,\text{and}\,\,\,\,\,d'(w^\ast)~=~\frac{\mathrm{d}}{\mathrm{d}w}d(w)\bigg|_{w=w^\ast}.$$
\hspace{-1.1mm}Looking for exponential solutions of \eqref{eq:Compound_RED_lin}, we get the following characteristic equation
\begin{align}
\lambda^3 + a_1 \lambda^2 + a_2 \lambda + a_3 + a_4 e^{-\lambda\tau} = 0,\label{eq:Compound_RED_char}
\end{align}
with
% \begin{align}
%  a_1 =&\, \gamma C - \Big(i'(w^\ast)-d'(w^\ast)p^\ast\Big)\frac{w^\ast}{\tau} > 0,\notag\\
%  a_2 =&\, \gamma C \Big(\rho-\big(i'(w^\ast) -d'(w^\ast)p^\ast\big)\Big)\frac{w^\ast}{\tau} > 0,\notag\\
%  a_3 =&\, - \rho\gamma C\Big(i'(w^\ast)-d'(w^\ast)p^\ast\Big)\left(\frac{w^{\ast}}{\tau}\right)^{2} > 0,\notag\\
%  a_4 =&\, \rho\gamma C\,d(w^\ast)(1-p^\ast)\frac{w^\ast}{\tau^2} > 0.\label{eq:abc3}
% \end{align}
 %%%%%%%%%%%%%%%%%%%%%%%%%%%%%%%%%%%%%%%%%%%%%% 1-p term %%%%%%%%%%%%%%%%%%%%%%%%%%%%%%%%%%%%%%%%%%%%%%
 \begin{align}
 a_1 =&\, \gamma \tilde{C} - \Big(i'(w^\ast)\big(1-p^\ast\big)-d'(w^\ast)p^\ast\Big)\frac{w^\ast}{\tau} =\, \gamma \tilde{C}+(2-k)i(w^\ast)(1-p^\ast)\frac{1}{\tau} >0,\notag\\
 a_2 =&\, \gamma \tilde{C} \Big(\rho-\big(i'(w^\ast)\big(1-p^\ast\big) -d'(w^\ast)p^\ast\big)\Big)\frac{w^\ast}{\tau} =\,\gamma \tilde{C} \Big(\rho w^\ast + (2-k)i(w^\ast)(1-p^\ast)\Big)\frac{1}{\tau} > 0,\notag\\
 a_3 =&\, - \rho\gamma \tilde{C}\Big(i'(w^\ast)\big(1-p^\ast\big)-d'(w^\ast)p^\ast\Big)\left(\frac{w^{\ast}}{\tau}\right)^{2} =\, \rho\gamma \tilde{C}^2 (2-k)i(w^\ast)\frac{1}{\tau} > 0,\notag\\
 a_4 =&\, \rho\gamma \tilde{C}\,\big(i(w^\ast)+d(w^\ast)\big)(1-p^\ast)\frac{w^\ast}{\tau^2} =\, \rho\gamma \tilde{C}^2 i(w^\ast)\frac{1}{p^\ast \tau} > 0.\label{eq:Compound_RED_abc3}
\end{align}
The positivity of the coefficients can be verified by substituting the functions $i(w), d(w)$. 
From the characteristic equation \eqref{eq:Compound_RED_char}, we may derive the loop transfer function for system in \eqref{eq:Compound_RED_lin} as 
\begin{align}
 L(\lambda) =\, \frac{a_4e^{-\lambda\tau}}{\lambda^3 + a_1\lambda^2 + a_2\lambda + a_3}.\label{eq:Compound_RED_loop_transfer}
\end{align}
It can be verified, using the Routh stability criterion~\cite{Franklin1995}, that the loop transfer function does not have any poles in the right half of the complex plane, 
and is hence stable. We first seek the cross-over frequency, i.e. the frequency for which the loop transfer has a phase of $\pi$. This cross-over frequency, denoted 
as $\omega_c$, satisfies
\begin{align}
 \tan(\omega_c\tau) =&\, \frac{\omega_c(a_2-\omega_c^2)}{a_1\omega_c^2-a_3}.
\end{align}
According to the Nyquist stability criterion, the system is stable if $|L(j\omega_c)| < 1$~\cite{Franklin1995}. 
This gives us the following sufficient condition for stability
\begin{align}
 \frac{a_4}{|a_2 \omega_c - \omega_c^3|}|\sin(\omega_c\tau)| < 1.\label{eq:Compound_RED_suff}
\end{align}
Substituting $a_2$ and $a_4$ from~\eqref{eq:Compound_RED_abc3}, we obtain
\begin{align}
 \frac{\rho\,\gamma\,\tilde{C}\,i(w^\ast)/(p^\ast\tau)}{\frac{\gamma \tilde{C}}{\tau}\big(\rho w^\ast + (2-k)i(w^\ast)(1-p^\ast)\big)\omega_c-\omega_c^3} \sin(\omega_c\tau)< 1.\label{eq:Compound_RED_suff_intermediate}
\end{align}
Condition~\eqref{eq:Compound_RED_suff_intermediate} is a sufficient condition for local stability for the Compound TCP-RED system~\eqref{eq:Compound_RED}. Observe that the function on the LHS of this condition is dependent on the system equilibrium, which in turn has a non-linear relationship with the system parameters and the feedback delay. Moreover, this function also depends on the cross-over frequency $\omega_c$ which depends on system parameters and equilibrium. It is rather difficult to obtain a qualitative understanding of the interplay of various system parameters from the above condition.
Therefore, we now simplify it using some approximations and the equilibrium conditions given by~\eqref{eq:Compound_RED_equilibrium}. Though such simplifications yield constrained conditions, they may aid qualitative understanding of the system stability.

The LHS of~\eqref{eq:Compound_RED_suff_intermediate} attains the maximum when $\sin(\omega_c\tau) = 1$, and this implies that $\omega_c\tau = \pi/2$.  Using this argument, substituting the functions $i(w^\ast)$ and $d(w^\ast)$ and simplifying~\eqref{eq:Compound_RED_suff_intermediate}, we obtain
% \begin{small}
\begin{align}
 \frac{\rho\,\gamma\,\alpha\, (w^\ast)^{k}\, \tilde{C}\,\tau/p^\ast}{\bigg(\gamma\Big(\rho (w^\ast)^2+(2-k)\beta (w^\ast)^2 p^\ast\Big)-\pi^2/4(1-p^\ast)\bigg)} < \frac{\pi}{2}.\label{eq:Compound_RED_suff_final}
\end{align}
We now make the following observations from condition~\eqref{eq:Compound_RED_suff_final}:
\begin{enumerate}
 \item [(i)] Stability is sensitive to the round-trip time $\tau$.
 \item [(ii)] Compound parameter $\alpha$ and RED parameter $\rho$ could be tuned to ensure stability.
 \item [(iii)] Queue weighting parameter $\gamma$, has a significant impact on system stability.
\end{enumerate}
The sufficient condition~\eqref{eq:Compound_RED_suff_final} helps tune TCP and AQM parameters such that local stability is ensured, and hence aids design. However, such a condition is restrictive as it characterises only a subset of the stable region in the parameter space. In order to characterise the entire stable region, one would have to study stability crossing curves which mark the edge of the stable region, at which the system transits from a locally stable to a locally unstable regime.

It can be verified that, in the absence of feedback delay, characteristic equation~\eqref{eq:Compound_RED_char} has all its roots in the left half of the Argand plane. Time-delayed systems are prone to instability as feedback delay increases~\cite{Franklin1995}. Therefore, as the round-trip time increases, at least one pair of characteristic roots would cross over the imaginary axis into the right Argand plane, and the system would become unstable. Such a topological change in system dynamics is termed as bifurcation. Further, if such a topological change occurs when a pair of complex conjugate roots of the characteristic equation crosses over the imaginary axis, the system is said to undergo a Hopf bifurcation~\cite{hassard1981theory}. In non-linear systems, occurrence of a Hopf bifurcation indicates the emergence of a limit cycle from an equilibrium. 
For details of Hopf bifurcation and its types, the reader is referred to~\cite{hassard1981theory,kuznetsov2013elements}.
We now analytically establish that Compound TCP-RED transits into instability via a Hopf bifurcation as system parameters are varied. We also numerically characterise the Hopf condition, which marks the boundary of the stable region, in terms of various system parameters.

In order to establish the existence of a Hopf bifurcation, we must first choose a bifurcation parameter that can be varied to study the change in system dynamics. As seen from the sufficient condition~\eqref{eq:Compound_RED_suff_final}, numerous system parameters affect system stability and can hence act as the bifurcation parameter. However, variation in any of the system parameters would affect the equilibrium, thus making it rather cumbersome to study the topological change in the system dynamics. Further, varying any of these parameters might affect other system parameters as well. Therefore, we introduce an exogenous non-dimensional parameter $\kappa$ which may act as the bifurcation parameter. Introducing this parameter, system~\eqref{eq:Compound_RED} becomes
\begin{align}
  \dot{w}(t) =&\, \kappa\Big(i\big(w(t)\big)\big(1-p(t-\tau)\big) - d\big(w(t)\big) p(t-\tau)\Big)\frac{w(t-\tau)}{\tau},\notag\\
 \dot{q}(t) =&\, \kappa\Big(\big(1-p(t)\big)\frac{w(t)}{\tau}-\tilde{C}\Big),\notag\\
 \dot{p}(t) =&\, -\kappa \gamma \tilde{C} \Big(p(t)+\rho \underline{b}-\rho q(t)\Big).\label{eq:Compound_RED_kappa}
\end{align}
Note that, when $\kappa = 1$, system~\eqref{eq:Compound_RED_kappa} reduces to the original system~\eqref{eq:Compound_RED}. Also, it is evident that $\kappa$ does not impact the system equilibrium. In order to use $\kappa$ as the bifurcation parameter, we tune the system parameters such that the roots of the characteristic equation cross over the imaginary axis at $\kappa=1$. With such a design of system parameters, the system is at the edge of the stability boundary at $\kappa=1$. We then marginally increase $\kappa$ to drive the system into a locally unstable state, and study the system dynamics. In essence, the parameter $\kappa$ aids analytical tractability and enables us to establish the occurrence of a Hopf bifurcation in the Compound TCP-RED system. We now prove that system~\eqref{eq:Compound_RED_kappa} loses stability via a Hopf bifurcation as $\kappa$ is varied.

 Linearising system~\eqref{eq:Compound_RED_kappa} as before, and looking for exponential solutions yields the following characteristic equation
\begin{align}
\lambda^3 + \kappa a_1 \lambda^2 + \kappa^2 a_2 \lambda + \kappa^3 a_3 + \kappa^3 a_4 e^{-\lambda\tau} = 0,\label{eq:Compound_RED_char_kappa}
\end{align}
where $a_1$, $a_2$, $a_3$ and $a_4$ are as defined in~\eqref{eq:Compound_RED_abc3}. To characterise the stability crossing curves, \emph{i.e.}, the condition at which~\eqref{eq:Compound_RED_char_kappa} has purely imaginary roots, we substitute $\lambda = j\omega$ in~\eqref{eq:Compound_RED_char_kappa} and separate the real and imaginary parts to obtain
\begin{align}
 \kappa^2 a_2 \omega - \omega^3 =&\, \kappa^3 a_4\sin(\omega\tau),&
 \kappa a_1\omega^2-\kappa^3 a_3 =&\, \kappa^3 a_4\cos(\omega\tau).\label{eq:Compound_RED_real_imag3}
\end{align}
Upon squaring and adding the equations in~\eqref{eq:Compound_RED_real_imag3}, we obtain
\begin{align}
 \omega^6 + \kappa^2\omega^4(a_1^2-2a_2)+\kappa^4\omega^2(a_2^2-2a_1a_3)+&\kappa^6(a_3^2-a_4^2)=0,\label{eq:Compound_RED_omega_equation_3}
\end{align}
whose solution gives the cross-over frequency, \emph{i.e.}, the frequency at which at least one pair of characteristic roots (roots of equation~\eqref{eq:Compound_RED_char_kappa}) crosses over the imaginary axis. Note that the system is at the edge of stability when~\eqref{eq:Compound_RED_real_imag3} is satisfied.

To establish that the system undergoes a Hopf bifurcation at the edge of stability, we need to prove the transversality condition of the Hopf spectrum which is given by~\cite{hassard1981theory} $$\text{Re}\bigg(\frac{\mathrm{d}\lambda}{\mathrm{d}\kappa}\bigg)_{\kappa=\kappa_c} \neq 0,$$ where $\kappa_c$ is the critical value of $\kappa$ that satisfies the equations in~\eqref{eq:Compound_RED_real_imag3}. Differentiating~\eqref{eq:Compound_RED_char_kappa} with respect to $\kappa$ yields
\begin{align}
\frac{\mathrm{d}\lambda}{\mathrm{d}\kappa} = \frac{-a_1\lambda^2-2\kappa a_2\lambda-3\kappa^2 a_3 - 3\kappa^2 a_4 e^{-\lambda\tau}}{3\lambda^2 + 2\kappa a_1\lambda+\kappa^2 a_2-\kappa^3a_4\tau e^{-\lambda\tau}}.\label{eq:Compound_RED_3rd_order_transversality_intermediate}
\end{align}
Let $\mathcal{N}$ and $\mathcal{D}$ denote the numerator and denominator of the RHS of equation~\eqref{eq:Compound_RED_3rd_order_transversality_intermediate} respectively. Then
\begin{align*}
 \text{sign}\bigg(\text{Re}\bigg(\frac{\mathrm{d}\lambda}{\mathrm{d}\kappa}\bigg)\bigg) = \text{sign}\Big(\text{Re}(\mathcal{N})\text{Re}(\mathcal{D})+\text{Im}(\mathcal{N})\text{Im}(\mathcal{D})\Big).
\end{align*}
Thus, to prove the transversality condition, we may show that, at $\kappa=\kappa_c$, $\text{Re}(\mathcal{N})\text{Re}(\mathcal{D})+\text{Im}(\mathcal{N})\text{Im}(\mathcal{D}) > 0.$ Substituting $\lambda = j\omega$ in the RHS of~\eqref{eq:Compound_RED_3rd_order_transversality_intermediate}, and simplifying yields
\begin{align}
 \text{Re}(\mathcal{N})\text{Re}(\mathcal{D})+\text{Im}(\mathcal{N})\text{Im}(\mathcal{D}) =&\, \frac{6\omega^6\tau}{\kappa_c}+2\kappa_c\omega^4\tau(a_1^2-2a_2) + \kappa_c^3\omega^2\tau(a_2^2-2a_1a_3).\label{eq:Compound_RED_3rd_order_transversality_simple}
\end{align}
The RHS of~\eqref{eq:Compound_RED_3rd_order_transversality_simple} can be simplified using~\eqref{eq:Compound_RED_omega_equation_3} to obtain
\begin{align*}
 \Big(5\omega^6+\kappa_c^2\omega^4(a_1^2-2a_2)-\kappa_c^6(a_3^2-a_4^2)\Big)\frac{\tau}{\kappa_c}.
\end{align*}
Let us now examine the signs of $(a_1^2-2a_2)$ and $(a_3^2-a_4^2)$. Consider $a_1^2-2a_2$, using the definitions in~\eqref{eq:Compound_RED_abc3}, 
\begin{align*}
%  =&\bigg(\gamma C+(2-k)i(w^\ast)(1-p^\ast)\frac{1}{\tau}\bigg)^2 - 2\gamma C \Big(\rho w^\ast + (2-k)i(w^\ast)(1-p^\ast)\Big)\frac{1}{\tau}\notag\\
%  =&\,\gamma^2C^2 + (2-k)^2\beta^2 (p^\ast)^2 \frac{(w^\ast)^2}{\tau^2} - 2\gamma\rho(1-p^\ast)\frac{(w^\ast)^2}{\tau^2}\notag\\
a_1^2-2a_2 =&\,\gamma^2\tilde{C}^2 + 2\gamma\rho p^\ast \frac{(w^\ast)^2}{\tau^2} + \bigg((2-k)^2\beta^2 (p^\ast)^2-2\gamma\rho\bigg)\frac{(w^\ast)^2}{\tau^2}.
\end{align*}
It can be shown that for the permissible ranges of Compound TCP and RED parameters $(2-k)^2\beta^2 (p^\ast)^2-2\gamma\rho > 0$, which implies $a_1^2-2a_2 > 0$.
Now let us deduce the sign of $a_3^2 - a_4^2$. Note that
$
 \text{sign}(a_3^2 - a_4^2) = \text{sign}(a_3 - a_4).
$
Using the definitions in~\eqref{eq:Compound_RED_abc3}
\begin{align*}
 a_3-a_4 =&\,\rho\gamma \tilde{C}^2 (2-k)i(w^\ast)\frac{1}{\tau} - \rho\gamma \tilde{C}^2 i(w^\ast)\frac{1}{p^\ast \tau}=\, \rho\gamma \tilde{C}\beta\Big((2-k)p^\ast -1\Big)\frac{(w^\ast)^2}{\tau^2}.
\end{align*}
Using the default values of the parameters, we may say that $(2-k)p^\ast \ll 1$ for the permissible ranges of these parameters. Therefore, $a_3 - a_4 < 0$. Given that $a_1^2 -2a_2~>~0$ and $a_3^2 - a_4^2~<~0$, 
equation~\eqref{eq:Compound_RED_3rd_order_transversality_simple} yields $$\text{Re}(\mathcal{N})\text{Re}(\mathcal{D})+\text{Im}(\mathcal{N})\text{Im}(\mathcal{D}) > 0,$$ which implies that $\text{Re}(\mathrm{d}\lambda/\mathrm{d}\kappa)_{\kappa=\kappa_c}> 0,$ thus proving the transversality condition. Thus, the equations in~\eqref{eq:Compound_RED_real_imag3} represent the Hopf condition.
The positivity of $\text{Re}(\mathrm{d}\lambda/\mathrm{d}\kappa)$ also proves that the roots of the characteristic equation would move to the right half of the Argand plane when $\kappa > \kappa_c$. Thus, $\kappa < \kappa_c$ is the necessary and sufficient condition for local stability of Compound TCP-RED. Deriving this condition analytically is rather difficult, as obtaining a closed form expression for the cross-over frequency $\omega$, which is given by the solution of equation~\eqref{eq:Compound_RED_omega_equation_3}, is cumbersome. Thus, we proceed to numerically compute the Hopf condition in terms of various system parameters. This allows us to define a region of local stability in the parameter space, and understand the trade-offs between these parameters for stable operation.

\begin{figure}
\captionsetup[subfigure]{labelformat=empty}
  \begin{center}
  \subfloat[]{
  \psfrag{100}{\scriptsize{$100$}}
  \psfrag{300}{\scriptsize{$300$}}
  \psfrag{200}{\scriptsize{$200$}}
  \psfrag{400}{\scriptsize{$400$}}
  \psfrag{500}{\scriptsize{$500$}}
    \psfrag{10}{\scriptsize{$10$}}
  \psfrag{70}{\scriptsize{$70$}}
  \psfrag{30}{\scriptsize{$30$}}
  \psfrag{90}{\scriptsize{$90$}}
  \psfrag{50}{\scriptsize{$50$}}
  \psfrag{p1}{\scriptsize{$\times 10^{-3}$}}
  \psfrag{C}{\small{Link capacity, $\tilde{C}$}}
  \psfrag{T}{\hspace{1.5mm}\small{Round-trip time, $\tau$}}
  \psfrag{tc}{\small{Hopf condition}}
  \psfrag{St}{\small{stable region}}
  \includegraphics[width=1.5in,height=1.75in,angle=270]{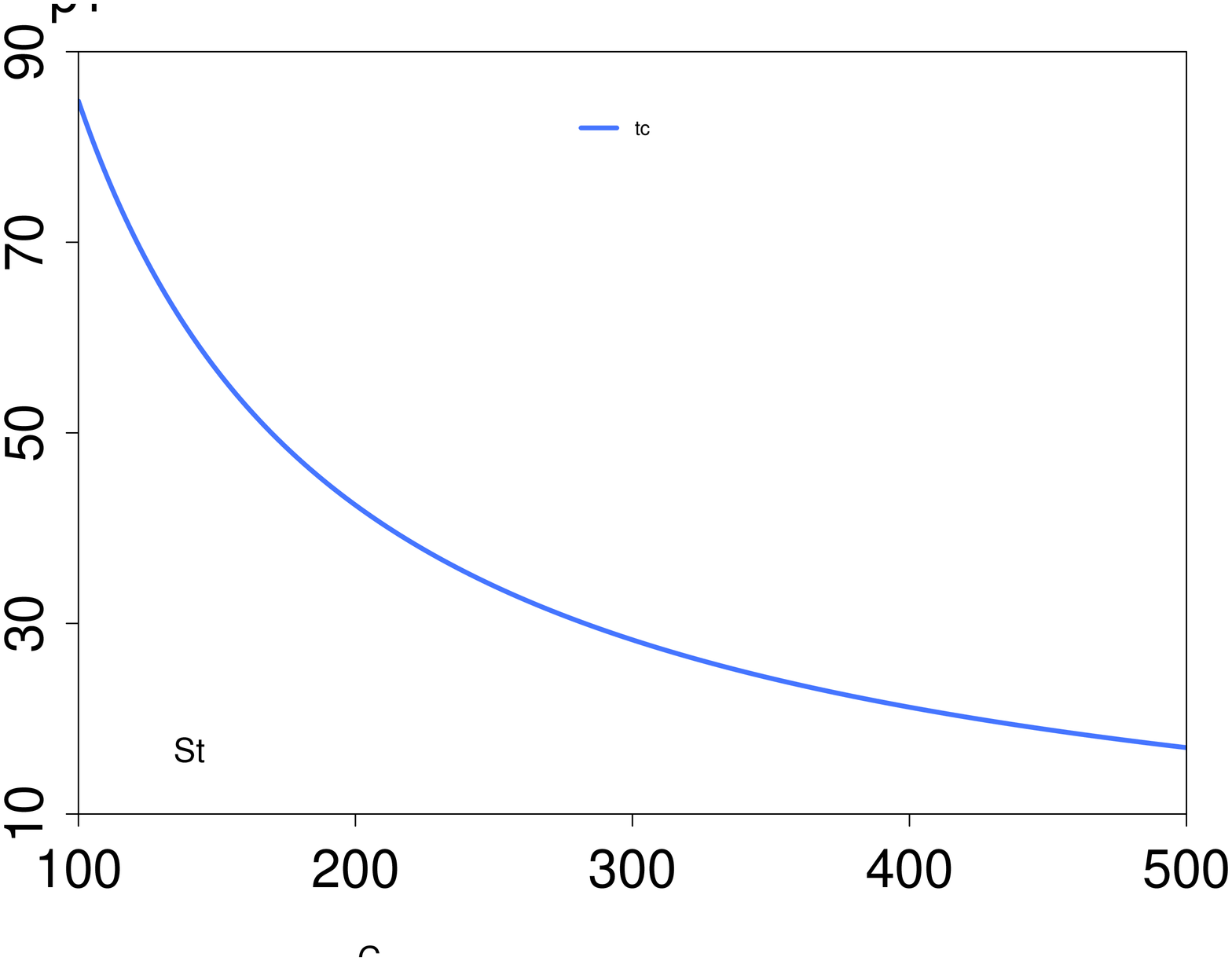}
  \label{fig:RED_stability_chart_3rd_order_C_tau}
  }
  \quad
    \subfloat[]{
  \psfrag{0.01}{\scriptsize{$0.01$}}
  \psfrag{0.03}{\scriptsize{$0.03$}}
  \psfrag{0.05}{\scriptsize{$0.05$}}
    \psfrag{90}{\scriptsize{$90$}}
  \psfrag{120}{\scriptsize{$120$}}
  \psfrag{150}{\scriptsize{$150$}}
    \psfrag{180}{\scriptsize{$180$}}
  \psfrag{210}{\scriptsize{$210$}}
  \psfrag{g}{\small{RED parameter, $\gamma$}}
  \psfrag{T}{\hspace{1.5mm}\small{Round-trip time, $\tau$}}
    \psfrag{p1}{\scriptsize{$\times 10^{-3}$}}
  \psfrag{tc}{\small{Hopf condition}}
  \psfrag{St}{\small{stable region}}
  \includegraphics[width=1.5in,height=1.75in,angle=270]{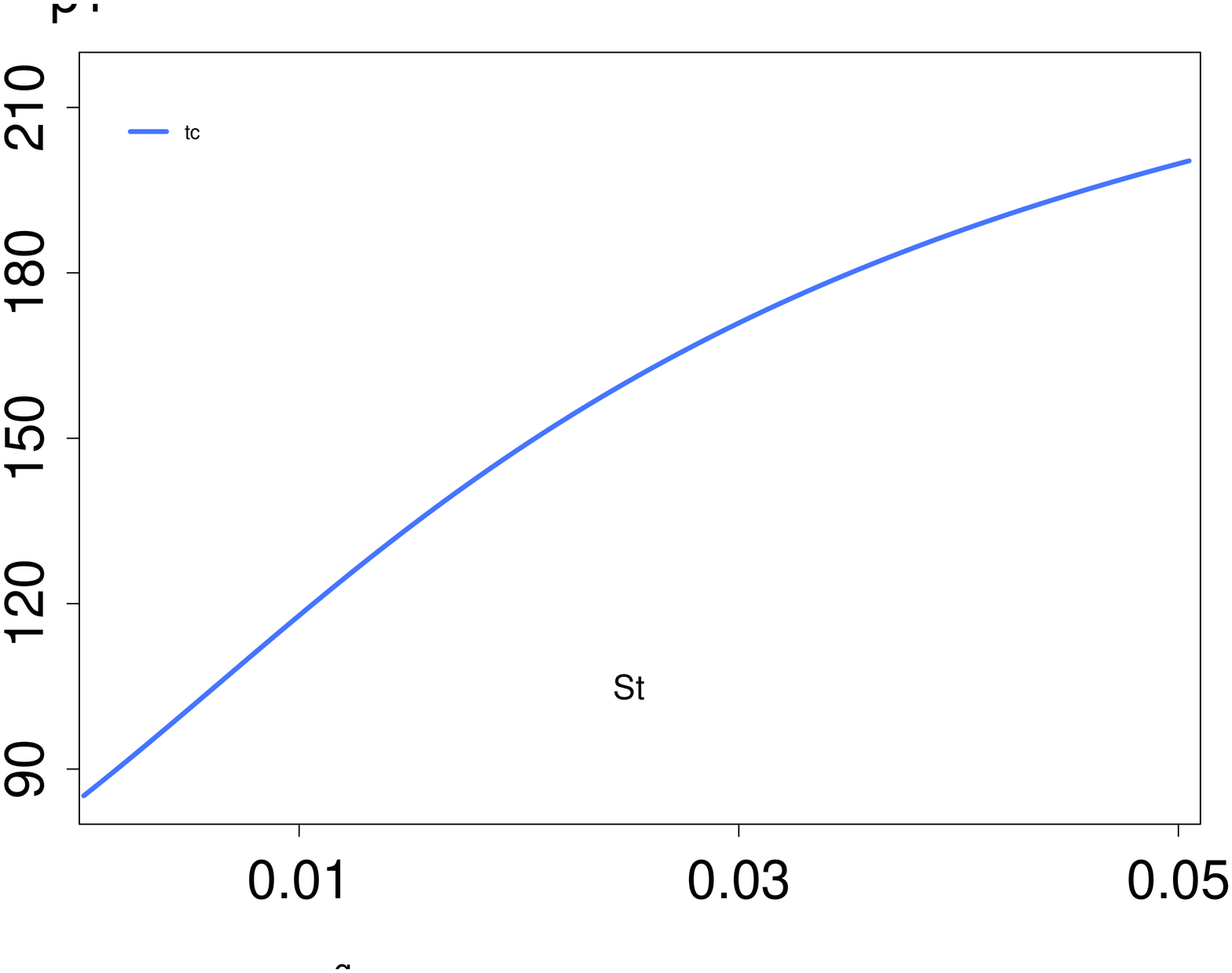}
  \label{fig:RED_stability_chart_3rd_order_gamma_tau}
  }
  \quad
\subfloat[]{
  \psfrag{0.125}{\scriptsize{$0.125$}}
  \psfrag{0.130}{\scriptsize{$0.130$}}
  \psfrag{0.135}{\scriptsize{$0.135$}}
  \psfrag{0.140}{\scriptsize{$0.140$}}
    \psfrag{50}{\scriptsize{$50$}}
  \psfrag{100}{\scriptsize{$100$}}
  \psfrag{150}{\scriptsize{$150$}}
  \psfrag{b}{\small{RED threshold, $\underline{b}$}}
  \psfrag{a}{\hspace{1mm}\small{Compound parameter, $\alpha$}}
  \psfrag{tc}{\small{Hopf condition}}
  \psfrag{St}{\small{stable region}}
  \includegraphics[width=1.5in,height=1.75in,angle=270]{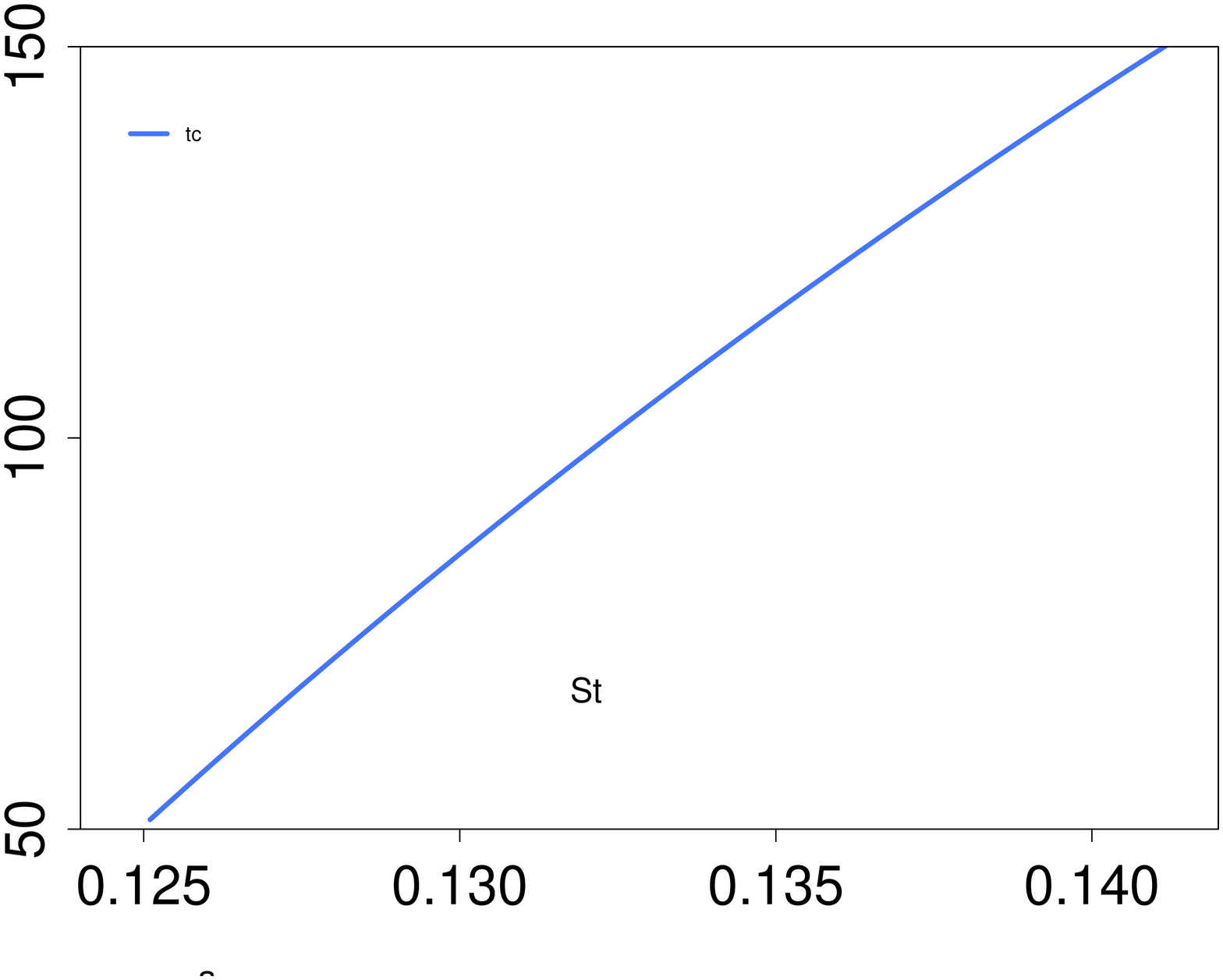}
  \label{fig:RED_stability_chart_3rd_order_blow_alpha}
  }
  \caption{Local stability charts, for Compound TCP with RED, showing the Hopf condition and the stable region in the parameter space. Observe the trade-offs between system parameters for local stability.}
  \label{fig:RED_stability_chart_3rd_order}
\end{center}
 \end{figure}
 
\subsection{Computations}
We now use DDEBIFTOOL~\cite{engelborghs2002numerical,engelborghs2001dde}, a package in the scientific computing software MATLAB, to compute the Hopf condition in terms of system parameters. We first plot the Hopf condition in terms of round-trip time $\tau$ (secs) and the per-flow link capacity $\tilde{C}$ (pkts/sec). We fix the rest of the system parameters as follows: $\alpha = 0.125, k=0.75, \beta = 0.5$ (default), $\gamma=10^{-4}, \underline{b}=50$ pkts, $\overline{b}=550$ pkts, $\overline{p}=0.1$, as suggested in~\cite{low2002dynamics}, and the exogenous parameter $\kappa=1$. We then define the range for $\tilde{C}$ as $[100,500]$ pkts/sec. DDEBIFTOOL computes the value of $\tau$ for which the system undergoes Hopf bifurcation, for each value of $\tilde{C}$ in the defined range. This plot is shown in Figure~\ref{fig:RED_stability_chart_3rd_order_C_tau}. Observe that as link capacity $\tilde{C}$ increases, round-trip time $\tau$ would have to necessarily reduce for the system to remain locally stable.
 
Recall that, sufficient condition~\eqref{eq:Compound_RED_suff_final} suggests that the RED parameter $\gamma$ which is the queue weighting parameter influences system stability. In order to examine this, we now plot the Hopf condition in terms of $\gamma$ and $\tau$. For this we fix the link capacity at $\tilde{C} = 100$ pkts/sec, the rest of the parameters are fixed as mentioned above. We define the range for parameter $\gamma$ as $[1,500]\times 10^{-4}$. The Hopf condition in terms of $\gamma$ and $\tau$ is presented in Figure~\ref{fig:RED_stability_chart_3rd_order_gamma_tau}. It is seen that as the RTT of the TCP flows increases, the value of $\gamma$ would have to be increased to ensure stability. System stability is influenced by the choice of RED parameter $\rho$ and Compound TCP parameter $\alpha$, as per the sufficient condition~\eqref{eq:Compound_RED_suff_final}. The parameter $\rho$ depends on the RED thresholds $\overline{b}$ and $\underline{b}$. In order to understand this trade-off, we plot the 
Hopf 
condition 
in terms of $\alpha$ and $\underline{b}$. To plot this we fix $\tilde{C} = 100$ pkts/sec and $\tau = 84.8\times10^{-3}$ secs, which is a point on the Hopf condition presented in Figure~\ref{fig:RED_stability_chart_3rd_order_C_tau}. The range for $\underline{b}$ is set as $[50,150]$ pkts. This plot is shown in Figure~\ref{fig:RED_stability_chart_3rd_order_blow_alpha}. RED threshold $\underline{b}$ and the Compound TCP parameter $\alpha$ are required to be co-designed to ensure stable operation. 

\begin{figure}
  \begin{center}
  \subfloat[Stable equilibrium]{
  \psfrag{0}{\scriptsize{$0$}}
  \psfrag{15}{\scriptsize{$15$}}
  \psfrag{30}{\scriptsize{$30$}}
  \psfrag{45}{\scriptsize{$45$}}
  \psfrag{60}{\scriptsize{$60$}}
  \psfrag{wt}{$w(t)$}
  \psfrag{wT}{$w(t-\tau)$}
  \includegraphics[width=1.4in,height=2in,angle=270]{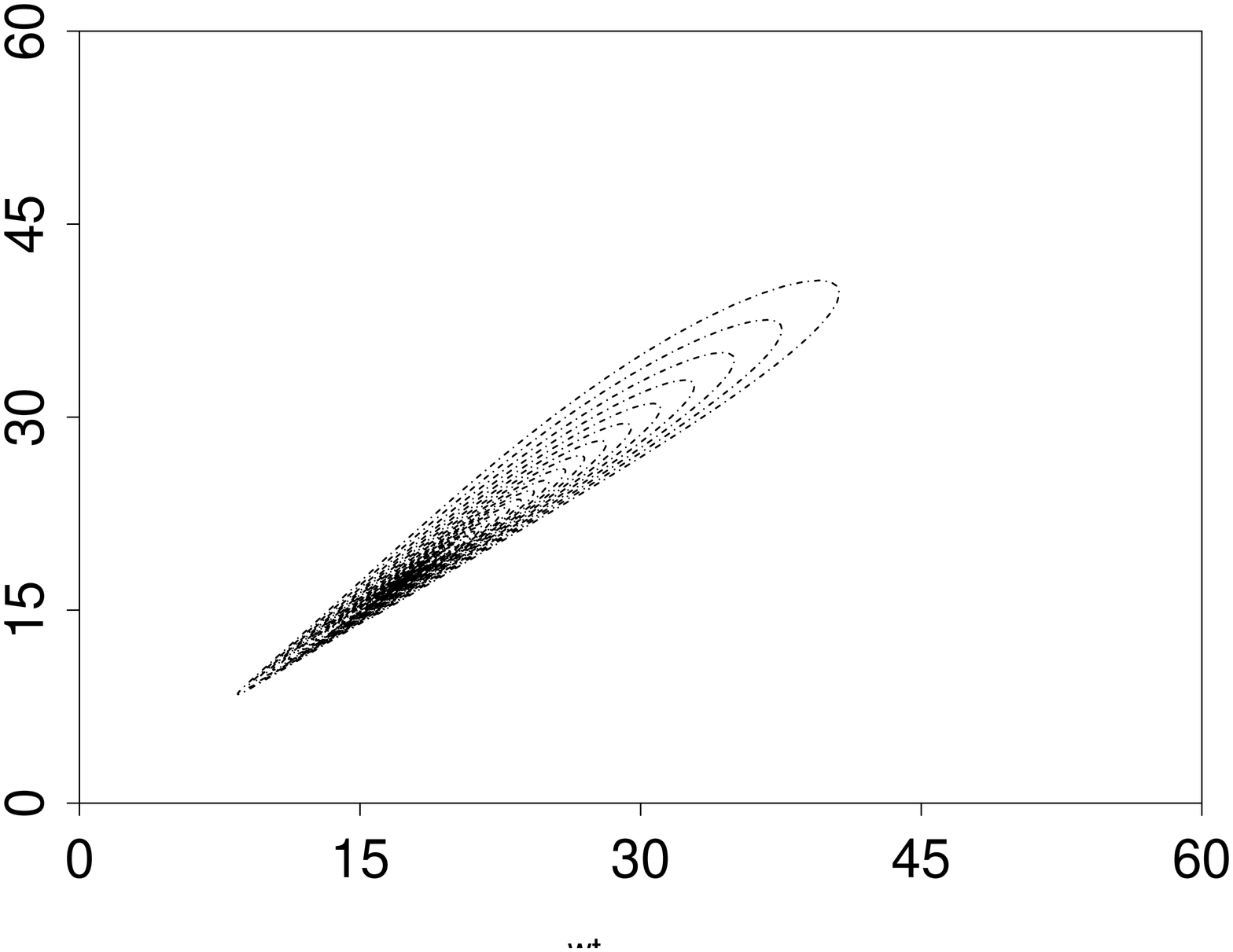}
  \label{fig:RED_convergence_3rd_order}
  }
  \qquad
  \subfloat[Limit cycle]{
  \psfrag{0}{\scriptsize{$0$}}
  \psfrag{15}{\scriptsize{$15$}}
  \psfrag{30}{\scriptsize{$30$}}
  \psfrag{45}{\scriptsize{$45$}}
  \psfrag{60}{\scriptsize{$60$}}
  \psfrag{wt}{$w(t)$}
  \psfrag{wT}{$w(t-\tau)$}
  \psfrag{e}{\scriptsize{Equilibrium}}
  \includegraphics[width=1.4in,height=2in,angle=270]{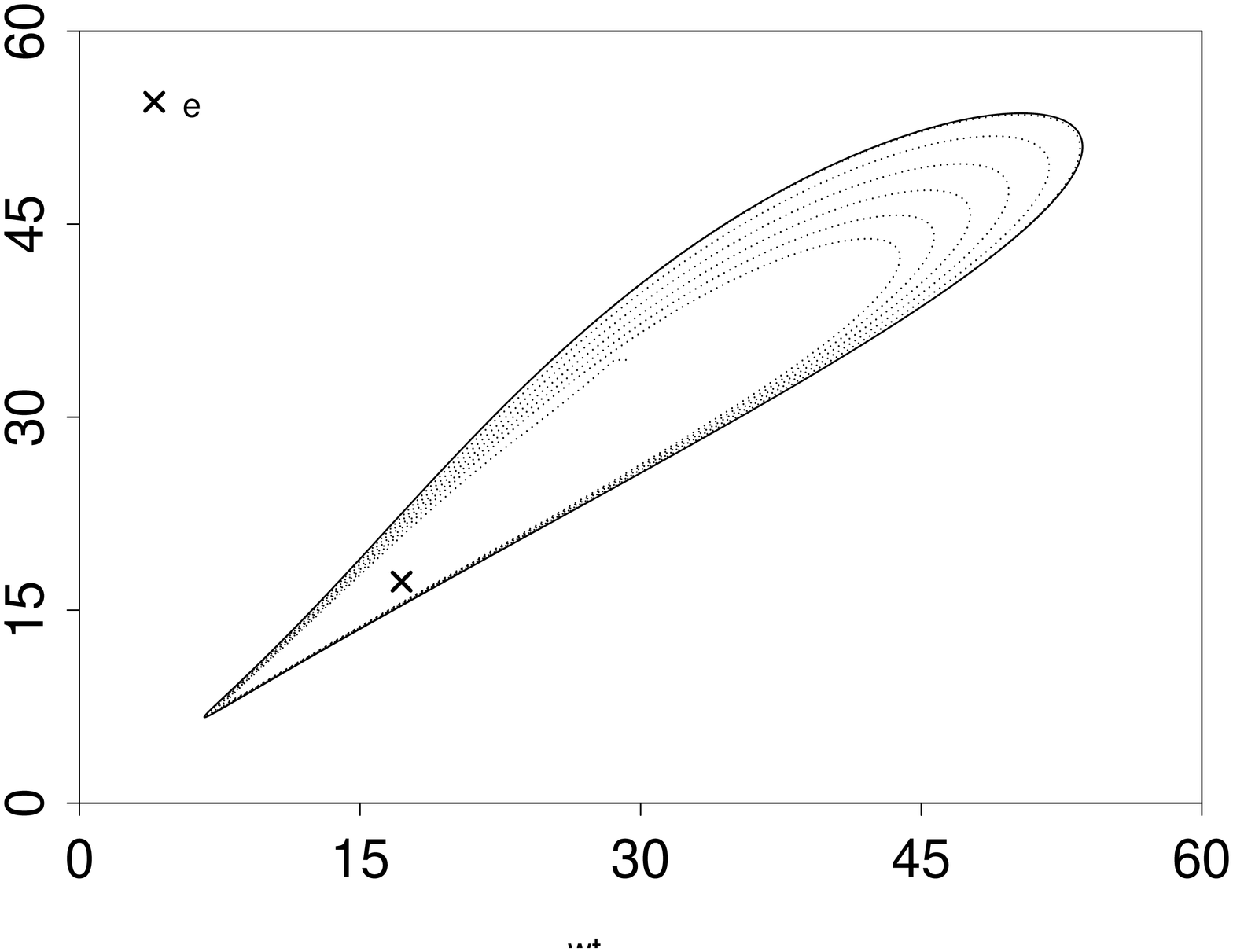}
  \label{fig:RED_limit_cycle_3rd_order}
  }
  \caption{Phase portraits for Compound TCP-RED: (a) convergence of trajectories to stable equilibrium for $\gamma = 0.032$, (b) emergence of limit cycle for $\gamma=0.028$. Notably the system undergoes a Hopf bifurcation at $\gamma = 0.03$ for a round-trip time of $\tau=171\times10^{-3}$. For a given round-trip time, smaller values of the queue weighting parameter can be destabilising.}
  \label{fig:RED_phase_portraits_3rd_order}
\end{center}
 \end{figure}

When the system parameters satisfy the Hopf condition, the system undergoes a Hopf bifurcation. Any further variation in any of the parameters pushes the system into a locally unstable. As discussed earlier, a Hopf bifurcation leads to the emergence of limit cycles. Therefore, we expect the system to undergo a topological transformation from a stable equilibrium to a limit cycle as any system parameter is varied. We now present phase portraits to exhibit this transition. Observe from the plot shown in Figure~\ref{fig:RED_stability_chart_3rd_order_gamma_tau}, that for a given round-trip time smaller values of the queue weighting parameter could render the system unstable. From this plot, we find a point on the Hopf condition, namely $\gamma=0.03,\tau=171\times10^{-3}$ secs. We first increase the value of $\gamma$ to $0.032$. The phase portrait, for the window-size dynamics, for this setting is shown in~Figure~\ref{fig:RED_convergence_3rd_order}. We see that the trajectories 
converge to a stable equilibrium $w^\ast \approx 17$ pkts. We then reduce the value of $\gamma$ to $0.028$. This operating point is located in the locally unstable region, as seen from Figure~\ref{fig:RED_stability_chart_3rd_order_gamma_tau}. As expected, we see the emergence of a limit cycle in the window-size dynamics in the phase portrait shown in Figure~\ref{fig:RED_limit_cycle_3rd_order}. Similar phase portraits can be obtained for the state variables $q(\cdot)$ and $p(\cdot)$.
\subsection{Discussions}
We have seen that local stability of Compound TCP-RED depends on network parameters (link capacity and RTT), protocol parameters ($\alpha, k, \beta$), and RED parameters: the queue weighting parameter ($\gamma$), maximum packet-drop probability ($\overline{p}$) and thresholds ($\overline{b},\underline{b}$). When the stability conditions are violated, the system loses local stability via a Hopf bifurcation, leading to the emergence of limit cycles in the system dynamics. These limit cycles could manifest as non-linear oscillations in the queue size that could cause loss of link utilisation, periodic packet loss and synchronisation of TCP flows, and could hence degrade network performance. Therefore, ensuring system stability is crucial. 

Analysis and computations suggest that large RTTs of TCP flows could be potentially destabilising unless the queue weighting parameter is set to a large enough value. In essence, large values of the queue weighting parameter aids stability. Increasing the queue weighting parameter $\gamma$ implies putting more weight on the instantaneous queue size in the averaging process. This leads us to the following question: what would happen if $\gamma$ was set to the maximum permissible value? In doing so, we would set the queue weighting parameter $w_q$ in the RED averaging algorithm (refer Section~\ref{sec:RED}) to $1$. This implies that the entire weight is put on the instantaneous queue size itself, and past samples are not considered. This would effectively degenerate to RED \emph{without} the averaging process. Our local stability analysis predicts that, under such a setting, owing to the large value of $\gamma$, the system must be stable for comparatively larger RTTs, refer to Figure~\ref{fig:RED_stability_chart_3rd_order_gamma_tau} for qualitative understanding. We proceed to analyse such a regime in the next section.

\section{Compound TCP with RED in the absence of averaging}
\label{sec:CompoundTCP-RED_no_averaging}
If averaging over the queue size is neglected, the packet-drop probability would have to be decided based on the instantaneous queue size itself. The equation for the packet-drop probability~\eqref{eq:prob_raw_RED} would then change to
\begin{align}
 p(t) =& \rho\,(q(t) - \underline{b}),\label{eq:prob_RED_nav}
\end{align}
where the average queue size is replaced by the instantaneous queue size, as queue size averaging is no longer performed.
The model for Compound TCP-RED now becomes 
\begin{align}
 \dot{w}(t) =& \kappa\Big(i\big(w(t)\big)\big(1-p(t-\tau)\big) - d\big(w(t)\big)p(t-\tau)\Big)\frac{w(t-\tau)}{\tau},\notag\\
 \dot{q}(t) =& \kappa\Big(\big(1-p(t)\big)\frac{w(t)}{\tau}-\tilde{C}\Big),\label{eq:Compound_RED_nav_eta}
\end{align}
where $i\big(w(t)\big)=\alpha w(t)^{k-1}, d\big(w(t)\big)=\beta w(t).$
We now proceed to analyse the local stability of system~\eqref{eq:Compound_RED_nav_eta}. Note the inclusion of the exogenous parameter $\kappa$, the premise remains the same as above.
We now proceed to derive the necessary and sufficient condition for local stability of system~\eqref{eq:Compound_RED_nav_eta}. Further, we establish that the system loses local stability via a Hopf bifurcation when this condition is violated.

 Equilibrium $(w^\ast,q^\ast)$ of system~\eqref{eq:Compound_RED_nav_eta} satisfies
% \begin{align*}
% w^\ast &= \frac{C\tau}{1-f(q^\ast)},\qquad \text{and}\qquad f(q^\ast) = \frac{i(w^\ast)}{d(w^\ast)}.
% \end{align*}
%%%%%%%%%%%%%%%%%%%%%%%%%%%%%%%%%%%%%%%%%%% 1- p term %%%%%%%%%%%%%%%%%%%%%%%%%%%%%%%%%%%%%%%%%%%%%%%%%%%%%%%%%%%%%%%%%%
\begin{align*}
i(w^\ast)\big(1-p^\ast\big) = d(w^\ast)p^\ast, && w^\ast\big(1-p^\ast\big) = \tilde{C}\tau,
\end{align*}
where $p^\ast$ represents the packet-drop probability at equilibrium queue size $q^\ast$, and is given by
$ p^\ast = \rho\,(q^\ast - \underline{b}).$ Consider the perturbations $u_1(t)~=~w(t) - w^\ast$ and $u_2(t)~=~q(t) - q^\ast$. Linearising system \eqref{eq:Compound_RED_nav_eta} about $(w^\ast,q^\ast)$, we obtain
\begin{align}
 \dot{u}_1(t) =& \,\kappa\big(i'(w^\ast)(1-p^\ast)-d'(w^\ast)p^\ast\big)\frac{w^\ast}{\tau} u_1(t)- \kappa \rho \big(i(w^\ast) + d(w^\ast)\big)\frac{w^\ast}{\tau}u_2(t-\tau),\notag\\
 \dot{u}_2(t) =&\, \kappa(1-p^\ast)\frac{1}{\tau} u_1(t) - \kappa \rho \frac{w^\ast}{\tau} u_2(t).\label{eq:Compound_RED_nav_lin}
\end{align}
Looking for exponential solutions of \eqref{eq:Compound_RED_nav_lin}, we get
\begin{align}
 \lambda^2 + \kappa a_1 \lambda + \kappa^2 a_2 +\kappa^2 a_3 e^{-\lambda\tau} = 0,\label{eq:Compound_RED_nav_char}
\end{align}
where
\begin{align}
a_1 =&\, \Big(\rho - \big(i'(w^\ast)(1-p^\ast) - d'(w^\ast)p^\ast\big)\Big)\frac{w^\ast}{\tau}=\, \Big(\rho w^\ast  + (2-k)i(w^\ast)(1-p^\ast)\Big)\frac{1}{\tau}>0,\notag\\
 a_2 =&\, -\rho\Big(i'(w^\ast)(1-p^\ast) - d'(w^\ast)p^\ast\Big)\bigg(\frac{w^\ast}{\tau}\bigg)^2 =\, \rho \tilde{C} (2-k)i(w^\ast)\frac{1}{\tau}>0,\notag\\
 a_3 =&\, \rho\big(i(w^\ast)+d(w^\ast)\big)(1-p^\ast)\frac{w^\ast}{\tau^2}=\, \rho \tilde{C} i(w^\ast) \frac{1}{p^\ast\tau}>0.\label{eq:Compound_RED_nav_abc}
\end{align}
If feedback is assumed to be instantaneous, the characteristic equation~\eqref{eq:Compound_RED_nav_char} reduces to a third-order polynomial in $\lambda$.
% \begin{align}
%  \lambda^2 + \kappa a_1 \lambda + \kappa^2 a_2 +\kappa^2 a_3= 0.\label{eq:Compound_RED_nav_char_no_delay}
% \end{align}
Using the Routh stability criteria~\cite{Franklin1995}, it can be shown that the roots of this polynomial equation have negative real parts. When feedback is delayed, the system is prone to instability. Therefore, as the round-trip time increases, the roots of the characteristic equation would cross over to the right half of the Argand plane. 
We now seek the condition, on the round-trip time and system parameters, at which at least one root of the characteristic equation crosses over to the right half of the 
Argand plane, and renders the system unstable. This transition would be marked by the roots crossing over the imaginary axis. Hence, we substitute $\lambda = j\omega$ to 
find the condition for the cross-over. From \eqref{eq:Compound_RED_nav_char}, we have
\begin{align*}
 -\omega^2 + j\kappa a_1 \omega + \kappa^2 a_2 + \kappa^2 a_3 \big(\cos(\omega\tau)-j\sin(\omega\tau)\big) = 0.
\end{align*}
Separating the real and imaginary parts, we get
\begin{align}
 \kappa^2 a_3 \cos(\omega\tau) =&\, \omega^2 - \kappa^2 a_2, & \kappa^2 a_3 \sin(\omega\tau) =&\, \kappa a_1\omega.\label{eq:Compound_RED_nav_real_imag}
\end{align}
Squaring and adding the equations in \eqref{eq:Compound_RED_nav_real_imag}, we get
\begin{align}
 \omega^4 + \kappa^2\omega^2 ( a_1^2 - 2 a_2) + \kappa^4(a_2^2-a_3^2) = 0.\label{eq:Compound_RED_nav_omega4}
\end{align}
Note that equation \eqref{eq:Compound_RED_nav_omega4} is a quadratic equation in $\omega^2$. Solving for $\omega^2$, we get two solutions which we denote as $\omega^2_\pm$
\begin{align}
 \omega^2_{\pm} =\, \frac{-\kappa^2(a_1^2 - 2a_2) \pm \kappa^2 \sqrt{(a_1^2-2a_2)^2 - 4(a_2^2-a_3^2)}}{2}.
\end{align}
Substituting $a_1,a_2$ and $a_3$ from~\eqref{eq:Compound_RED_nav_abc}, and simplifying the expression for $\omega^2_{\pm}$, we get
\begin{align*}
 \omega^2 =&\, \frac{\kappa^2 (w^\ast)^2}{2\tau^2}\bigg(-\big(\rho^2 + (2-k)^2\beta^2 (p^\ast)^2\big)\pm\sqrt{\big(\rho^2 -(2-k)^2\beta^2(p^\ast)^2\big)^2+4\rho^2\beta^2}\bigg).
\end{align*}
From the above, the only real value of $\omega$ is $\omega_0 = \sqrt{\omega^2_{+}}$, which is given by
\begin{align*}
 \omega_0 =&\, \kappa\, \Omega\, \frac{w^\ast}{\tau},
\end{align*}
where 
\begin{align*}
 \Omega =&\, \Bigg(\bigg(-\rho^2-(2-k)^2\beta^2 (p^\ast)^2+\sqrt{\Big(\rho^2 - (2-k)^2\beta^2(p^\ast)^2\Big)^2+\rho^2\beta^2}\bigg)\bigg/2\Bigg)^{1/2}.
\end{align*}
From \eqref{eq:Compound_RED_nav_real_imag}, we find that the cross over takes place when
\begin{align}
 \tau = \frac{1}{\omega_0}\sin^{-1}\bigg(\frac{a_1\omega_0}{\kappa_c a_3}\bigg),\label{eq:Compound_RED_nav_hopf}
\end{align}
where $\kappa_c$ is the critical value of $\kappa$ at which the characteristic equation~\eqref{eq:Compound_RED_nav_char} has a pair of imaginary roots $\lambda=j\omega_0$. Note that equation~\eqref{eq:Compound_RED_nav_hopf} defines the stability crossing curve that marks the boundary of the stable region in the parameter space.

The existence of purely imaginary roots indicates the occurrence of a Hopf bifurcation. This can be verified using the transversality condition of the 
Hopf spectrum~\cite{hassard1981theory}, \emph{i.e}. $\text{Re}(\mathrm{d}\lambda/\mathrm{d}\kappa)_{\kappa=\kappa_c} \neq 0$. Equivalently we could show that 
$\text{Re}(\mathrm{d}\lambda/\mathrm{d}\kappa)^{-1}_{\kappa=\kappa_c} > 0$. Differentiating equation \eqref{eq:Compound_RED_nav_char}, we get
\begin{align}
 \text{Re}\bigg(\frac{\mathrm{d}\lambda}{\mathrm{d}\kappa}\bigg)^{-1}_{\kappa=\kappa_c} =\, \frac{2\omega_0^2\kappa_c\tau +  \tau\kappa_c^3 (a_1^2 - 2a_2)}{4\omega_0^2\tau + \kappa_c^2 a_1^2}.
\end{align}
Using the expressions for $a_1$ and $a_2$ given in~\eqref{eq:Compound_RED_nav_abc}, we obtain
\begin{align*}
 a_1^2 - 2a_2 = \Big(\rho^2 + (2-k)^2\beta^2(p^\ast)^2\Big)\bigg(\frac{w^\ast}{\tau}\bigg)^2 > 0.
\end{align*}
Therefore, $\text{Re}(\mathrm{d}\lambda/\mathrm{d}\kappa)^{-1}_{\kappa = \kappa_c} > 0$. Thus, one could expect the emergence of limit cycles in the system dynamics when condition~\eqref{eq:Compound_RED_nav_hopf} is met.

The inequality $\text{Re}(\mathrm{d}\lambda/\mathrm{d}\kappa)^{-1}_{\kappa = \kappa_c} > 0$ implies that $\kappa < \kappa_c$ is the necessary and sufficient condition for local stability of system~\eqref{eq:Compound_RED_nav_eta}. Using the expressions in~\eqref{eq:Compound_RED_nav_abc} and the window increase and decrease functions for Compound TCP, this necessary and sufficient condition can be written as
% \begin{small}
\begin{align}
\kappa_c\frac{a_3}{\omega_0\,a_1}\sin(\omega_0\tau) <&\, 1,\notag\\
  \frac{\rho\, \alpha(w^\ast)^{(k-3)}\tilde{C}\tau}{\Omega p^\ast \big(\rho + (2-k)\beta  p^\ast\big) }\sin(\kappa_c\, w^\ast\, \Omega)<&\,1.\label{eq:Compound_RED_nav_necc}
\end{align}
System~\eqref{eq:Compound_RED_nav_eta} is locally stable as long as the necessary and sufficient condition~\eqref{eq:Compound_RED_nav_necc} is satisfied. We now make the following observations regarding the local stability of Compound TCP-RED in the absence of averaging, from condition~\eqref{eq:Compound_RED_nav_necc}:
\begin{enumerate}
 \item [(i)] Large values of $\tau$ could destabilise the system.
 \item [(ii)] Small values of AQM parameter $\rho$ could aid stability. Recall that $\rho=\overline{p}/(\overline{b}-\underline{b})$. Hence, system stability is sensitive to RED thresholds $\overline{b}$ and $\underline{b}$.
 \item [(iii)] TCP and AQM parameters need to be co-designed to ensure stable operation.
\end{enumerate}
 When the inequality in~\eqref{eq:Compound_RED_nav_necc} is replaced with an equality, we obtain the Hopf condition, which is an equivalent representation of condition~\eqref{eq:Compound_RED_nav_hopf}. This condition represents the boundary of the stable region in the parameter space. We now present some graphical representations of the Hopf condition derived above, which enable us to better understand the trade-offs in system parameters for local stability.

\begin{figure}
\captionsetup[subfigure]{labelformat=empty}
  \begin{center}
  \subfloat[]{
  \psfrag{100}{\scriptsize{$100$}}
  \psfrag{300}{\scriptsize{$300$}}
  \psfrag{200}{\scriptsize{$200$}}
  \psfrag{400}{\scriptsize{$400$}}
  \psfrag{500}{\scriptsize{$500$}}
    \psfrag{0.1}{\scriptsize{$0.1$}}
  \psfrag{0.2}{\scriptsize{$0.2$}}
  \psfrag{0.3}{\scriptsize{$0.3$}}
  \psfrag{p1}{\scriptsize{$\times 10^{-3}$}}
  \psfrag{C}{\small{Link capacity, $\tilde{C}$}}
  \psfrag{T}{\small{Round-trip time, $\tau$}}
  \psfrag{tc}{\small{Hopf condition}}
  \psfrag{St}{\small{stable region}}
  \includegraphics[width=1.4in,height=1.75in,angle=270]{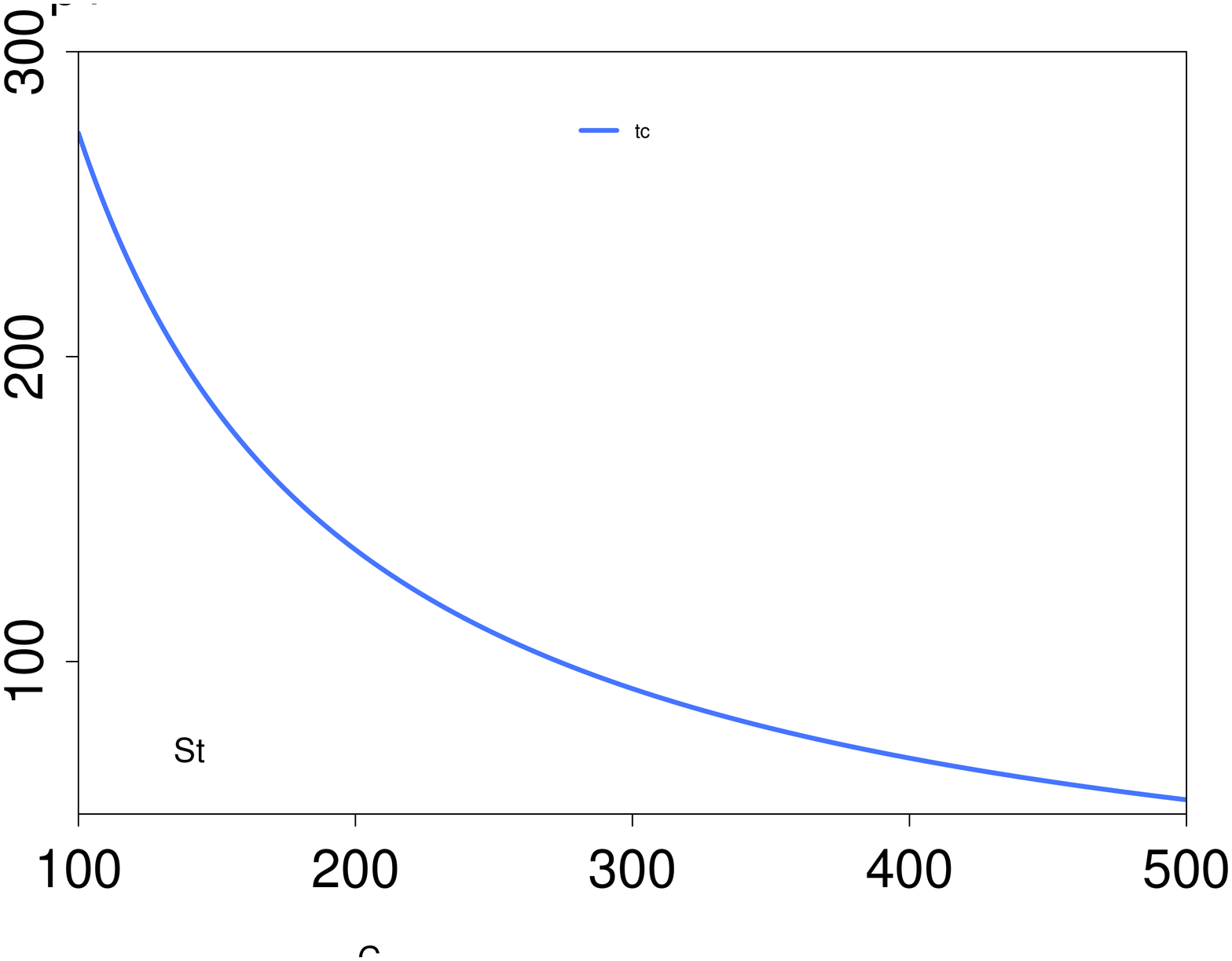}
  \label{fig:RED_stability_chart_2nd_order_C_tau}
  }
  \quad
    \subfloat[]{
  \psfrag{245}{\scriptsize{$245$}}
  \psfrag{260}{\scriptsize{$260$}}
  \psfrag{275}{\scriptsize{$275$}}
    \psfrag{50}{\scriptsize{$50$}}
  \psfrag{100}{\scriptsize{$100$}}
  \psfrag{150}{\scriptsize{$150$}}
  \psfrag{p1}{\scriptsize{$\times 10^{-3}$}}
  \psfrag{b}{\small{RED threshold, $\underline{b}$}}
  \psfrag{T}{\small{Round-trip time, $\tau$}}
  \psfrag{tc}{\small{Hopf condition}}
  \psfrag{St}{\small{stable region}}
  \includegraphics[width=1.4in,height=1.75in,angle=270]{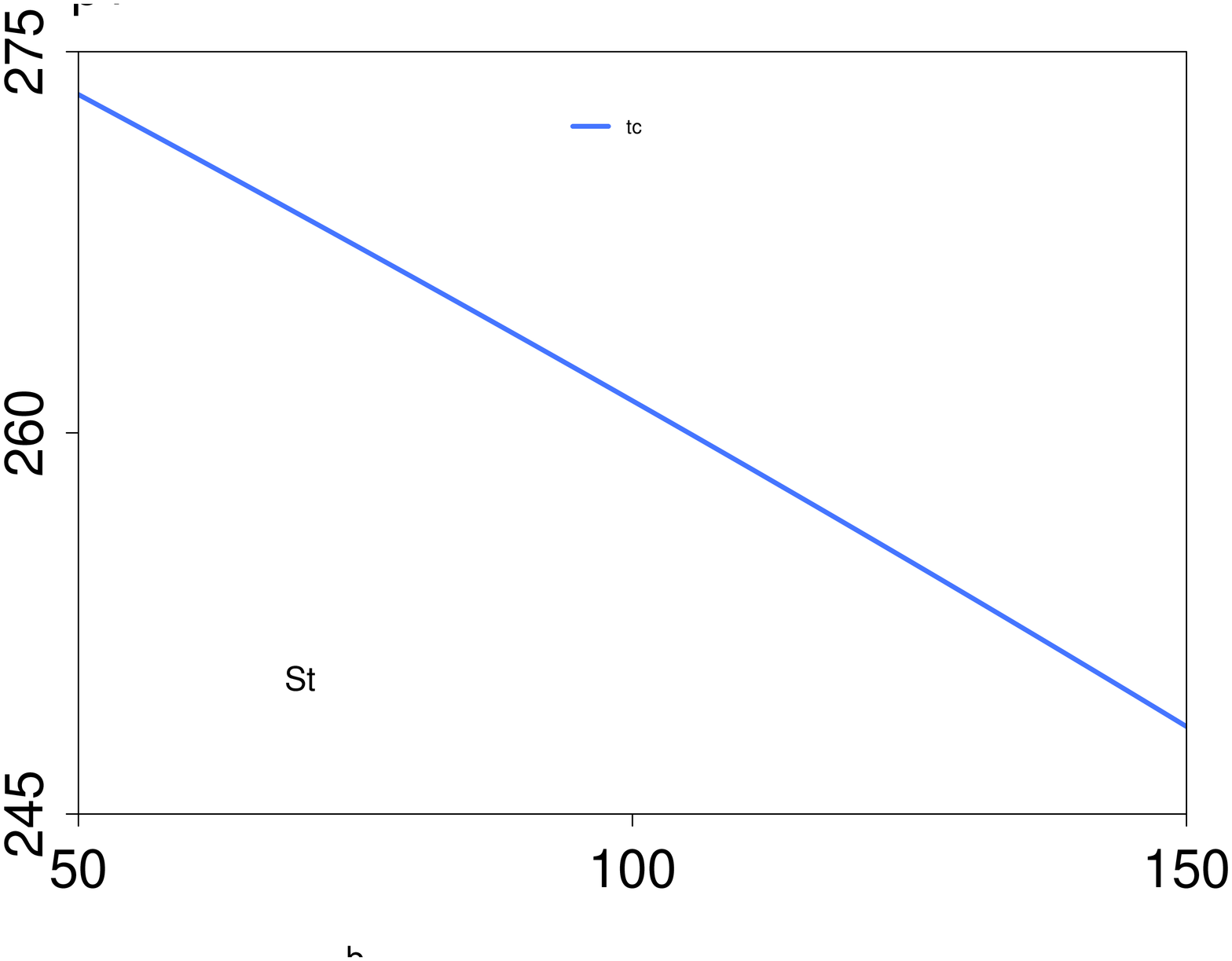}
  \label{fig:RED_stability_chart_2nd_order_blow_tau}
  }\quad
\subfloat[]{
  \psfrag{0.125}{\scriptsize{$0.125$}}
  \psfrag{0.135}{\scriptsize{$0.135$}}
  \psfrag{0.145}{\scriptsize{$0.145$}}
  \psfrag{0.155}{\scriptsize{$0.155$}}
    \psfrag{50}{\scriptsize{$50$}}
  \psfrag{100}{\scriptsize{$100$}}
  \psfrag{150}{\scriptsize{$150$}}
  \psfrag{p1}{\scriptsize{$\times 10^{-2}$}}
  \psfrag{b}{\small{RED threshold, $\underline{b}$}}
  \psfrag{a}{\small{Compound parameter, $\alpha$}}
  \psfrag{tc}{\small{Hopf condition}}
  \psfrag{St}{\small{stable region}}
  \includegraphics[width=1.4in,height=1.75in,angle=270]{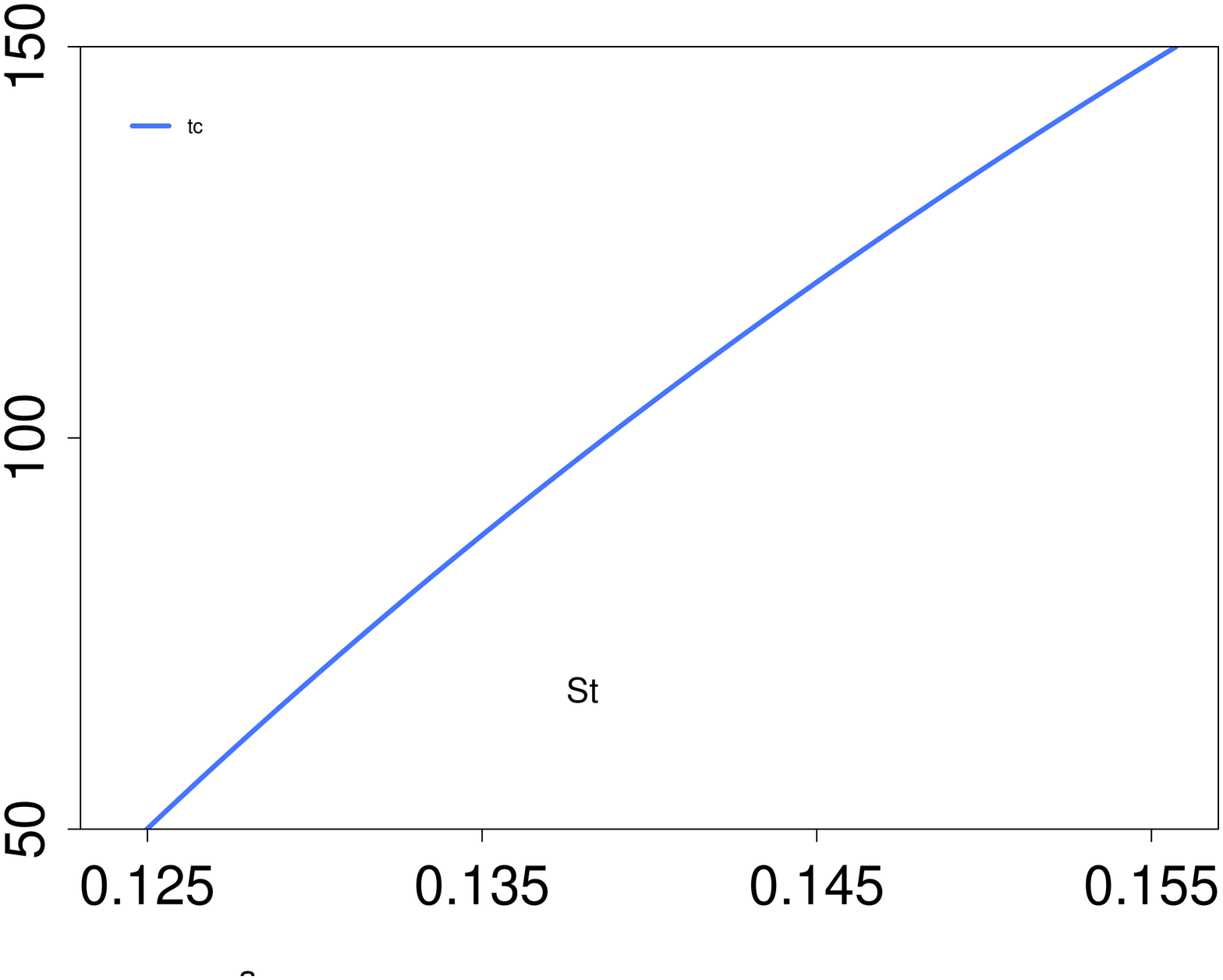}
  \label{fig:RED_stability_chart_2nd_order_blow_alpha}
  }
  \caption{Local stability charts, for Compound TCP-RED in the absence of averaging over queue size, showing the Hopf condition and the stable region in the parameter space. Observe the trade-offs between system parameters.}
  \label{fig:RED_stability_chart_2nd_order}
\end{center}
 \end{figure}

\subsection{Computations}
Condition~\eqref{eq:Compound_RED_nav_necc} indicates that the product $\tilde{C}\times\tau$ impacts system stability. In order to examine this, we first plot the Hopf condition in terms of the per-flow link capacity $\tilde{C}$ and the round-trip time $\tau$. We fix the Compound TCP parameters at their default values $(\alpha=0.125,k=0.75,\beta=0.5)$. The RED parameters are fixed as: $\gamma=~10^{-4}, \overline{b}=550$ pkts, $\underline{b}=50$ pkts and $\overline{p}=0.1$~\cite{low2002dynamics}. The exogenous parameter $\kappa$ is fixed at $1$. We vary the link capacity $\tilde{C}$ in the range $[100,500]$ pkts/sec, and for each value of $\tilde{C}$ we find the corresponding value of round-trip time $\tau$ (secs) that satisfies the Hopf condition~\eqref{eq:Compound_RED_nav_hopf}. This plot is shown in Figure~\ref{fig:RED_stability_chart_2nd_order_C_tau}. It can be seen that as $\tilde{C}$ increases, round-trip time $\tau$ would have to reduce to ensure system stability. This is similar 
to the conclusion drawn from the stability charts presented in Section~\ref{sec:CompoundTCP-RED}.  We next examine the trade-off between the round-trip time and RED threshold $\underline{b}$. For this we fix the per-flow link capacity as $\tilde{C}=100$ pkts/sec, and vary the threshold in the range $[50,150]$ pkts. Upon computing the value of $\tau$ (secs) that satisfies the Hopf condition for each value of $\underline{b}$ (pkts), we obtain the plot shown in Figure~\ref{fig:RED_stability_chart_2nd_order_blow_tau}. One may observe that as threshold $\underline{b}$ is increased, the system may become unstable for relatively smaller round-trip times. Similarly, we plot the Hopf condition in terms of Compound parameter $\alpha$ and RED threshold $\underline{b}$ (Figure~\ref{fig:RED_stability_chart_2nd_order_blow_alpha}). For this plot, we fix $\tilde{C}=100$ pkts/sec, $\tau=0.273$ secs (computed using the Hopf condition~\eqref{eq:Compound_RED_nav_hopf}, for default $\alpha$). This plot depicts the relationship 
between RED 
threshold $\underline{b}$ 
and parameter $\alpha$ for system stability.

Observe that the plot in Figure~\ref{fig:RED_stability_chart_2nd_order_C_tau} is qualitatively similar to the one presented in Figure~\ref{fig:RED_stability_chart_3rd_order_C_tau} for Compound TCP-RED system with averaging. By comparing the two plots, one may observe that for any given value of link capacity $\tilde{C}$, the system remains locally stable for comparatively larger RTTs when averaging is not performed. This indicates that averaging over queue size may not be beneficial to system stability.

\subsection{Remarks}
\label{sec:key_insights_for_validation}
Through the analysis and computations described in Sections~\ref{sec:CompoundTCP-RED} and \ref{sec:CompoundTCP-RED_no_averaging}, we have gained some insight regarding the stability of Compound TCP-RED. The key inferences are:
\begin{itemize}
 \item [(i)] Large round-trip time can destabilise the system.
 \item [(ii)] Without queue size averaging, instability sets in for comparatively larger round-trip time.
 \item [(iii)] System stability is sensitive to packet-dropping thresholds.
\end{itemize}
We have also established that, loss of stability occurs via a Hopf bifurcation in both the regimes. A detailed Hopf bifurcation analysis that gives the analytical framework to characterise the type of the Hopf bifurcation and determine the asymptotic orbital stability of the emergent limit cycles in system~\eqref{eq:Compound_RED_nav_eta} can be found in~\ref{sec:appendix}.
The existence of the Hopf bifurcation guarantees the emergence of limit cycles in system dynamics, as instability sets in. In Compound TCP-RED system these limit cycles manifest in the form of non-linear oscillations in the queue size dynamics, which could be detrimental to network performance. Therefore, the above insight may be used to guide design of Compound TCP-RED parameters such that system stability is ensured. However, before that, one needs to verify this analytical insight using some simulations. 
 
\section{Packet-level simulations}
 \label{sec:sims}
The results outlined, so far, are obtained by analysing fluid approximations of Compound TCP-RED, which is a packet-level system. Therefore, they must be validated through packet-level simulations, before they can guide design principles. We now present some packet-level simulations conducted using NS2, to examine the impact of RTT, queue size averaging, packet-dropping thresholds on stability. 

\begin{figure}
\newcommand{\myarrowlength}{1.5}
 \newcommand{\myarrowsize}{0.08cm 5.0}
 \newcommand{\mylinewidth}{0.06}
 \begin{center}
 \scalebox{0.5} % Change this value to rescale the drawing.
{
\begin{pspicture}(0,-3.5)(14.98,3.5)
\psframe[linewidth=\mylinewidth,dimen=outer](11.62,0.75)(5.56,-0.75) %queue box
\psframe[linewidth=\mylinewidth,dimen=outer](3.2,0.4)(2.7,0.9) % top box 3
\psline[linewidth=\mylinewidth,arrowsize=\myarrowsize,arrowlength=\myarrowlength,arrowinset=0.4]{->}(3.2,0.65)(5.58,0.3)
\psframe[linewidth=\mylinewidth,dimen=outer](3.2,1.0)(2.7,1.5) % top box 2
\psline[linewidth=\mylinewidth,arrowsize=\myarrowsize,arrowlength=\myarrowlength,arrowinset=0.4]{->}(3.2,1.25)(5.58,0.4)
\psframe[linewidth=\mylinewidth,dimen=outer](3.2,1.6)(2.7,2.1) % top box 1
\psline[linewidth=\mylinewidth,arrowsize=\myarrowsize,arrowlength=\myarrowlength,arrowinset=0.4]{->}(3.2,1.85)(5.58,0.5)

\psline[linestyle=dashed,linecolor=black](2.95,0.4)(2.95,-0.4)
 \psframe[linewidth=\mylinewidth,dimen=outer](3.2,-0.4)(2.7,-0.9) % bottom box 3
\psline[linewidth=\mylinewidth,arrowsize=\myarrowsize,arrowlength=\myarrowlength,arrowinset=0.4]{->}(3.2,-0.65)(5.58,-0.3)
\psframe[linewidth=\mylinewidth,dimen=outer](3.2,-1.0)(2.7,-1.5) % bottom box 2
\psline[linewidth=\mylinewidth,arrowsize=\myarrowsize,arrowlength=\myarrowlength,arrowinset=0.4]{->}(3.2,-1.25)(5.58,-0.4)
\psframe[linewidth=\mylinewidth,dimen=outer](3.2,-1.6)(2.7,-2.1) % bottom box 1
\psline[linewidth=\mylinewidth,arrowsize=\myarrowsize,arrowlength=\myarrowlength,arrowinset=0.4]{->}(3.2,-1.85)(5.58,-0.5)
\pscircle[linewidth=\mylinewidth,dimen=outer](12.23,0.0){0.6} %queue circle
% \psline[linewidth=\mylinewidth,arrowsize=\myarrowsize,arrowlength=\myarrowlength,arrowinset=0.4]{-}(12.8,0.0)(13.66,0.0) %middle line
% \psline[linewidth=\mylinewidth](0.8,2.5)(13.6,2.5) %top big line
\psline[linewidth=\mylinewidth](0.8,-2.5)(15.7,-2.5) % bottom big line
% \psarc[linewidth=\mylinewidth](13.6,1.25){1.25}{-90.0}{90.0} %top right semi-cicle
\psarc[linewidth=\mylinewidth](15.7,-1.25){1.25}{-90.0}{90.0} %bottom right semi-circle
% text input
% \rput(2.5,0){\begin{huge}$\Vast\{$\end{huge}}
% \rput(2.5,-1.2){\begin{huge}$\bigg\{$\end{huge}}
\rput(12.18,0.0){\textbf{\begin{huge}$C$\end{huge}}}

% \rput(1.32,0.7){\textbf{\begin{huge}{\color{blue}  $x_1(t)$}\end{huge}}}
\rput(4,2.2){\textbf{\begin{huge}{$w(t)$}\end{huge}}}

% \rput(7.8,3.0){ \begin{huge} {\color{blue} Round-trip time, $\tau_1$} \end{huge}}
\rput(8.25,-3.25){\begin{huge} {Average round-trip time, $\tau$}\end{huge}}

% \psarc[linewidth=\mylinewidth](0.8,1.85){0.65}{90.0}{-90.0} %left top semi circle
\psarc[linewidth=\mylinewidth](0.8,-1.25){1.25}{90.0}{-90.0} %left bottom semi circle
\psline[linewidth=\mylinewidth,arrowsize=\myarrowsize,arrowlength=\myarrowlength,arrowinset=0.4]{->}(0.8,0)(2.2,0) % bottom left small horizontal line
% \psline[linewidth=\mylinewidth,arrowsize=\myarrowsize,arrowlength=\myarrowlength,arrowinset=0.4]{->}(0.8,1.2)(2.2,1.2) % top left small horizontal line
\psline[linewidth=\mylinewidth](8.6,0.75)(8.6,-0.75) %queue box - 2nd line
\psline[linewidth=\mylinewidth](9.6,0.75)(9.6,-0.75) %queue box - 3rd line
\psline[linewidth=\mylinewidth](10.6,0.75)(10.6,-0.75) %queue box - 4th line
\rput(7.60,0.0){\huge{$\cdots$}} %dots inside queue box
\psline[linewidth=\mylinewidth](6.4,0.75)(6.4,-0.75) %queue box - 1th line
\rput(6.00,-1.54){\textbf{\begin{huge}${B}$\end{huge}}}
% %%% sinks %%%%%
\psframe[linewidth=\mylinewidth,dimen=outer](15.2,0.4)(15.7,0.9) % top box 3
\psline[linewidth=\mylinewidth,arrowsize=\myarrowsize,arrowlength=\myarrowlength,arrowinset=0.4]{->}(12.83,0)(15.2,0.65)
\psframe[linewidth=\mylinewidth,dimen=outer](15.2,1.0)(15.7,1.5) % top box 2
\psline[linewidth=\mylinewidth,arrowsize=\myarrowsize,arrowlength=\myarrowlength,arrowinset=0.4]{->}(12.83,0)(15.2,1.25)
\psframe[linewidth=\mylinewidth,dimen=outer](15.2,1.6)(15.7,2.1) % top box 1
\psline[linewidth=\mylinewidth,arrowsize=\myarrowsize,arrowlength=\myarrowlength,arrowinset=0.4]{->}(12.83,0)(15.2,1.85)
\psline[linestyle=dashed,linecolor=black](15.45,0.4)(15.45,-0.4)
 \psframe[linewidth=\mylinewidth,dimen=outer](15.2,-0.4)(15.7,-0.9) % bottom box 3
\psline[linewidth=\mylinewidth,arrowsize=\myarrowsize,arrowlength=\myarrowlength,arrowinset=0.4]{->}(12.83,0)(15.2,-0.65)
\psframe[linewidth=\mylinewidth,dimen=outer](15.2,-1.0)(15.7,-1.5) % bottom box 2
\psline[linewidth=\mylinewidth,arrowsize=\myarrowsize,arrowlength=\myarrowlength,arrowinset=0.4]{->}(12.83,0)(15.2,-1.25)
\psframe[linewidth=\mylinewidth,dimen=outer](15.2,-1.6)(15.7,-2.1) % bottom box 1
\psline[linewidth=\mylinewidth,arrowsize=\myarrowsize,arrowlength=\myarrowlength,arrowinset=0.4]{->}(12.83,0)(15.2,-1.85)
\end{pspicture}
}\caption{Single bottleneck dumbbell topology, showing many end systems sharing a router with buffer size $B$ and service capacity $C$. The flows have an average round-trip time of $\tau$.}\label{fig:single_bottleneck_dumbbell}
\end{center}
\end{figure}
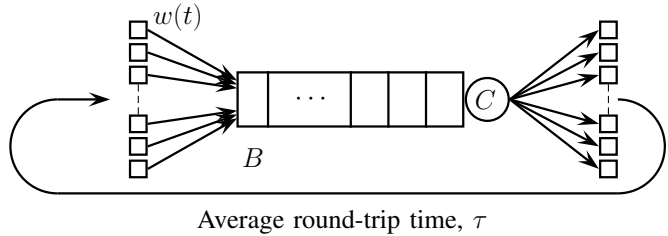

For these simulations, we consider a single bottleneck, dumbbell topology (shown in Figure~\ref{fig:single_bottleneck_dumbbell}). There are $60$ end systems sharing a single bottleneck link at the router, to transfer data to another $60$ end systems on the other side of the router. Each end system feeds into the router using an access link of $2$ Mbps. The link capacity is fixed at $100$~Mbps. Note that, this is the total link capacity, and not per-flow as considered in the analysis. The RED policy is used for queue management at the router. 

The end-user traffic is generated such that a total of $120$~Mbps ($20\%$ more than the service capacity) is fed into the queue, thus simulating a congested link. We consider two cases for traffic mix. We first consider a traffic setup that is aligned with the assumptions of the fluid model outlined in Section~\ref{sec:models}, \emph{i.e.}, only long-lived Compound TCP flows. We then deviate from these assumptions and consider a setup of mixed traffic where a bunch of Compound TCP, CUBIC (default TCP in Linux), UDP and HTTP flows share a bottleneck link that uses the RED policy for queue management. Such a setup enables us to verify if the analytical insight can be extended to scenarios that are not aligned with the model assumptions and may hence be closer to real-world scenarios. The packet size for TCP and UDP flows is fixed at $1500$ bytes.

The RED parameters are fixed at their default values, \emph{i.e.}, $q_{weight} = -1, linterm = 10$ as given in the NS2 implementation, unless specifically mentioned. The parameter $q_{weight}$ decides how RED chooses the weight on the instantaneous queue size for the averaging process, and the parameter $linterm$ is the reciprocal of the maximum packet-drop probability $\overline{p}$. The RED thresholds for dropping packets ($\underline{b}$ and $\overline{b}$) are fixed as per the requirement of the simulations. Compound TCP parameters are fixed as $\alpha=0.125,\beta=0.5,k=0.75$ (default). 
The buffer at the router is sized according to the bandwidth-delay product rule, $C\times\text{RTT}$~\cite{raina2005buffer}. Often for buffer sizing, an average round-trip time of $250$ ms is used. With this, for a link of $100$ Mbps capacity and a packet size of $1500$ bytes, the buffer size turns out to be $B = 2084$ pkts~\cite{raina2005buffer}. 

We aim to validate results obtained from fluid model analysis using simulations conducted on a discrete-event simulator. While the analysis yields asymptotic results for the system in equilibrium, simulation traces obtained on NS2 are likely to have some transient behaviour. In order to reconcile the two, it is imperative that the simulations are run for a long enough time to ensure that the transient behaviour settles down and the traces obtained represent the system dynamics in steady state. These traces would then be comparable to the dynamics of the fluid model. Therefore, we run the simulations up to $500$ seconds and study the traces corresponding to the last $25$ seconds. We now describe the two traffic mixes in greater detail, and discuss the results observed. 

\subsection{Homogeneous traffic, RED queue policy}
\label{sec:RED_sims_homo}
In this subsection, we consider $60$ long-lived Compound TCP flows, each with $2$~Mbps access speed. Note that, $60$ flows each fed through a $2$~Mbps link add up to a total of $120$~Mbps. Each long-lived flow is started at a random time instant within the first $10$ seconds of the simulation, and lasts for the entire duration of the simulation.
This scenario of only long-lived Compound TCP flows sharing a single bottleneck adheres to the assumptions of the fluid model for Compound TCP-RED~\eqref{eq:Compound_RED}, and is hence an appropriate setting for validating the analytical insight obtained. We conduct three sets of simulations to validate the key insights highlighted in Section~\ref{sec:key_insights_for_validation}.

\subsubsection{Impact of variation in RTT}
\begin{figure*}[t]
% \captionsetup[subfigure]{labelformat=empty}
\psfrag{t}[b][b]{\small{Time (sec)}}
      \psfrag{q}{\small{Queue size (pkts)}}
      \psfrag{a}{\hspace{-1mm}\small{Average queue size (pkts)}}
      \psfrag{u}{\small{Link utilisation (\%)}}
      \psfrag{w}{\hspace{-1mm}\small{Mean window size (pkts)}}
      \psfrag{0}[b][b]{\scriptsize{$0$}}
      \psfrag{10}[b][b]{\scriptsize{$10$}}
      \psfrag{20}[b][b]{\scriptsize{$20$}}
      \psfrag{40}[b][b]{\scriptsize{$40$}}
      \psfrag{50}[b][b]{\scriptsize{$50$}}
      \psfrag{100}[b][b]{\scriptsize{$100$}}
      \psfrag{200}[b][b]{\scriptsize{$200$}}
      \psfrag{475}[c][t]{\scriptsize{$475$}}
      \psfrag{500}[c][t]{\scriptsize{$500$}}
 \begin{center}
     \subfloat[\normalsize{Round-trip time $= 10$ ms}]{\includegraphics[width=1.6in,height=5.2in,angle=270]{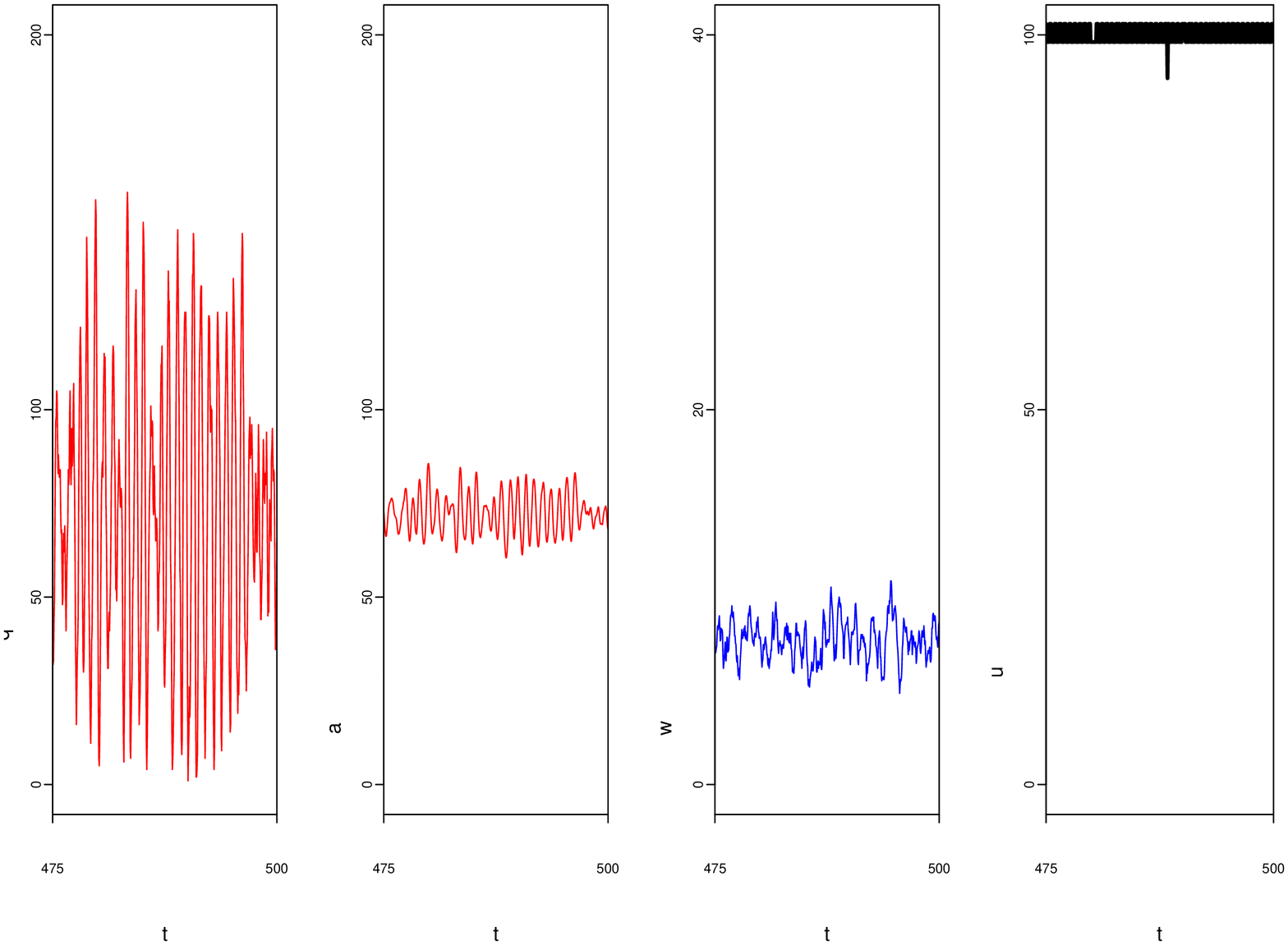}}\\
%      \subfloat[\normalsize{Round-trip time $=$ 25 ms}]{\includegraphics[width=1.6in,height=6.8in,angle=270]{RTT_effect_25_.eps}}\\
  \subfloat[\normalsize{Round-trip time $= 200$ ms}] {\includegraphics[width=1.6in,height=5.2in,angle=270]{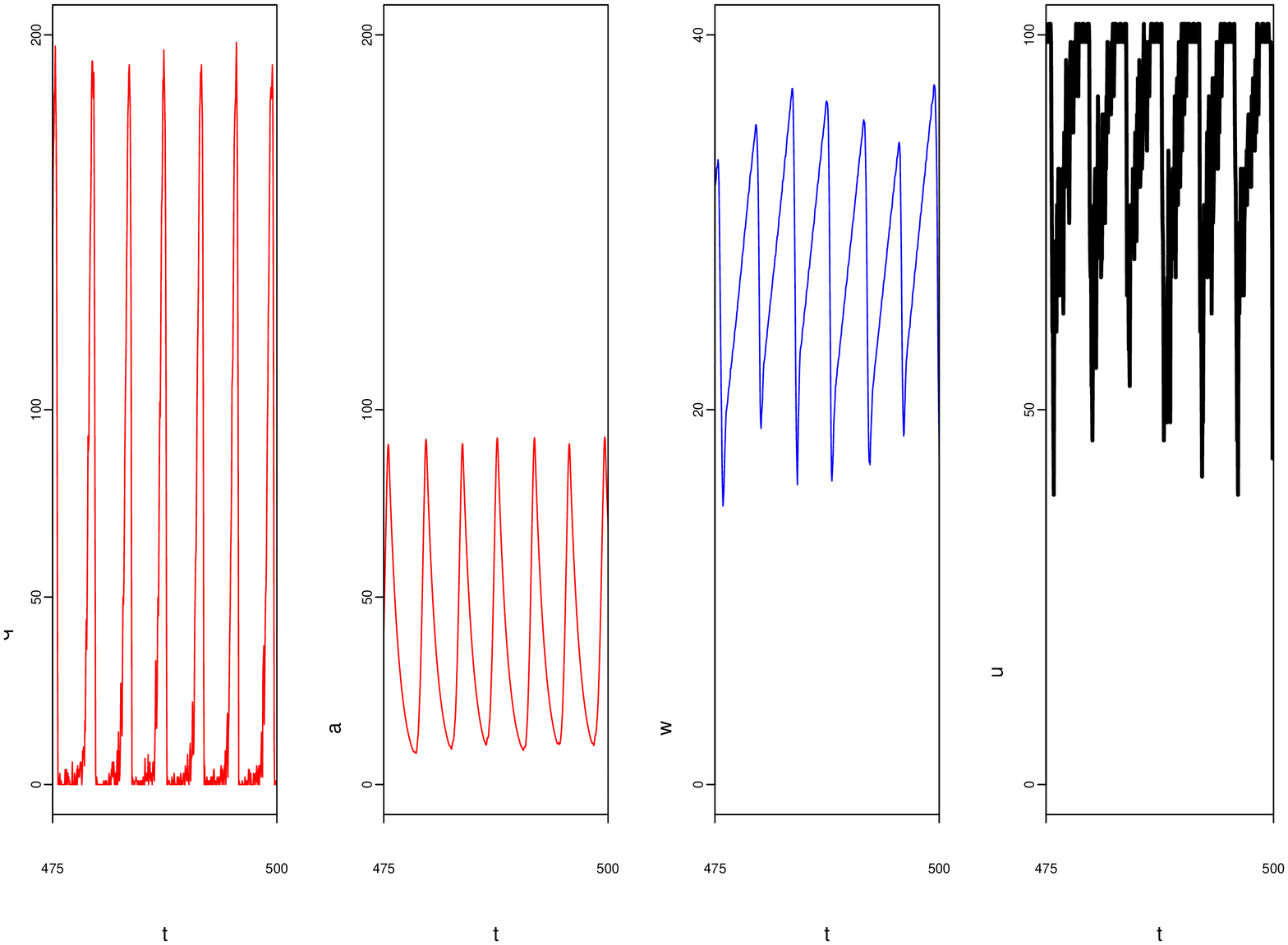}}
  \caption{Impact of RTT variation (Packet-level simulations. Homogeneous traffic with RED policy). $B~=~2084, \overline{b}~=~100, \underline{b}~=~50$~pkts. Compound and RED parameters are retained at their default values. Queue size, average queue size exhibit oscillations when round-trip time is increased. This leads to synchronisation of TCP flows and loss of link utilisation.}
  \label{fig:RED_RTT_sensitivity}
 \end{center}
\end{figure*}

Stability of the Compound TCP-RED system is seen to be sensitive to round-trip time of the TCP flows. The transition into instability is shown to occur via a Hopf bifurcation, which could lead to the emergence of limit cycles in the queue size. In order to validate this we conduct two sets of simulations designed such that the average RTT of the TCP flows is (i) $10$ ms, and (ii) $200$ ms. The RED thresholds are fixed as $\underline{b} = 50$ pkts and $\overline{b}=100$ pkts. The traces of the queue size observed at the link, average queue size as computed by RED, average window size of $15$ randomly chosen TCP flows and the link utilisation observed are plotted in Figure~\ref{fig:RED_RTT_sensitivity}. 

For RTT $= 10$ ms, RED maintains the average queue size between the thresholds, as intended. The average window size also appears to be randomly varying around $10$ packets, this implies that the TCP flows are desynchronised and the system is stable. The link utilisation is seen to be $100$ \%. When the RTT is 
increased to $200$ ms, we observe a qualitative change in the queue size dynamics. The queue size and the average queue size begin to oscillate. This leads to synchronisation of TCP flows, which implies that the window sizes of all the TCP flows reach their respective peaks and troughs at the same time. Hence, the average window size of the TCP flows appears to be a saw-tooth wave, which is the expected behaviour of the sending window of a single Compound TCP flow. Owing to the oscillations in the queue size, the link utilisation is expected to drop. Indeed this is seen in the trace of the link utilisation. Thus, large RTTs are detrimental to network performance, as predicted by analysis. 

\begin{figure*}[tbh]
% \captionsetup[subfigure]{labelformat=empty}
      \psfrag{t}[b][b]{\small{Time (sec)}}
      \psfrag{q}{\small{Queue size (pkts)}}
      \psfrag{a}{\hspace{-1mm}\small{Average queue size (pkts)}}
      \psfrag{u}{\small{Link utilisation (\%)}}
      \psfrag{w}{\hspace{-1mm}\small{Mean window size (pkts)}}
      \psfrag{0}[b][b]{\scriptsize{$0$}}
      \psfrag{10}[b][b]{\scriptsize{$10$}}
      \psfrag{20}[b][b]{\scriptsize{$20$}}
      \psfrag{40}[b][b]{\scriptsize{$40$}}
      \psfrag{50}[b][b]{\scriptsize{$50$}}
      \psfrag{100}[b][b]{\scriptsize{$100$}}
      \psfrag{200}[b][b]{\scriptsize{$200$}}
      \psfrag{475}[c][t]{\scriptsize{$475$}}
      \psfrag{500}[c][t]{\scriptsize{$500$}}
 \begin{center}
     \subfloat[\normalsize{Round-trip time $= 5$ ms and $15$ ms}]{\includegraphics[width=1.6in,height=5.2in,angle=270]{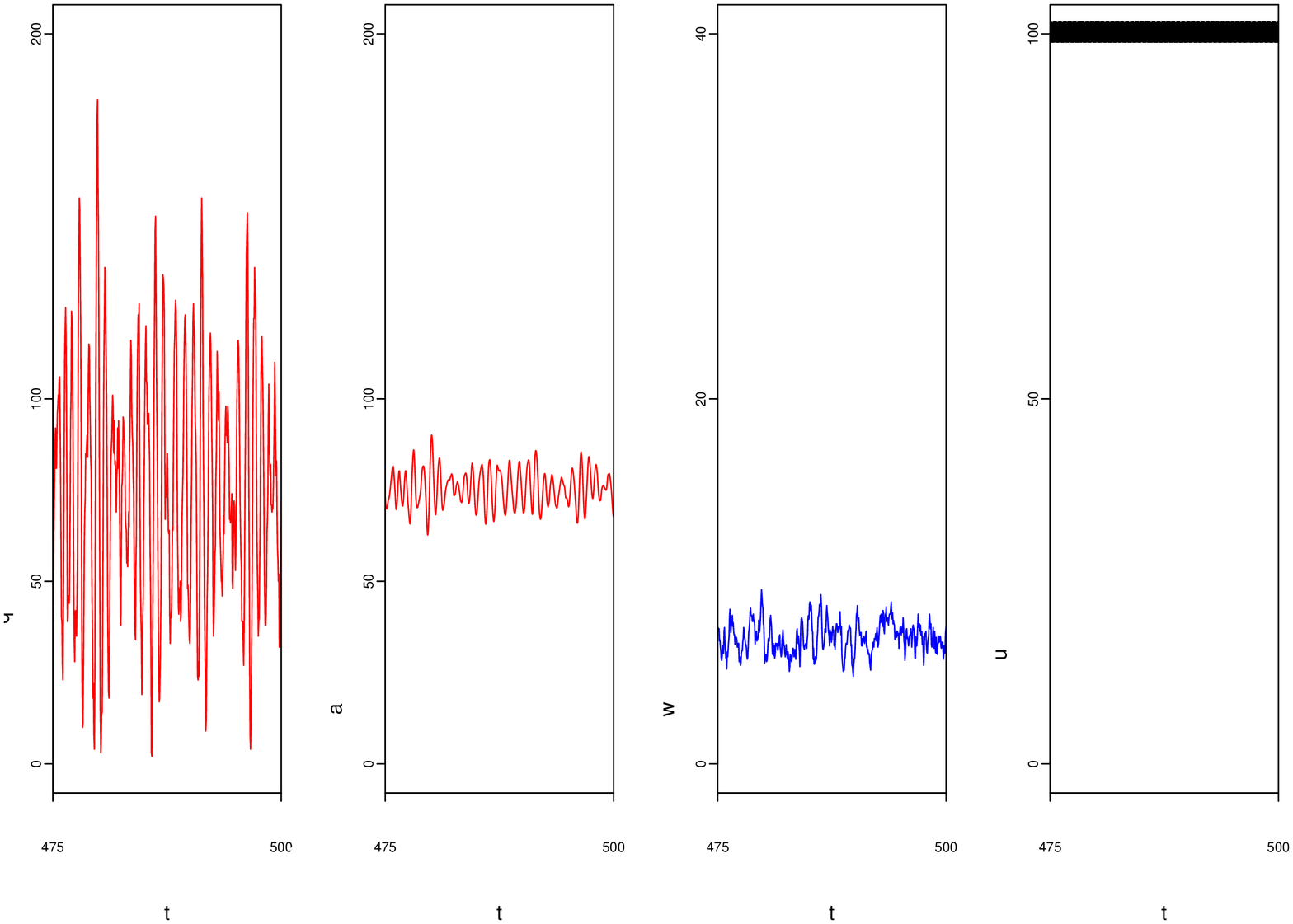}}\\
%      \subfloat[\normalsize{Round-trip time $=$ 25 ms}]{\includegraphics[width=1.6in,height=6.8in,angle=270]{RTT_effect_25_.eps}}\\
  \subfloat[\normalsize{Round-trip time $= 100$ ms and $300$ ms}] {\includegraphics[width=1.6in,height=5.2in,angle=270]{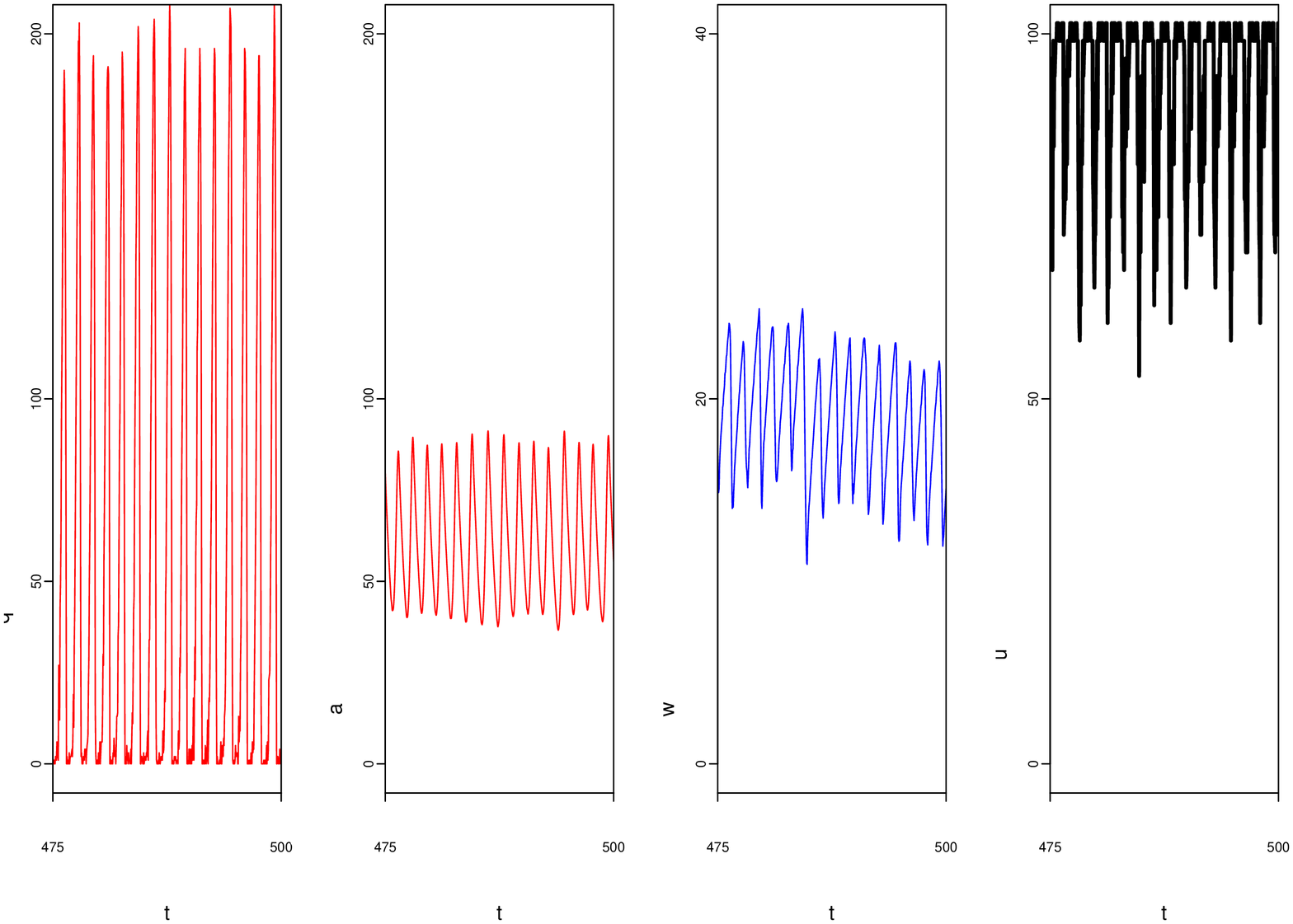}}
  \caption{Impact of RTT variation (Packet-level simulations. Homogeneous traffic with RED policy, multiple RTTs). $B~=~2084, \overline{b}~=~100, \underline{b}~=~50$~pkts. Compound and RED parameters are retained at their default values. Queue size, average queue size exhibit oscillations when round-trip time is increased. This leads to synchronisation of TCP flows, loss of link utilisation.}
  \label{fig:RED_2RTT_sensitivity}
 \end{center}
\end{figure*}

In order to ensure that the impact of increase in RTT is not an artefact of the Compound TCP flows having a single average RTT, we considered a scenario with heterogeneous RTTs. For these simulations, the TCP flows are divided into two bunches. The first bunch is assigned an average round-trip time of RTT$_1$ and the second one is assigned RTT$_2$. With this setting, we conduct two sets of simulations (i) RTT$_1 = 5$ ms, RTT$_2 = 15$ ms, and (ii) RTT$_1 = 100$ ms, RTT$_2 = 300$ ms. These simulations are presented in Figure~\ref{fig:RED_2RTT_sensitivity}. The qualitative change in queue size, average queue size, average window size and link utilisation is the same as seen in Figure~\ref{fig:RED_RTT_sensitivity} for the case of single round-trip time. This shows that the insight obtained from the analysis can also be extended to a network scenario with multiple round-trip times. 

\subsubsection{Impact of queue averaging}
\begin{figure}
% \captionsetup[subfigure]{labelformat=empty}
\psfrag{t}[t][t]{\small{Time (sec)}}
      \psfrag{q}{\small{Queue size (pkts)}}
      \psfrag{0}[b][b]{\scriptsize{$0$}}
      \psfrag{10}[b][b]{\scriptsize{$10$}}
      \psfrag{50}[b][b]{\scriptsize{$50$}}
      \psfrag{150}[b][b]{\scriptsize{$150$}}
      \psfrag{100}[b][b]{\scriptsize{$100$}}
      \psfrag{200}[b][b]{\scriptsize{$200$}}
      \psfrag{475}[t][t]{\scriptsize{$475$}}
      \psfrag{500}[t][t]{\scriptsize{$500$}}
  \psfrag{m}{\scriptsize{Min. utilisation (\%)}}
  \psfrag{T}{\small{Round-trip time (ms)}}
 \begin{center}
   \subfloat[Comparison of queue size dynamics]{ 
   \psfrag{A}[t][t]{\hspace{15mm}\small{With averaging}}
   \psfrag{N}[t][t]{\hspace{20mm}\small{Without averaging}}
   \psfrag{T}{\small{Round-trip time (ms)}}
   \includegraphics[width=1.6in,height=2.6in,angle=270]{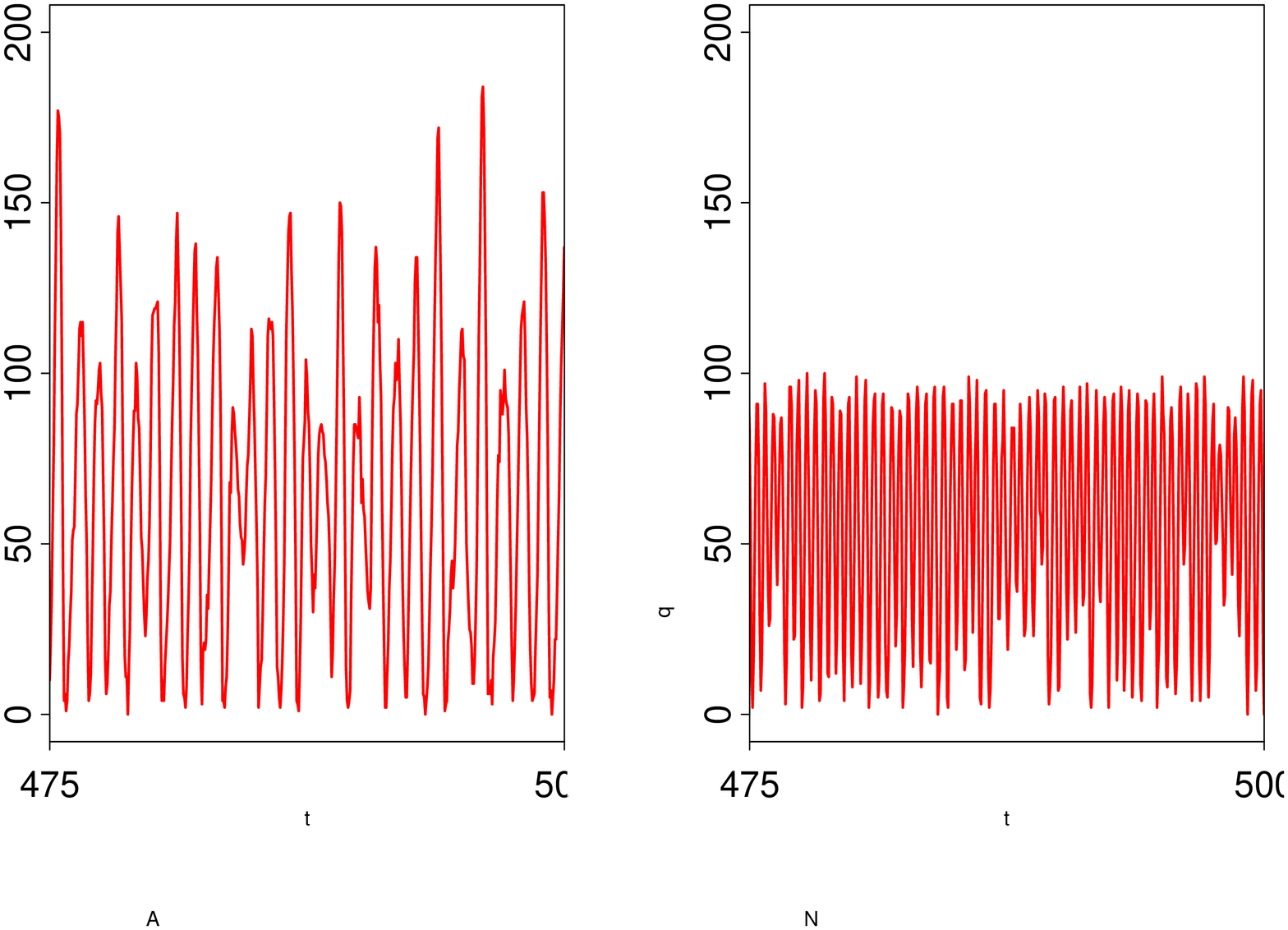}\label{fig:RED_averaging_effect_q}
   }\qquad
   \subfloat[Comparison of utilisation]{
   \psfrag{10}[t][t]{\scriptsize{$10$}}
   \psfrag{50}[t][t]{\scriptsize{$50$}}
   \psfrag{150}[t][t]{\scriptsize{$150$}}
   \psfrag{100}[t][t]{\scriptsize{$100$}}
   \psfrag{200}[t][t]{\scriptsize{$200$}}
   \psfrag{A}{\small{with averaging}}
   \psfrag{N}{\small{without averaging}} 
   \psfrag{T}[t][t]{\hspace{25mm}\small{Round-trip time (ms)}}
   \psfrag{m}{\small{Min. utilisation (\%)}}
   \includegraphics[width=1.6in,height=2.2in,angle=270]{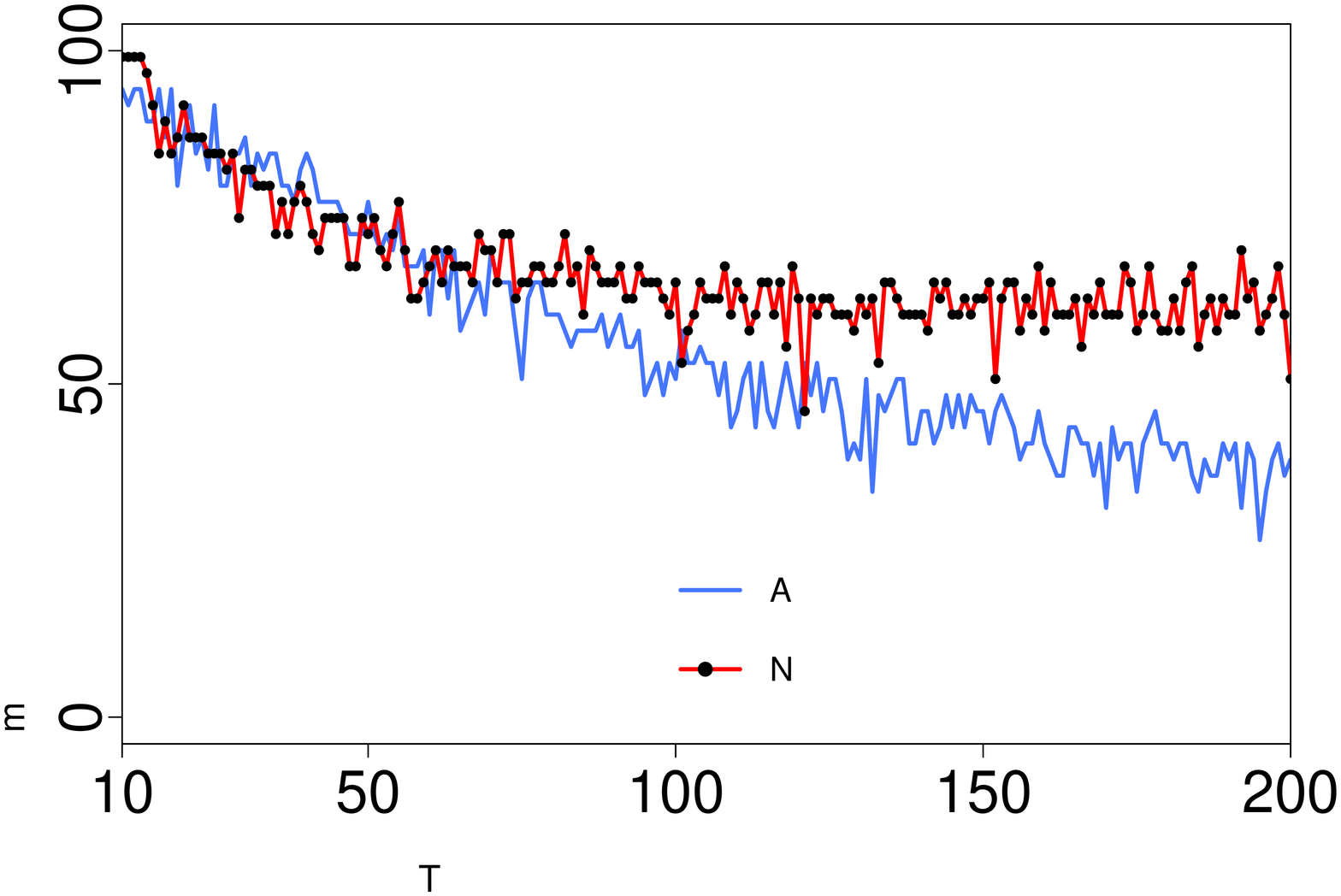}\label{fig:RED_utilisation_AV_NAV}
   }
  \caption{Impact of averaging (Packet-level simulations. Homogeneous traffic with RED policy). RED thresholds set at $100$ and $50$ pkts, RTT $=25$ ms. There is a qualitative change in the queue sizes observed. Averaging over the queue size causes lower link utilisation.}
 \end{center}
 \end{figure}

The traces observed for RTT $=200$ ms in Figure~\ref{fig:RED_RTT_sensitivity} indicate that for large RTTs both the queue size and the average queue size would exhibit oscillations. Therefore, it may not make a significant difference whether the packet-dropping decisions are based on the queue size or the exponentially weighted moving average of the queue size computed by RED. The insight obtained from fluid model analysis suggests that queue size averaging may not be beneficial. Further, analysis also suggests that Compound TCP-RED remains stable for larger round-trip times when queue size averaging is not performed, compared to the regime where averaging is performed. To verify this, we now compare the queue dynamics of Compound TCP-RED with averaging and without averaging. To simulate the regime where RED does not perform the queue size averaging, we modified the source code for the NS2 implementation of RED such that packet-dropping decisions are based on the instantaneous queue size instead of the 
average queue size. The rest of the parameters are same as mentioned above. The queue size traces obtained for average RTT = $25$ ms, with and without averaging, are plotted in Figure~\ref{fig:RED_averaging_effect_q}. A qualitative difference in the queue size behaviour can be observed in the two cases. Notably, when there is no averaging, the queue size does not increase beyond the upper threshold $100$ pkts.  

Our analysis predicts that, in both the regimes of Compound TCP-RED, \emph{i.e.} with and without queue size averaging, the system loses stability for large RTT, and loss of stability is predicted to occur for comparatively larger RTT when averaging is not performed. In order to study this change in system dynamics, we may vary RTT and observe the queue dynamics in both the regimes. However, queue size plots in Figure~\ref{fig:RED_averaging_effect_q} alert us to the fact that, this change may not be easily noticeable owing to the minute qualitative difference in the queue dynamics observed in both the regimes. As observed in Figure~\ref{fig:RED_RTT_sensitivity}, queue size oscillations have an adverse impact on the link utilisation. Therefore, we plot the minimum link utilisation observed in the two regimes, as a function of RTT. For this plot, we varied the average RTT of the Compound TCP flows in the range $[10,200]$ ms. From the simulation traces obtained for each RTT, we recorded the minimum link 
utilisation observed in the last $25$ seconds of 
the simulation. This plot is presented in Figure~\ref{fig:RED_utilisation_AV_NAV}. It can be observed that, as RTT increases, the link utilisation falls in both the regimes. However, the minimum link utilisation is higher when queue size averaging is not performed, as compared to the case where averaging is performed.

 \subsubsection{Impact of packet-dropping threshold}
\begin{figure}[tbh]
% \captionsetup[subfigure]{labelformat=empty}
      \psfrag{t}[b][b]{\small{Time (sec)}}
      \psfrag{q}{\hspace{-1mm}\small{Queue size (pkts)}}
      \psfrag{a}{\hspace{-1mm}\small{Average queue size (pkts)}}
      \psfrag{0}[b][b]{\scriptsize{$0$}}
      \psfrag{8}[b][b]{\scriptsize{$8$}}
      \psfrag{10}[b][b]{\scriptsize{$10$}}
      \psfrag{15}[b][b]{\scriptsize{$15$}}
      \psfrag{30}[b][b]{\scriptsize{$30$}}
      \psfrag{300}[b][b]{\scriptsize{$300$}}
      \psfrag{25}[b][b]{\scriptsize{$25$}}
      \psfrag{100}[b][b]{\scriptsize{$100$}}
      \psfrag{200}[b][b]{\scriptsize{$200$}}
      \psfrag{400}[b][b]{\scriptsize{$400$}}
      \psfrag{475}[c][t]{\scriptsize{$475$}}
      \psfrag{500}[c][t]{\scriptsize{$500$}}
 \begin{center}
    \subfloat[\normalsize{$\overline{b}=15, \underline{b}=8$ pkts}]{ \includegraphics[width=1.6in,height=2.6in,angle=270]{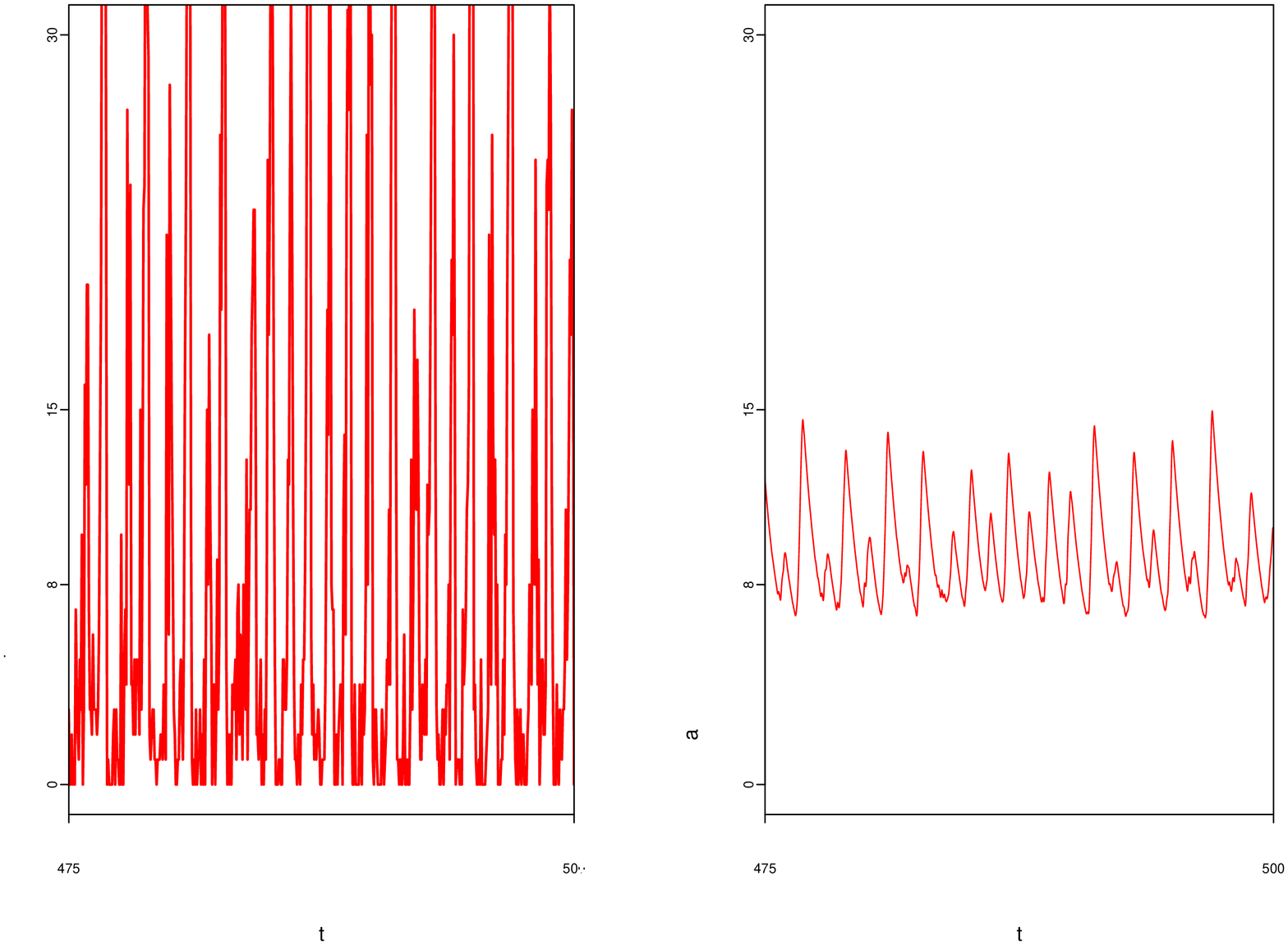}}\qquad
    \subfloat[\normalsize{$\overline{b}=200, \underline{b}=100$ pkts}]{ \includegraphics[width=1.6in,height=2.6in,angle=270]{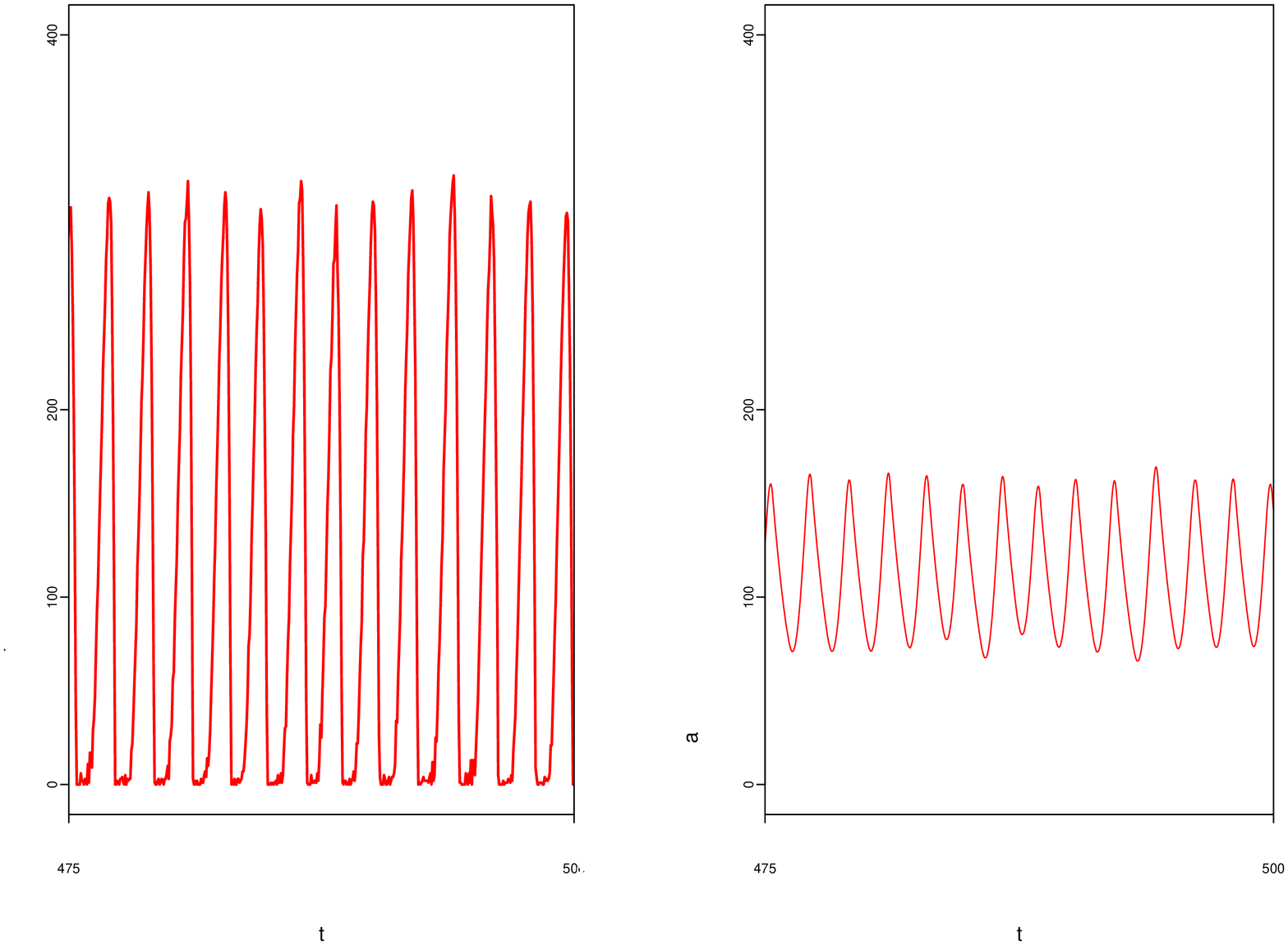}}
  \caption{Impact of packet-dropping threshold (Packet-level simulations of Homogeneous traffic with RED policy). $B = 2084$ pkts, RTT $=100$ ms. RED parameters and Compound parameters are set at default. Queue size oscillates for large thresholds.}\vspace{-5mm}
  \label{fig:RED_threshold_sensitivity}
 \end{center}
\end{figure}

Analysis predicts that Compound TCP-RED would become unstable as the packet-dropping thresholds are increased. In order to validate this, we conduct simulations for two threshold settings: $\overline{b}=15,\underline{b}=8$ pkts and $\overline{b}=200,\underline{b}=100$ pkts. The average RTT is fixed as $100$ ms. The rest of the Compound TCP and RED parameters are retained at their default values. The traces of the queue size obtained are plotted in Figure~\ref{fig:RED_threshold_sensitivity}.  For $\overline{b}=15,\underline{b}=8$ pkts, the queue appears to be varying randomly. As the thresholds are increased to $200$ and $100$ pkts respectively, the queue size exhibits oscillations. This indicates that small thresholds for dropping packets indeed aid system stability.

\subsection{Heterogeneous traffic, RED queue policy}
\label{sec:RED_sims_hetro}
The Internet comprises of multiple users with varied requirements, served by a variety of applications on the end systems. Applications like File Transfer Protocol (FTP) and HTTP use TCP at the transport layer, while applications such as Voice over IP use UDP. Therefore, it would be rather simplistic to assume that a router in the network would serve only Compound TCP flows. Hence, it would be useful to investigate if the analytical insight holds for a router carrying mixed traffic.  For this, we consider a setup with $55$ long-lived TCP flows generated by FTP clients (each with a link of $2$ Mbps), $8$ UDP flows (contributing a total of $8$ Mbps) and $50$ short-lived HTTP flows generated every second (contributing a total of $2$ Mbps). Among the FTP flows, we consider $27$ Compound TCP and $28$ CUBIC flows, in order to cater to traffic from both Windows and Linux users. The HTTP flows are generated using the PackMime\textemdash HTTP package in NS2, and use Tahoe TCP~\cite{cao2004stochastic}. 

\begin{figure*}[t]
% \captionsetup[subfigure]{labelformat=empty}
\psfrag{t}[b][b]{\small{Time (sec)}}
      \psfrag{q}{\small{Queue size (pkts)}}
      \psfrag{a}{\hspace{-1mm}\small{Average queue size (pkts)}}
      \psfrag{u}{\small{Link utilisation (\%)}}
      \psfrag{w}{\hspace{-1mm}\small{Mean window size (pkts)}}
      \psfrag{0}[b][b]{\scriptsize{$0$}}
      \psfrag{10}[b][b]{\scriptsize{$10$}}
      \psfrag{20}[b][b]{\scriptsize{$20$}}
      \psfrag{40}[b][b]{\scriptsize{$40$}}
      \psfrag{50}[b][b]{\scriptsize{$50$}}
      \psfrag{100}[b][b]{\scriptsize{$100$}}
      \psfrag{200}[b][b]{\scriptsize{$200$}}
      \psfrag{475}[c][t]{\scriptsize{$475$}}
      \psfrag{500}[c][t]{\scriptsize{$500$}}
 \begin{center}
     \subfloat[\normalsize{Round-trip time $= 10$ ms}]{\includegraphics[width=1.6in,height=5.2in,angle=270]{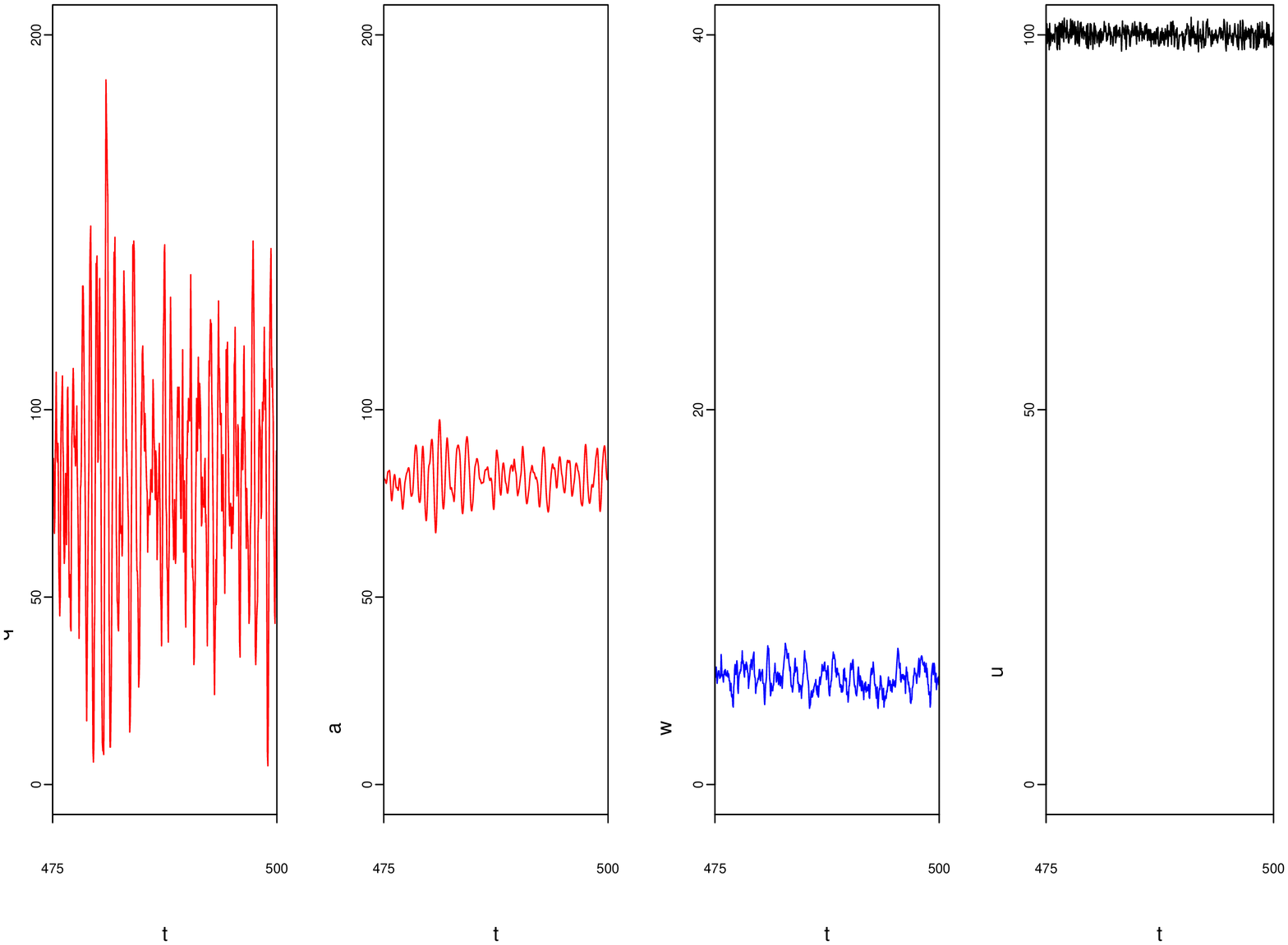}}\\
%      \subfloat[\normalsize{Round-trip time $=$ 25 ms}]{\includegraphics[width=1.6in,height=6.8in,angle=270]{RTT_effect_25_.eps}}\\
  \subfloat[\normalsize{Round-trip time $= 200$ ms}] {\includegraphics[width=1.6in,height=5.2in,angle=270]{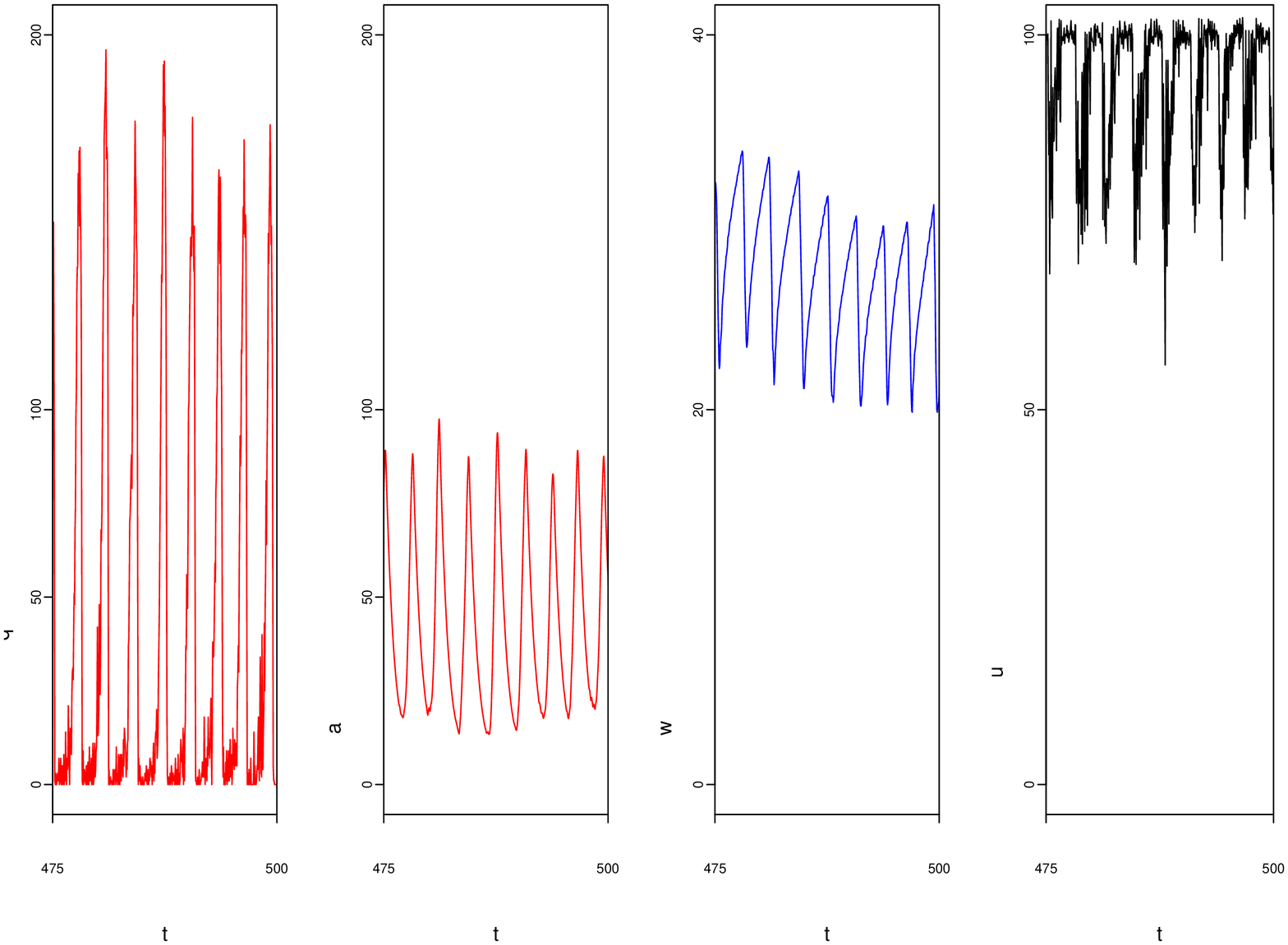}}
  \caption{Impact of RTT variation (Packet-level simulations. Heterogeneous traffic with RED policy). $B = 2084, \overline{b}= 100, \underline{b} = 50$ pkts. Compound and RED parameters are retained at their default values. Queue size, average queue size exhibit oscillations when round-trip time is increased. This leads to synchronisation of TCP flows, loss of link utilisation.}
  \label{fig:RED_RTT_sensitivity_MT}
 \end{center}
\end{figure*}

\begin{figure*}[tbh]
% \captionsetup[subfigure]{labelformat=empty}
\psfrag{t}[b][b]{\small{Time (sec)}}
      \psfrag{q}{\small{Queue size (pkts)}}
      \psfrag{a}{\hspace{-1mm}\small{Average queue size (pkts)}}
      \psfrag{u}{\small{Link utilisation (\%)}}
      \psfrag{w}{\hspace{-1mm}\small{Mean window size (pkts)}}
      \psfrag{0}[b][b]{\scriptsize{$0$}}
      \psfrag{10}[b][b]{\scriptsize{$10$}}
      \psfrag{20}[b][b]{\scriptsize{$20$}}
      \psfrag{40}[b][b]{\scriptsize{$40$}}
      \psfrag{50}[b][b]{\scriptsize{$50$}}
      \psfrag{100}[b][b]{\scriptsize{$100$}}
      \psfrag{200}[b][b]{\scriptsize{$200$}}
      \psfrag{475}[c][t]{\scriptsize{$475$}}
      \psfrag{500}[c][t]{\scriptsize{$500$}}
 \begin{center}
     \subfloat[\normalsize{Round-trip time $= 5$ ms and $15$ ms}]{\includegraphics[width=1.6in,height=5.2in,angle=270]{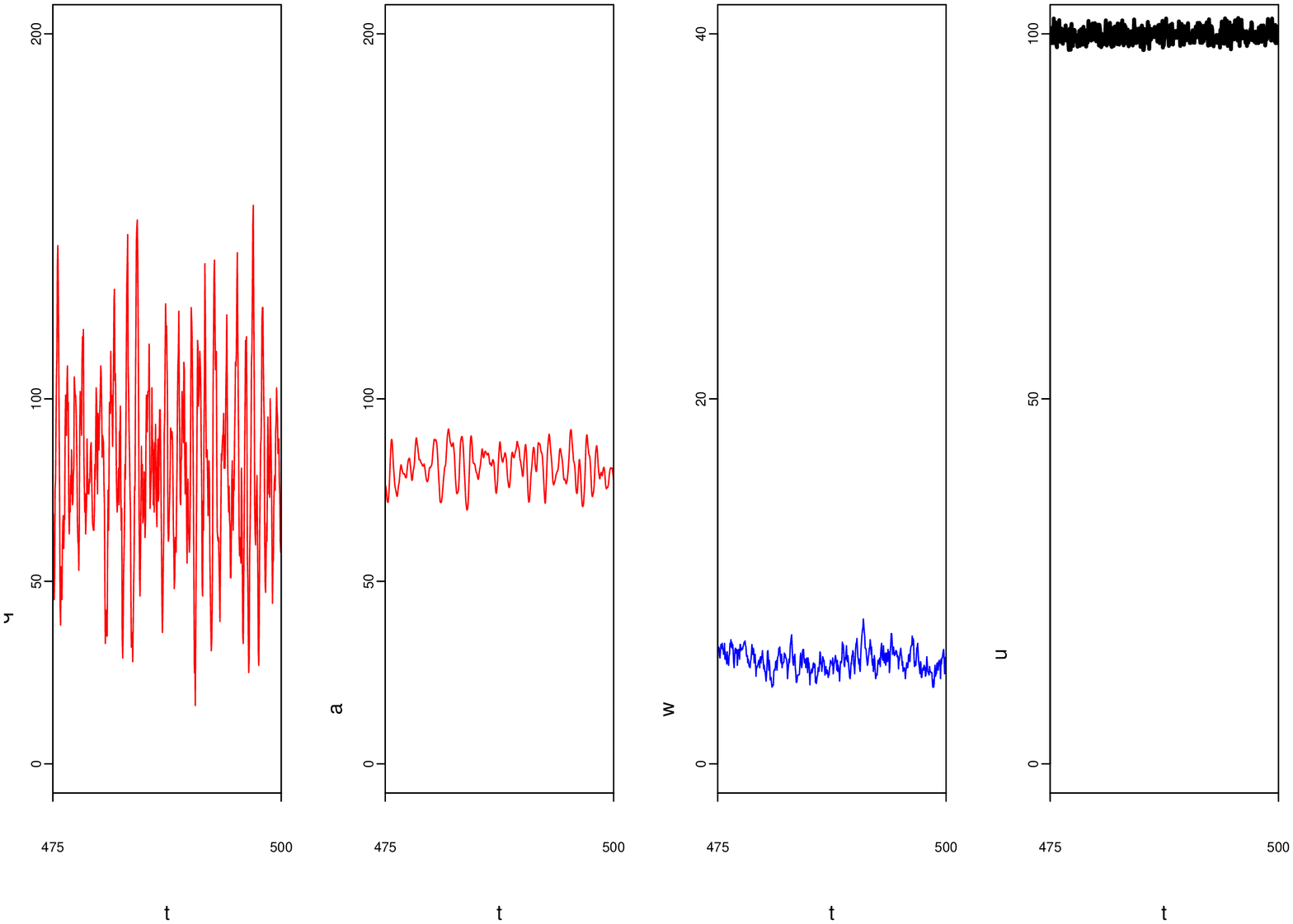}}\\
%      \subfloat[\normalsize{Round-trip time $=$ 25 ms}]{\includegraphics[width=1.6in,height=6.8in,angle=270]{RTT_effect_25_.eps}}\\
  \subfloat[\normalsize{Round-trip time $= 100$ ms and $300$ ms}] {\includegraphics[width=1.6in,height=5.2in,angle=270]{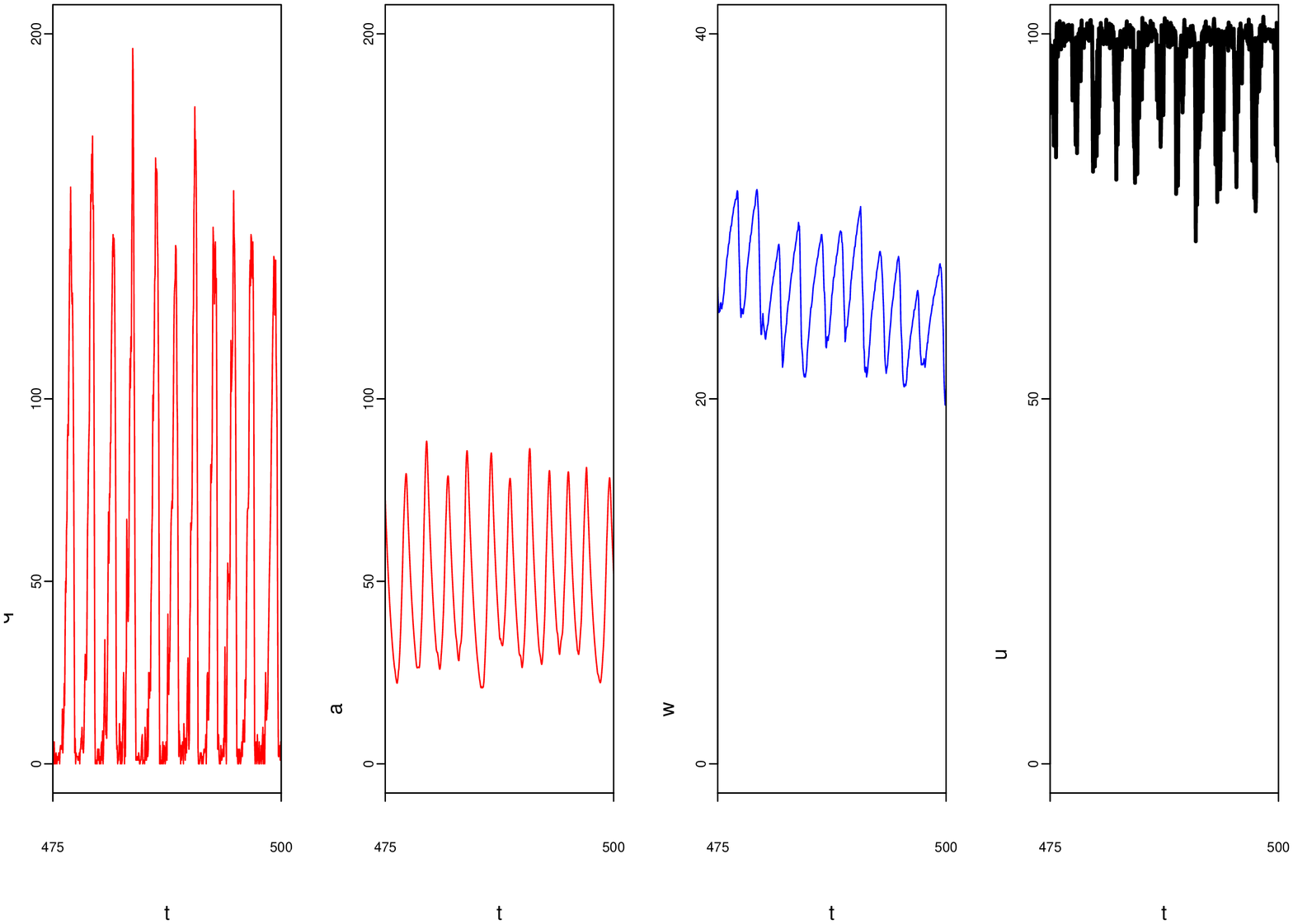}}
  \caption{Impact of RTT variation (Packet-level simulations. Heterogeneous traffic with RED policy, multiple RTTs). $B~=~2084, \overline{b}~=~100, \underline{b}~=~50$~pkts. Compound and RED parameters are retained at their default values. Queue size, average queue size exhibit oscillations when round-trip time is increased. This leads to synchronisation of TCP flows, loss of link utilisation.}
  \label{fig:RED_2RTT_sensitivity_MT}
 \end{center}
\end{figure*}

\begin{figure}[tbh]
% \captionsetup[subfigure]{labelformat=empty}
\psfrag{t}[t][t]{\small{Time (sec)}}
      \psfrag{q}{\small{Queue size (pkts)}}
      \psfrag{0}[b][b]{\scriptsize{$0$}}
      \psfrag{10}[b][b]{\scriptsize{$10$}}
      \psfrag{50}[b][b]{\scriptsize{$50$}}
      \psfrag{150}[b][b]{\scriptsize{$150$}}
      \psfrag{100}[b][b]{\scriptsize{$100$}}
      \psfrag{200}[b][b]{\scriptsize{$200$}}
      \psfrag{475}[t][t]{\scriptsize{$475$}}
      \psfrag{500}[t][t]{\scriptsize{$500$}}
  \psfrag{m}{\scriptsize{Min. utilisation (\%)}}
  \psfrag{T}{\small{Round-trip time (ms)}}
 \begin{center}
   \subfloat[Comparison of queue size dynamics]{ 
   \psfrag{A}[t][t]{\hspace{15mm}\small{With averaging}}
   \psfrag{N}[t][t]{\hspace{20mm}\small{Without averaging}}
   \psfrag{T}{\small{Round-trip time (ms)}}
   \includegraphics[width=1.6in,height=2.6in,angle=270]{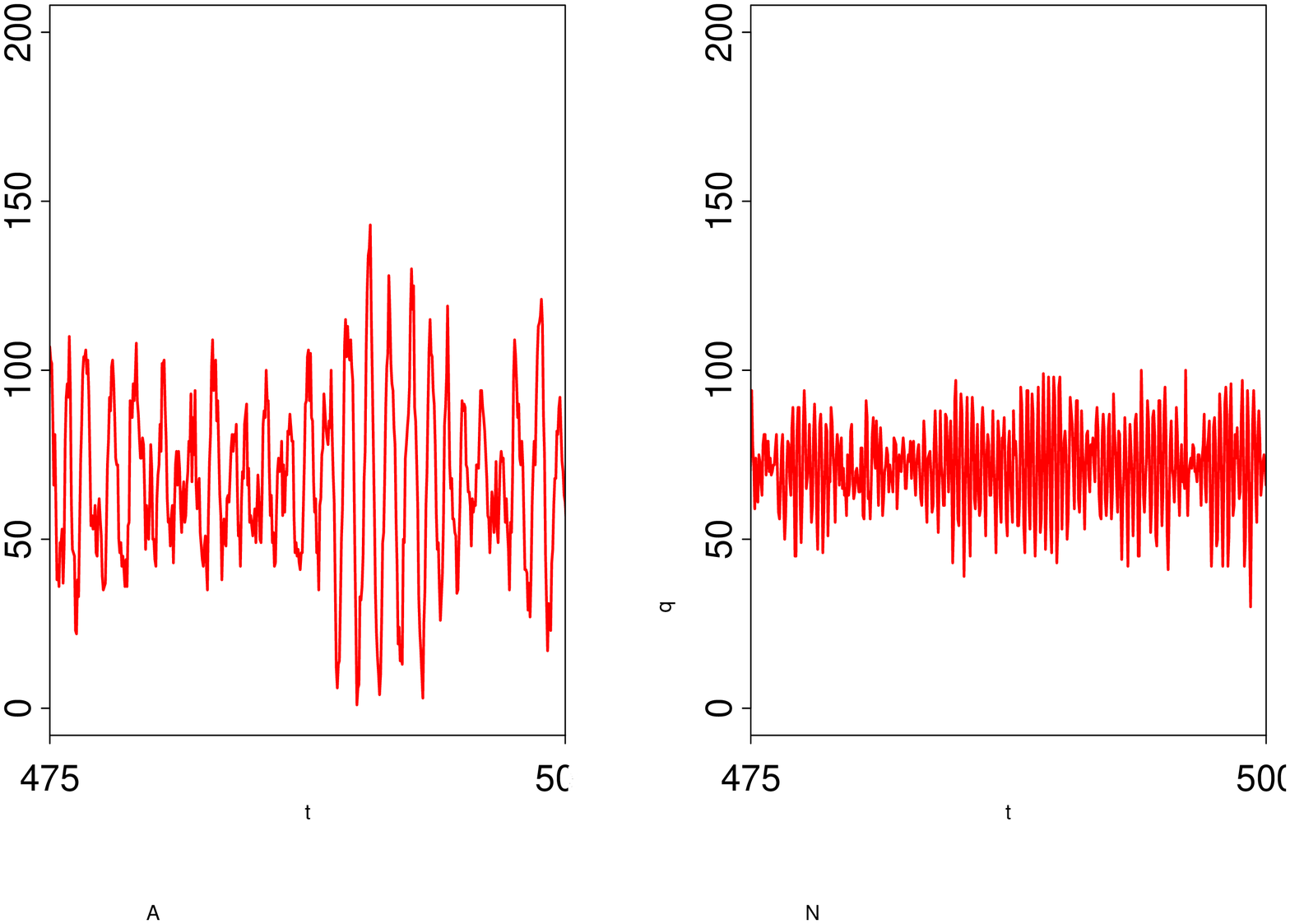}\label{fig:RED_averaging_effect_q_MT}
   }\qquad
   \subfloat[Comparison of utilisation]{
   \psfrag{10}[t][t]{\scriptsize{$10$}}
   \psfrag{50}[t][t]{\scriptsize{$50$}}
   \psfrag{150}[t][t]{\scriptsize{$150$}}
   \psfrag{100}[t][t]{\scriptsize{$100$}}
   \psfrag{200}[t][t]{\scriptsize{$200$}}
   \psfrag{A}{\small{with averaging}}
   \psfrag{N}{\small{without averaging}} 
   \psfrag{T}[t][t]{\hspace{25mm}\small{Round-trip time (ms)}}
   \psfrag{m}{\small{Min. utilisation (\%)}} 
   \includegraphics[width=1.6in,height=2.2in,angle=270]{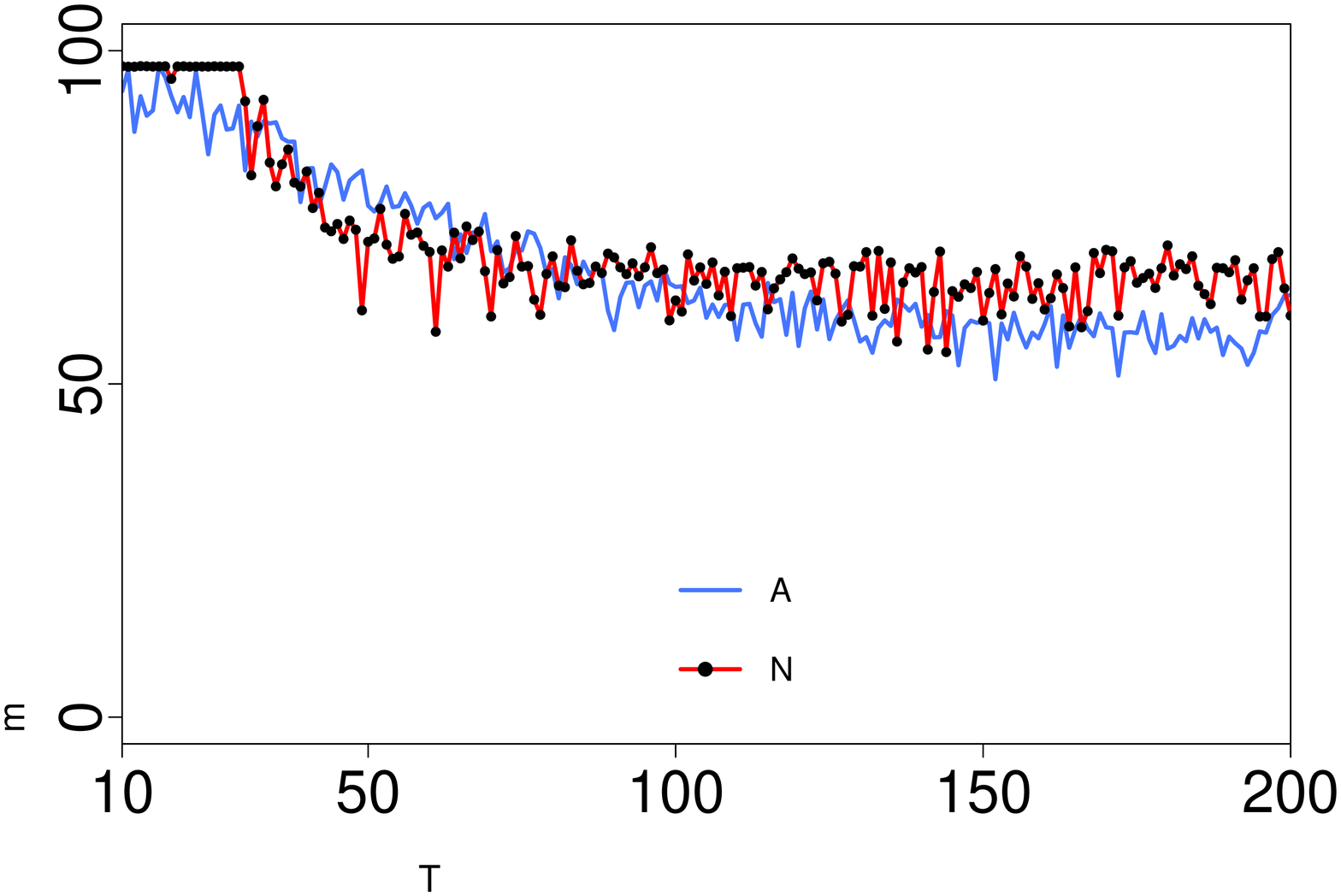}\label{fig:RED_utilisation_AV_NAV_MT}
   }
  \caption{Impact of averaging (Packet-level simulations. Heterogeneous traffic with RED policy). RED thresholds set at $100$ and $50$ pkts, RTT $=25$ ms. There is a qualitative change in the queue sizes observed. Averaging over the queue size causes lower link utilisation.}\label{fig:RED_averaging_effect_MT}
 \end{center}
 \end{figure}

\begin{figure}[tbh]
% \captionsetup[subfigure]{labelformat=empty}
      \psfrag{t}[b][b]{\small{Time (sec)}}
      \psfrag{q}{\hspace{-1mm}\small{Queue size (pkts)}}
      \psfrag{a}{\hspace{-1mm}\small{Average queue size (pkts)}}
      \psfrag{0}[b][b]{\scriptsize{$0$}}
      \psfrag{8}[b][b]{\scriptsize{$8$}}
      \psfrag{10}[b][b]{\scriptsize{$10$}}
      \psfrag{15}[b][b]{\scriptsize{$15$}}
      \psfrag{30}[b][b]{\scriptsize{$30$}}
      \psfrag{300}[b][b]{\scriptsize{$300$}}
      \psfrag{25}[b][b]{\scriptsize{$25$}}
      \psfrag{100}[b][b]{\scriptsize{$100$}}
      \psfrag{200}[b][b]{\scriptsize{$200$}}
      \psfrag{400}[b][b]{\scriptsize{$400$}}
      \psfrag{475}[c][t]{\scriptsize{$475$}}
      \psfrag{500}[c][t]{\scriptsize{$500$}}
 \begin{center}
    \subfloat[\normalsize{$\overline{b}=15, \underline{b}=8$ pkts}]{ \includegraphics[width=1.6in,height=2.6in,angle=270]{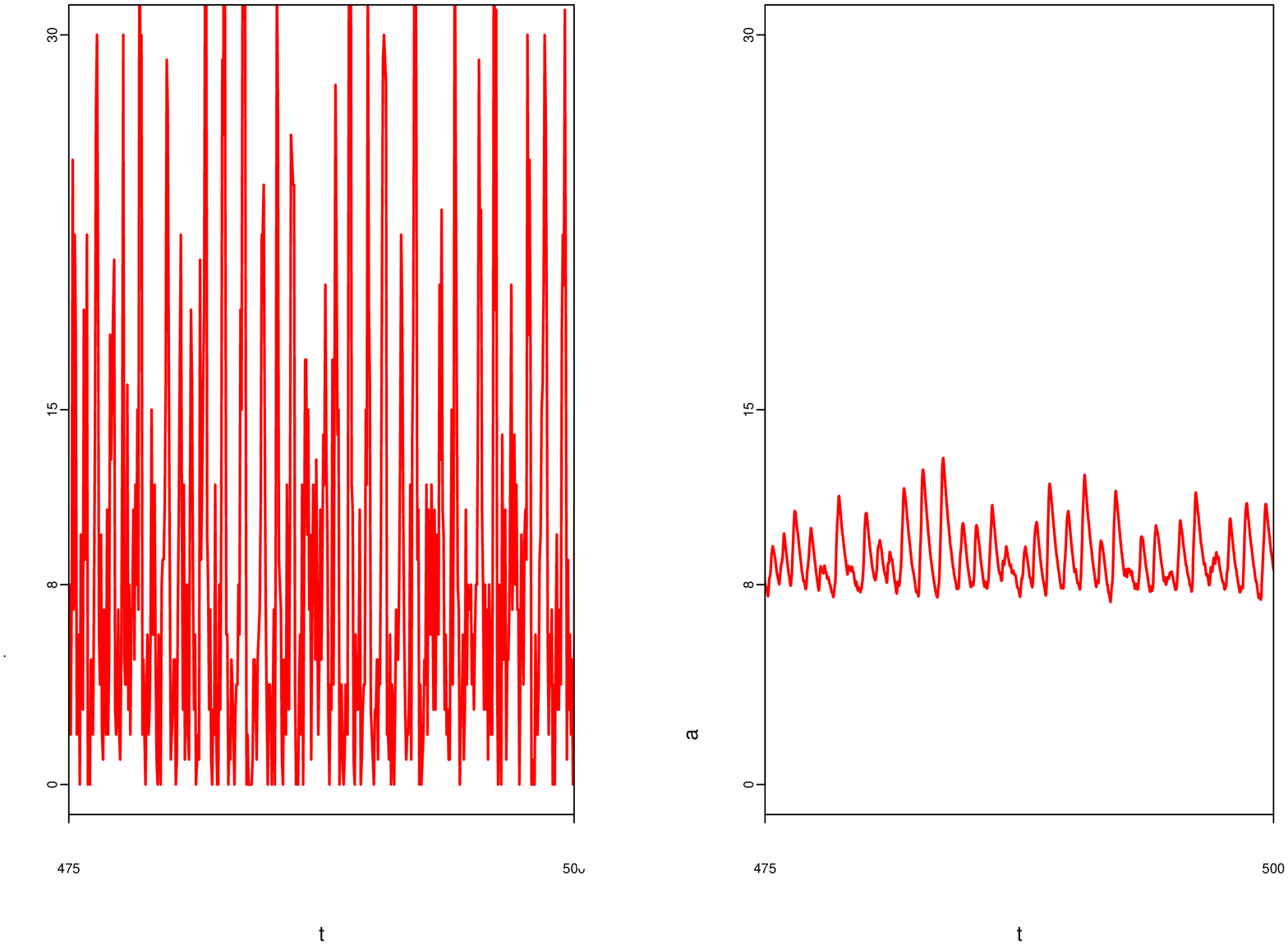}}\qquad
    \subfloat[\normalsize{$\overline{b}=200, \underline{b}=100$ pkts}]{ \includegraphics[width=1.6in,height=2.6in,angle=270]{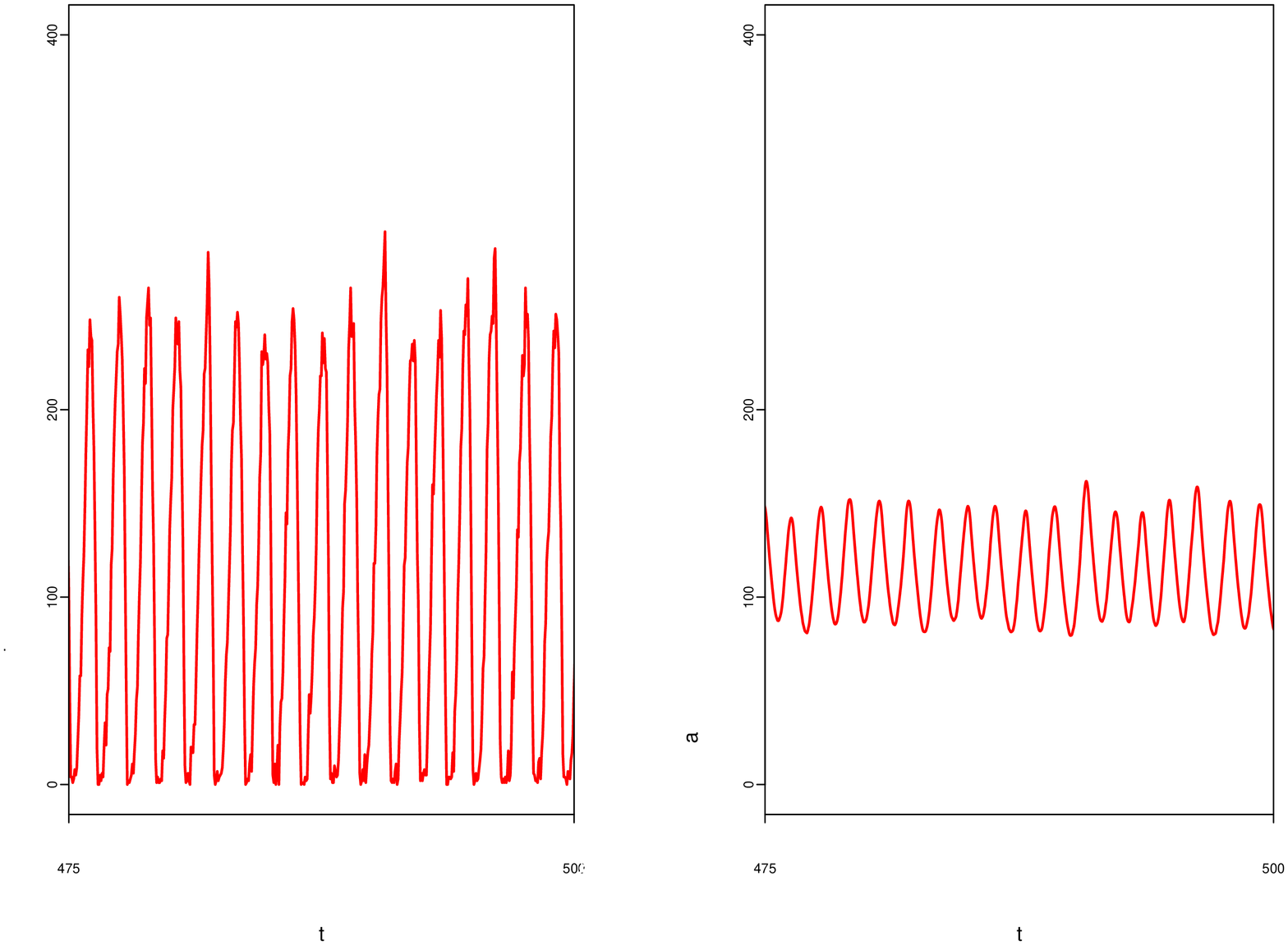}}
  \caption{Impact of packet-dropping threshold (Packet-level simulations. Heterogeneous traffic with RED policy). $B = 2084$ pkts, RTT $=100$ ms. RED and Compound parameters are set at default values. Queue size oscillates for large thresholds.}\vspace{-5mm}
  \label{fig:RED_threshold_sensitivity_MT}
 \end{center}
\end{figure}

In this traffic setting, we repeat the experiments conducted in Section~\ref{sec:RED_sims_homo}. The traces obtained for simulations conducted to examine RTT sensitivity (Figure~\ref{fig:RED_RTT_sensitivity_MT} and Figure~\ref{fig:RED_2RTT_sensitivity_MT}), impact of averaging (Figure~\ref{fig:RED_averaging_effect_MT}) and threshold sensitivity (Figure~\ref{fig:RED_threshold_sensitivity_MT}) appear qualitatively similar to the ones observed for Compound TCP traffic. This indicates that the insight obtained from the analysis may hold even as the underlying system deviates from model assumptions. 

So far, we have seen that large RTTs and large thresholds impact system stability. Our analysis and simulations also reveal that queue size averaging may not be beneficial to RED performance. Researchers have established that tuning RED parameters is not straight forward, and RED requires precise tuning in order to avoid instabilities~\cite{Bonald2000,christiansen2000tuning,may1999reasons}. In light of this, one may argue that simple queue policies that can give equivalent performance, and do not require extensive parameter tuning may be desirable. Additionally, such policies must be designed to ensure system stability irrespective of the round-trip time of the TCP flows. We outline one such queue policy in the next section.

\section{A threshold-based queue policy}
\label{sec:threshold_policy}

Motivated by the results obtained from our study of Compound TCP-RED, we now outline a simple threshold-based queue policy. This policy has a tunable packet-dropping threshold which can be tuned to ensure small queues and reduced queueing delay. Consider the following microscopic rule for the queue management policy 
\begin{align*}
 \bigg\{\begin{array}{l l}
   \text{if }q(t) \geq q_{th}, & \text{drop the packet},\\
   \text{else}, & \text{admit to the queue},\\
 \end{array}
\end{align*}
where $q_{th}$ is a tunable packet-dropping threshold, that must be fixed at a few tens of packets to ensure system stability. 

The above queue policy has only one parameter to tune, and is rather simple in terms of implementation. 
For a large number of users, this rule can be approximated as the marking probability of an M/M/1 queue, which is~\cite{raja2012delay}
\begin{align*}
 p(w(t)) = \bigg(\frac{w(t)}{\tilde{C}\tau}\bigg)^{q_{th}}.
\end{align*}
Using this model for the packet-drop probability, the model for Compound TCP with the threshold-based queue policy would be
\begin{align}
 \dot{w}(t) =& \Big(i\big(w(t)\big)\big(1-p(w(t-\tau))\big)-d\big(w(t)\big)p(w(t-\tau))\Big)\frac{w(t-\tau)}{\tau},\label{eq:Compound_Threshold}
\end{align}
where $i\big(w(t)\big) = \alpha w(t)^{k-1}$ and $d\big(w(t)\big)=\beta w(t)$. All the symbols have the same interpretation as before. We now outline some local stability results for system~\eqref{eq:Compound_Threshold}.

The equilibrium $w^\ast$ of system \eqref{eq:Compound_Threshold} satisfies
\begin{align}
 i(w^\ast)\big(1-p(w^\ast)\big) = d(w^\ast)p(w^\ast).\label{eq:Compound_Threshold_equilibrium}
\end{align}
Let $u(t)=w(t)-w^\ast$ be a perturbation about the equilibrium. Linearising~\eqref{eq:Compound_Threshold} about $w^\ast$ yields
\begin{align}
 \dot{u}(t) =& \Big(i'(w^\ast)\big(1-p(w^\ast)\big)-d'(w^\ast)p(w^\ast)\Big)\frac{w^\ast}{\tau} u(t)- p'(w^\ast)\big(i(w^\ast)+d(w^\ast)\big)\frac{w^\ast}{\tau}u(t-\tau).\label{eq:Compound_Threshold_lin}
\end{align}
Note that, $p'(w^\ast) = \big(\mathrm{d} p(w)/\mathrm{d}w\big)_{w=w^\ast}$. Looking for exponential solutions of \eqref{eq:Compound_Threshold_lin}, we would get
\begin{align}
 \lambda + a_1 + a_2 e^{-\lambda\tau} = 0,\label{eq:char_1st_order}
\end{align}
where 
\begin{align*}
 a_1 =&\, -\big(i'(w^\ast)(1-p(w^\ast))-d'(w^\ast)p(w^\ast)\big)\frac{w^\ast}{\tau}=\,(2-k)i(w^\ast)(1-p^\ast)\frac{1}{\tau} > 0,\notag\\
 a_2 =& \,p'(w^\ast)\big(i(w^\ast)+d(w^\ast)\big)\frac{w^\ast}{\tau} =\, q_{th}\, i(w^\ast)\frac{1}{\tau} > 0.
\end{align*}
Necessary and sufficient condition for local stability of system \eqref{eq:Compound_Threshold} is~\cite{raina2005local}
\begin{align}
 \tau\,\,\sqrt{a_2^2-a_1^2} < \cos^{-1}\big(-a_1/a_2\big).\label{eq:Compound_Threshold_ness}
\end{align}
When the inequality in \eqref{eq:Compound_Threshold_ness} is replaced with an equality, we obtain the corresponding Hopf condition.
A sufficient condition for stability of system~\eqref{eq:Compound_Threshold} is~\cite{raina2005local}
\begin{align}
 a_2\tau < \pi/2.\label{eq:Compound_Threshold_suff}
\end{align}
Condition~\eqref{eq:Compound_Threshold_ness} may be written in terms of the system parameters as
\begin{align}
\alpha (w^\ast)^{k-1}\sqrt{q_{th}^2-\big((k-2)(1-p(w^\ast))\big)^2} <\, \cos^{-1}\bigg(\frac{(k-2)\big(1-p(w^\ast)\big)}{q_{th}}\bigg).\label{eq:Compound_Threshold_necc_buffer}
\end{align} 
Similarly, the sufficient condition~\eqref{eq:Compound_Threshold_suff} can be written as
\begin{align}
  \alpha\, q_{th}\,(w^\ast)^{k-1} < \pi/2.\label{eq:Compound_Threshold_suff_buffer}
\end{align}
The above sufficient condition depends on the value of $(w^\ast)^{k-1}$. Upon the assumption that $1-p^\ast \approx 1$, one may derive the following expression for $(w^\ast)^{k-1}$ from equation~\eqref{eq:Compound_Threshold_equilibrium}
\begin{align*}
(w^\ast)^{k-1} = \big(\alpha\,(\tilde{C}\tau)^{q_{th}}/\beta\big)^{(k-1)/(q_{th}+2-k)}.
\end{align*}
As the Compound parameter $k < 1$, and the value of $q_{th}$ is to be fixed at a few tens of packets, $(w^\ast)^{k-1}$ reduces as $\tau$ increases. Therefore, if the threshold $q_{th}$ is fixed in accordance with the Compound parameter $\alpha$ to satisfy the sufficient condition~\eqref{eq:Compound_Threshold_suff}, the system will remain stable irrespective of the round-trip time of the Compound TCP flows.
% We now present some stability charts representing the Hopf condition and the sufficient condition for local stability. 
We now present some computations to illustrate these stability conditions.
\begin{figure}
  \begin{center}
  \subfloat[Stability chart]{
  \psfrag{20}{\scriptsize{$20$}}
  \psfrag{40}{\scriptsize{$40$}}
  \psfrag{60}{\scriptsize{$60$}}
  \psfrag{80}{\scriptsize{$80$}}
  \psfrag{100}{\scriptsize{$100$}}
  \psfrag{0}[c][t]{\scriptsize{$0$}}
  \psfrag{0.25}[c][t]{\scriptsize{$0.25$}}
  \psfrag{0.5}[c][t]{\scriptsize{$0.5$}}
  \psfrag{qth}{\small{Queue threshold, $q_{th}$}}
  \psfrag{a}{\small{Protocol parameter, $\alpha$}}
  \psfrag{st}{\small{stable region}}
  \psfrag{tc}[l][c][1][0]{\small{Hopf condition}}
  \psfrag{ts}[l][c][1][0]{\small{Sufficient condition}}
  \includegraphics[width=1.6in,height=2.2in,angle=270]{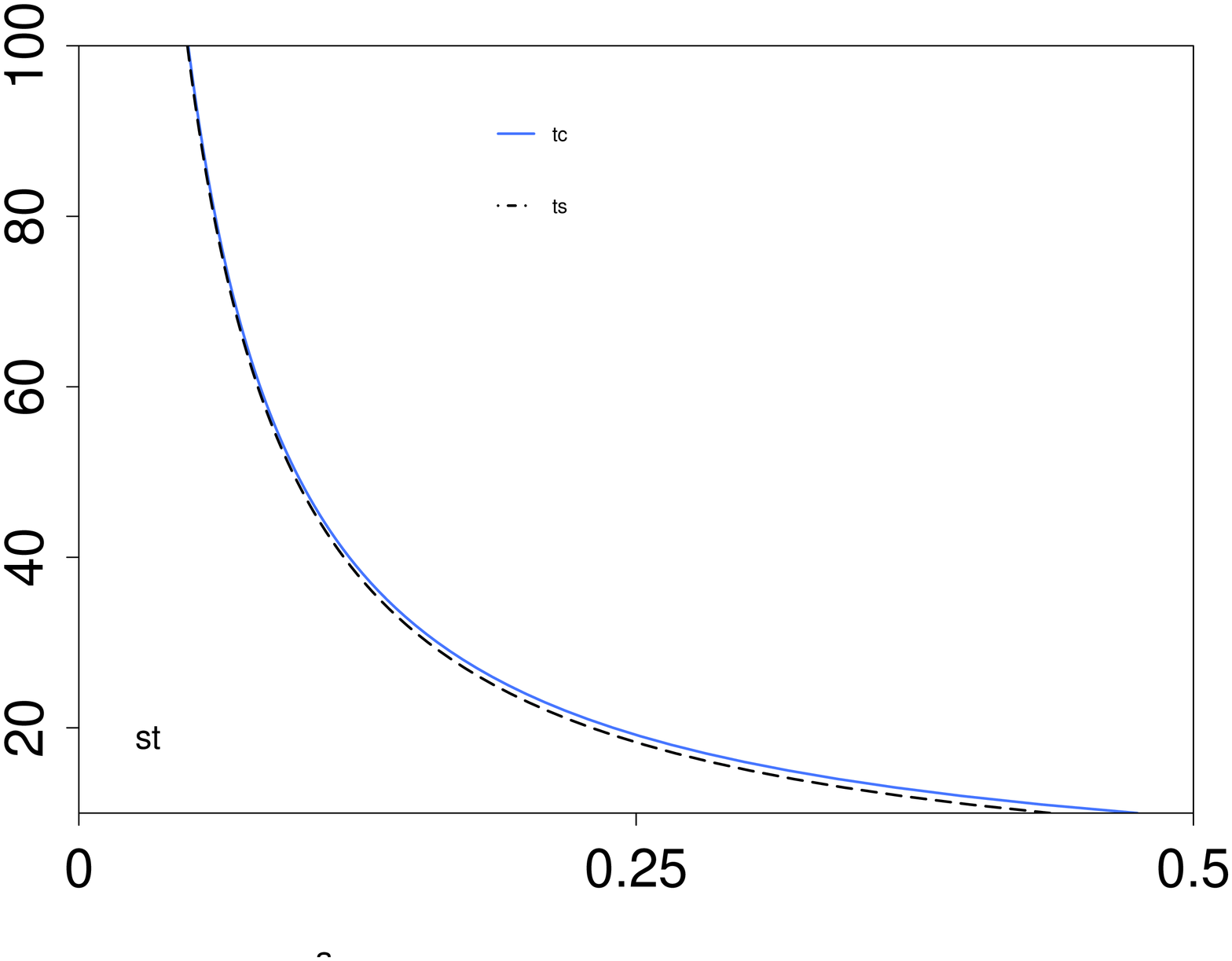}\label{fig:RED_stability_chart_1st_order}
  }\qquad
  \subfloat[Bifurcation diagram]{
% \end{center}
%  \end{figure}
%   \begin{figure}[tbh]
%   \begin{center}
  \psfrag{20}[c][t]{\scriptsize{$20$}}
  \psfrag{40}[c][t]{\scriptsize{$40$}}
  \psfrag{60}[c][t]{\scriptsize{$60$}}
  \psfrag{80}[c][t]{\scriptsize{$80$}}
  \psfrag{100}[c][t]{\scriptsize{$100$}}
  \psfrag{50}{\scriptsize{$50$}}
  \psfrag{75}{\scriptsize{$75$}}
  \psfrag{qth}{\small{Queue threshold, $q_{th}$}}
  \psfrag{w}[l][c][1][0]{\small{Window size, $w(t)$}}
  \includegraphics[width=1.6in,height=2.2in,angle=270]{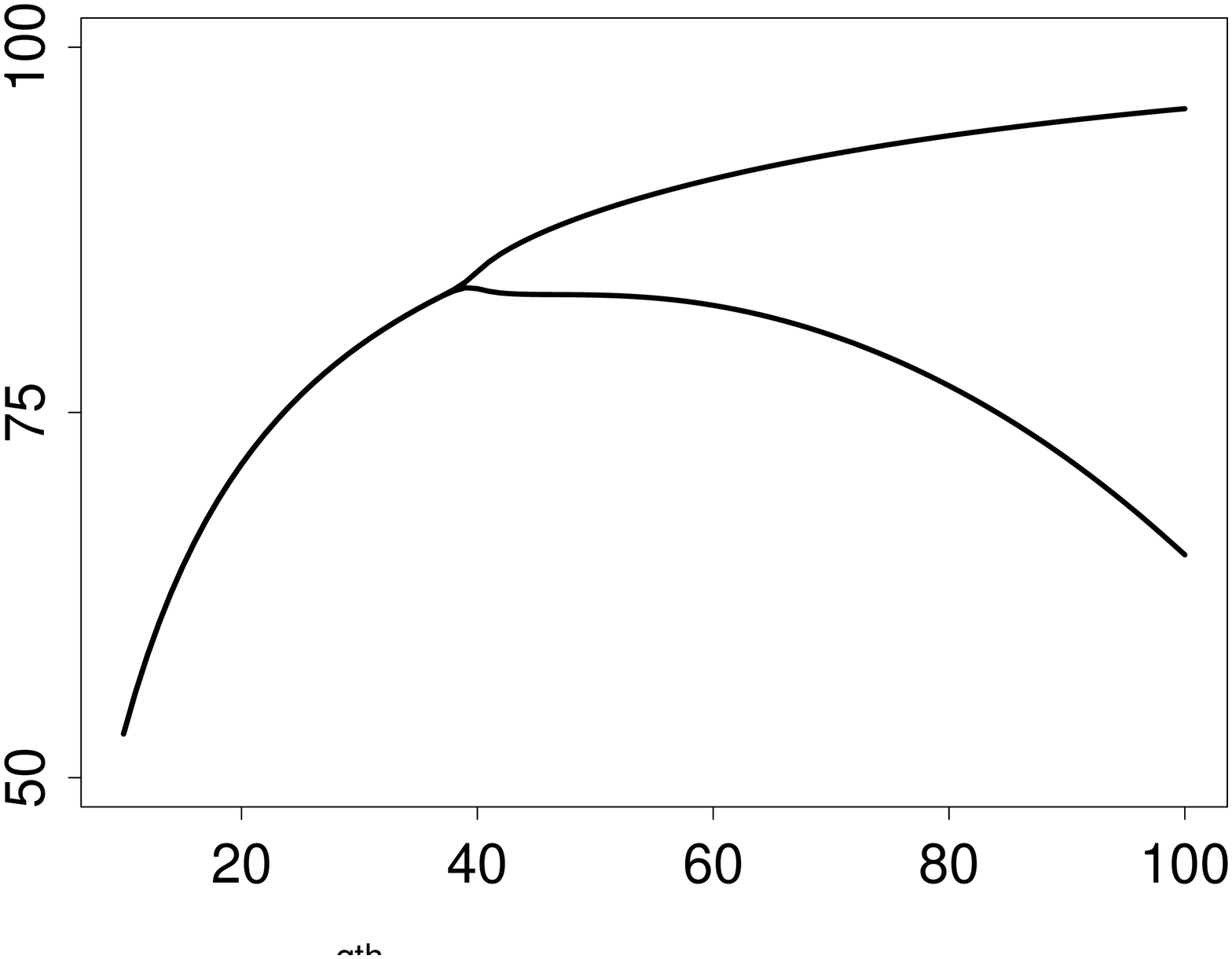}\label{fig:RED_bifurcation_diagram_1st_order}
  }
  \caption{Local stability and Hopf bifurcation in a system of Compound TCP flows operating over a router with the threshold-based queue policy. Observe that as the threshold $q_{th}$ increases, Compound TCP parameter $\alpha$ would have to be reduced to maintain stability. The system undergoes a Hopf bifurcation $q_{th}~=~39$, for default value of $\alpha$, as predicted from the stability chart. Observe the amplitude of the limit cycle increasing as $q_{th}$ increases.}
\end{center}
 \end{figure}
 
\subsection{Computations}
The sufficient condition~\eqref{eq:Compound_Threshold_suff_buffer} suggests that Compound TCP with threshold-based queue policy can be stabilised by tuning the protocol parameter $\alpha$ and the queue threshold $q_{th}$. Therefore, we plot the Hopf condition and the sufficient condition derived above in terms of $\alpha$ and $q_{th}$. System parameters are fixed as: $k=0.75,\beta=0.5,\tilde{C}=100$ pkts/sec, $\tau=1$ sec. We vary the threshold $q_{th}$ in the range $[10,100]$ pkts, and compute the value of $\alpha$ corresponding to the Hopf condition, and the boundary of the sufficient condition. The plot obtained is presented in Figure~\ref{fig:RED_stability_chart_1st_order}. As queue threshold increases, the value of $\alpha$ would have to be reduced to ensure stability. Observe that the sufficient condition is not very conservative, and could lead to accurate design. 

When the system parameters satisfy the Hopf condition, the system undergoes a Hopf bifurcation and we expect the emergence of limit cycles. To observe this transition in system dynamics, we plot the bifurcation diagram which represents the amplitude of the oscillations observed in the window dynamics. We fix $\alpha=0.125$ and the rest of the parameters as mentioned above, and vary the threshold $q_{th}$ in the range $[10,100]$ pkts. For $\alpha = 0.125$, the corresponding value of $q_{th}$ on the Hopf condition is $39$ pkts. Therefore, we expect the window dynamics to converge to equilibrium for $q_{th} < 39$, and break into a limit cycle for $q_{th} \geq 39$. This phenomenon can be observed in Figure~\ref{fig:RED_bifurcation_diagram_1st_order}. Observe that the amplitude of the limit cycle increases as the threshold is further increased. It is noteworthy that the equilibrium window size itself increases as $q_{th}$ is increased, which is 
expected from equilibrium condition. 

\subsection{Remarks}
We now make the following remarks regarding Compound TCP operating in conjunction with the threshold-based queue policy:
\begin{enumerate}
\item [(i)] System stability does not explicitly depend on RTT.
 \item [(ii)] System can be stabilised by tuning $\alpha$ and $q_{th}$ as per sufficient condition~\eqref{eq:Compound_Threshold_suff_buffer}.
 \item [(iii)] Smaller threshold (a few tens of packets) aids system stability.
\end{enumerate}
Owing to the simplicity of the sufficient condition~\eqref{eq:Compound_Threshold_suff_buffer}, we may conclude that Compound TCP may be easier to control, when operating in conjunction with this threshold-based queue policy. We now conduct some packet-level simulations to compare the performance of the threshold policy with that of RED.

 \subsection{Performance evaluation}
 \label{sec:performance_evaluation}

The threshold-based queue policy drops packets once the queue size reaches the packet-dropping threshold. When this threshold is fixed to a small value such as a few tens of packets, this queue policy could ensure reduced queueing delay. Additionally, this policy is also seen to ensure stability with Compound TCP flows, regardless of the round-trip time. To that end, it appears to be a promising alternative for queue management at the routers. However, a detailed performance evaluation of this policy is in order. We now present some packet-level simulations to compare and contrast the performance of RED and the threshold-based queue policy, in a regime of small thresholds. 

We use two different network settings for these simulations. The first is the single bottleneck dumbbell topology that is described in Section~\ref{sec:sims}. Then, we use a network topology called the parking-lot topology, that is often used in simulation-based study of TCP-AQM systems,~\cite{chen2007design,katabi2002congestion,park2004analysis}.

   \begin{figure}[t]
 \psfrag{T10}{\hspace{-12mm}Round-trip time = $10$ ms}
      \psfrag{T200}{\hspace{-12mm}Round-trip time = $200$ ms}
      \psfrag{t}[b][b]{\small{Time (sec)}}
      \psfrag{q}{\small{\hspace{-1mm}Queue size (pkts)}}
      \psfrag{u}{\small{\hspace{2.2mm}Link utilisation ($\%$)}}
%       \psfrag{p}{\begin{scriptsize}Packet loss (pkts)\end{scriptsize}}
      \psfrag{0}[b][b]{\scriptsize{0}}
      \psfrag{15}[b][b]{\scriptsize{15}}
      \psfrag{75}[b][b]{\scriptsize{75}}
      \psfrag{46}[b][b]{\scriptsize{46}}
      \psfrag{60}[b][b]{\scriptsize{60}}
      \psfrag{15}[b][b]{\scriptsize{15}}
      \psfrag{53}[b][b]{\scriptsize{53}}
      \psfrag{100}[b][b]{\scriptsize{100}}
      \psfrag{74}[b][b]{\scriptsize{74}}
      \psfrag{79}[b][b]{\scriptsize{79}}
      \psfrag{55}[b][b]{\scriptsize{55}}
      \psfrag{38}[b][b]{\scriptsize{38}}
      \psfrag{72}[b][b]{\scriptsize{72}}
       \psfrag{475}[c][t]{\scriptsize{475}}
      \psfrag{500}[c][t]{\scriptsize{500}}
      \psfrag{300}[b][b]{\scriptsize{300}}
     \begin{center}
     \begin{tabular}{c c}
     \toprule      RED policy with $\overline{b} = 15, \underline{b}=8$ pkts &  Threshold policy with $q_{th} = 15$ pkts\\ 
    \cmidrule(r){1-1}\cmidrule(lr){2-2}
     \raisebox{1mm}{\includegraphics[width=4.27in,height=2.75in,angle=270]{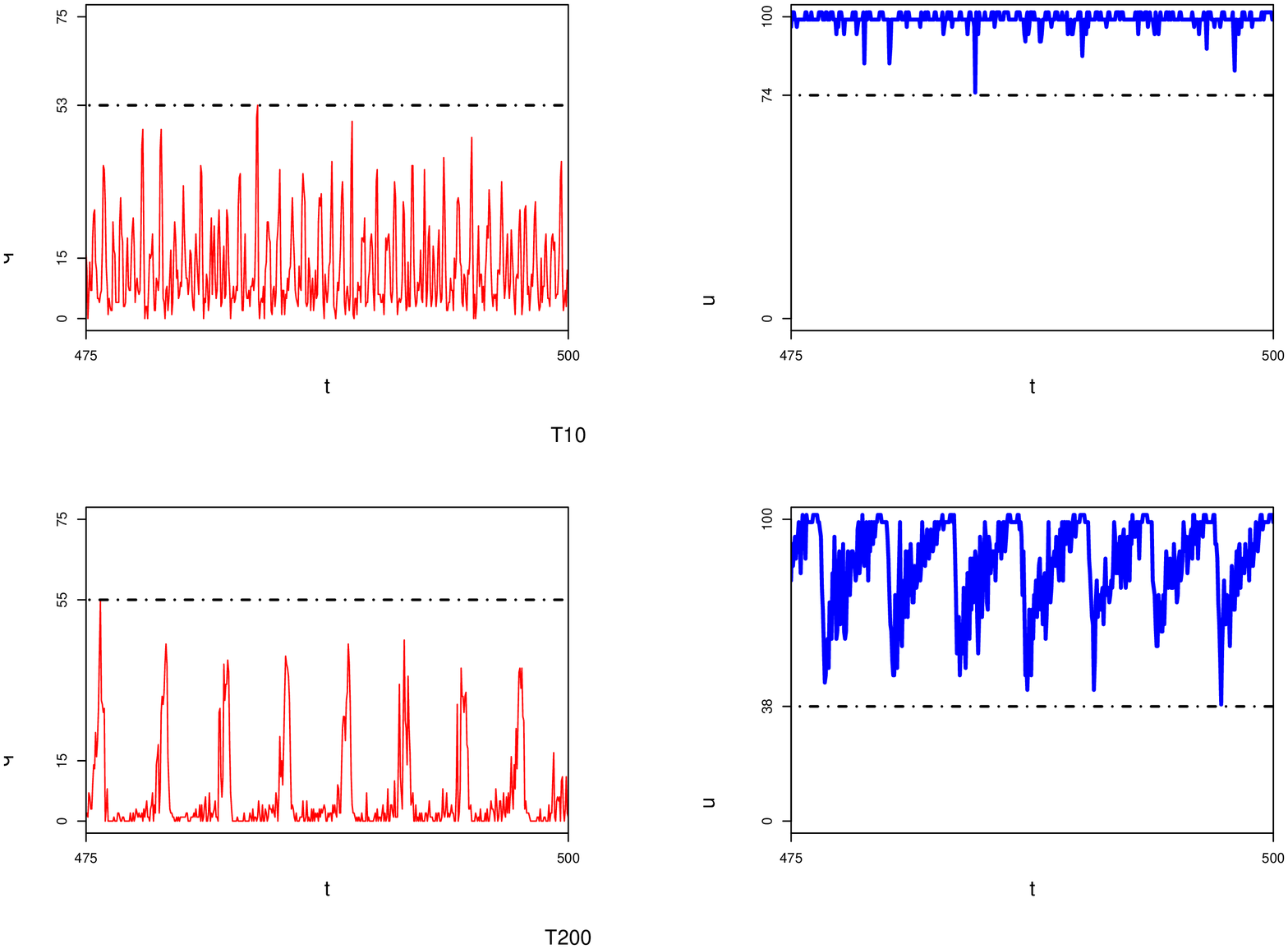}}
      & 
     \raisebox{1mm}{\includegraphics[width=4.27in,height=2.75in,angle=270]{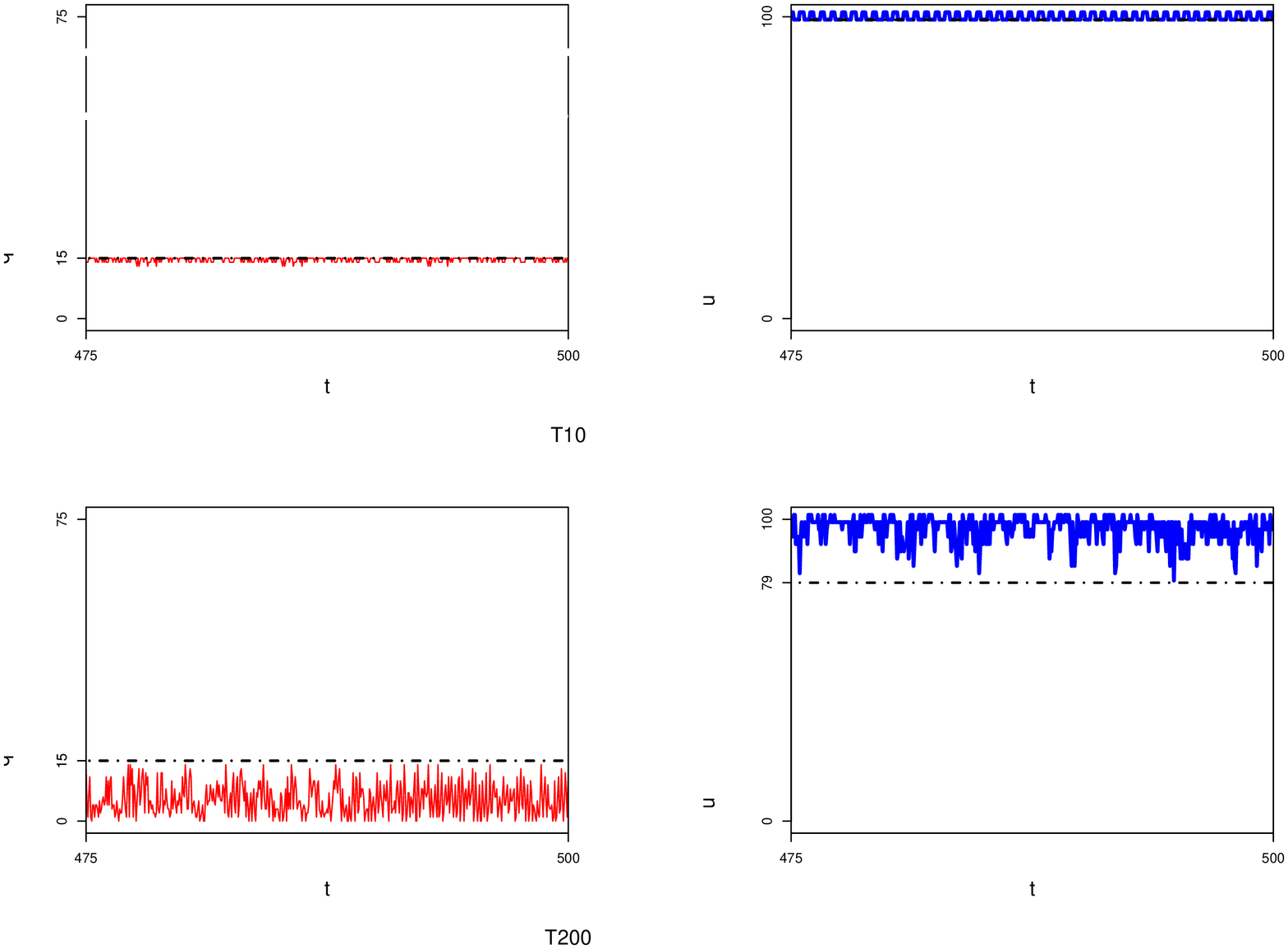}}
     \vspace{0.5mm} \\ 
      \bottomrule
      \end{tabular}
      \caption{Comparison of RED and threshold-based queue policy. The threshold policy successfully maintains the queue size below $15$ pkts, and ensures higher link utilisation for both RTTs. }\label{fig:comparison_RED_Threshold}
      \end{center}
\end{figure}
\subsubsection{Single bottleneck}
 
The network topology is the same as described in Section~\ref{sec:sims}. We consider the homogeneous traffic scenario, \emph{i.e.} 60 Compound TCP flows. For more details of the traffic setting, the reader is referred to Section~\ref{sec:RED_sims_homo}. For RED, we fix the thresholds as follows: $\overline{b}=15, \underline{b}=8$ pkts, aiming to maintain small queues and hence reduce queueing delay. The queue weight is chosen automatically as per the default configuration in the NS2 implementation of RED, in order to ensure that RED operates in its full capacity. The value of $\overline{p}$ is retained at its default value. For the threshold policy, $q_{th} = 15$ pkts.
 
We first consider two values for the average round-trip time, $10, 200$ ms. The traces of the queue size and link utilisation obtained for both the queue policies for these RTTs are presented in Figure~\ref{fig:comparison_RED_Threshold}. It is seen that for RTT $= 10$ ms, the queue size appears to randomly fluctuate in case of RED, but the threshold policy successfully maintains the queue at the desired value. The threshold policy ensures $100$ \% link utilisation, while in case of RED the link utilisation is seen to drop occasionally. When the round-trip time is increased to $200$ ms, both policies fail to maintain the queue size at $15$ pkts. This affects the link utilisation, which drops to a minimum of $79\%$ in case of threshold and $38\%$ in case of RED. It is noteworthy that, for both the RTTs, RED fails to maintain the queue size below the upper threshold $15$ pkts, and we see occasional bursts in the queue size which reaches upto $60$ pkts. Meanwhile, the threshold policy successfully ensures that the queue size remains below $15$ pkts for both round-trip times, owing to the deterministic dropping of packets. The implications of such deterministic dropping on other performance metrics needs to be investigated. 
 
We now consider the following range of RTTs: $[10,200]$ ms. With all the other parameters being the same, we conduct simulations for each RTT in the range. We record the mean queueing delay, mean packet loss percentage and mean throughput as observed from the traces corresponding to the last 25 seconds of each simulation. These quantities as a function of the round-trip time are plotted in Figures~\ref{fig:PE_RED_Th_1} and~\ref{fig:PE_RED_Th_2} . Observe that the threshold policy ensures a smaller queueing delay. The threshold policy is expected to drop more packets compared to RED owing to its deterministic dropping of packets once the queue size reaches the threshold $q_{th}$. However, traces of mean packet loss suggest that the packet loss is almost equal for the two policies. In fact, for large RTTs, the mean packet loss of the threshold policy drops below that of RED. Notably, the mean throughput is equal for both the policies.

We also conduct simulations to compare the Average Flow Completion Time (AFCT) of the two queue policies. AFCT is crucial to user experience and is possibly the most important metric from the users' perspective~\cite{dukkipati2006flow}. For these simulations, we use the same Compound TCP traffic, but each flow is required to transfer packets worth of $500$ MB total, after which the TCP flow is stopped. We vary the RTT in the range $[10,200]$ ms, and run each simulation until the last TCP flow is complete. For each simulation, we record the time required for each flow to complete, and then compute the average. Repeating this for each RTT in the specified range, we get a plot of the AFCT with respect to RTT, also presented in Figure~\ref{fig:PE_RED_Th_1}. It is observed that the threshold-based policy facilitates faster flow completion compared to RED. With both the queue policies, AFCT increases as RTT increases, but the increase in steeper in case of RED. For 200 ms RTT, the difference in the AFCT is seen to be about 5 mins. 

We conducted a similar set of simulations with a heterogeneous traffic mix of Compound TCP, CUBIC, UDP and HTTP flows. The reader is referred to Section~\ref{sec:RED_sims_hetro}, for the details of the traffic mix. The plots of mean queueing delay, mean packet loss percentage, mean throughput and average flow completion time for this traffic scenario is presented in Figures~\ref{fig:PE_RED_Th_1} and~\ref{fig:PE_RED_Th_2}. The threshold-based policy is observed to perform better than RED in this traffic scenario as well.

 \begin{figure}
 \begin{center}
 \newcommand{\myarrowlength}{1.5}
 \newcommand{\myarrowsize}{0.08cm 5.0}
 \newcommand{\mylinewidth}{0.06}
 \scalebox{0.59} % Change this value to rescale the drawing.
{
\begin{pspicture}(3.5,-3.5)(25,3.5)
\psframe[linewidth=\mylinewidth,dimen=outer](11.62,0.75)(5.56,-0.75) %queue box
\psframe[linewidth=\mylinewidth,dimen=outer](3.2,0.4)(2.7,0.9) % top box 3
\psline[linewidth=\mylinewidth,arrowsize=\myarrowsize,arrowlength=\myarrowlength,arrowinset=0.4]{->}(3.2,0.65)(5.58,0.3)
\psframe[linewidth=\mylinewidth,dimen=outer](3.2,1.0)(2.7,1.5) % top box 2
\psline[linewidth=\mylinewidth,arrowsize=\myarrowsize,arrowlength=\myarrowlength,arrowinset=0.4]{->}(3.2,1.25)(5.58,0.4)
\psframe[linewidth=\mylinewidth,dimen=outer](3.2,1.6)(2.7,2.1) % top box 1
\psline[linewidth=\mylinewidth,arrowsize=\myarrowsize,arrowlength=\myarrowlength,arrowinset=0.4]{->}(3.2,1.85)(5.58,0.5)

\psline[linestyle=dashed,linecolor=black](2.95,0.4)(2.95,-0.4)
 \psframe[linewidth=\mylinewidth,dimen=outer](3.2,-0.4)(2.7,-0.9) % bottom box 3
\psline[linewidth=\mylinewidth,arrowsize=\myarrowsize,arrowlength=\myarrowlength,arrowinset=0.4]{->}(3.2,-0.65)(5.58,-0.3)
\psframe[linewidth=\mylinewidth,dimen=outer](3.2,-1.0)(2.7,-1.5) % bottom box 2
\psline[linewidth=\mylinewidth,arrowsize=\myarrowsize,arrowlength=\myarrowlength,arrowinset=0.4]{->}(3.2,-1.25)(5.58,-0.4)
\psframe[linewidth=\mylinewidth,dimen=outer](3.2,-1.6)(2.7,-2.1) % bottom box 1
\psline[linewidth=\mylinewidth,arrowsize=\myarrowsize,arrowlength=\myarrowlength,arrowinset=0.4]{->}(3.2,-1.85)(5.58,-0.5)
\pscircle[linewidth=\mylinewidth,dimen=outer](12.23,0.0){0.6} %queue circle
% \psline[linewidth=\mylinewidth](0.8,-2.5)(22.17,-2.5) % bottom big line
% \psarc[linewidth=\mylinewidth](13.6,1.25){1.25}{-90.0}{90.0} %top right semi-cicle
% \psarc[linewidth=\mylinewidth](22.17,-1.25){1.25}{-90.0}{90.0} %bottom right semi-circle
% text input
% \rput(2.5,0){\begin{huge}$\Vast\{$\end{huge}}
% \rput(2.5,-1.2){\begin{huge}$\bigg\{$\end{huge}}
\rput(12.18,0.0){\textbf{\begin{huge}$C$\end{huge}}}
\psline[linewidth=\mylinewidth,arrowsize=\myarrowsize,arrowlength=\myarrowlength,arrowinset=0.4]{->}(12.83,0)(15.5,0) % horizontal line into second queue

\rput(4,2.2){\textbf{\begin{huge}$w(t)$\end{huge}}}

\rput(1.7,2.2){\textbf{\begin{huge}$S_{A1}$\end{huge}}}
\psline[linestyle=dotted,linewidth=\mylinewidth,linecolor=black](2,2)(2,0.6)
\rput(1.7,0.4){\textbf{\begin{huge}$S_{A30}$\end{huge}}}

\rput(1.7,-0.4){\textbf{\begin{huge}$S_{A31}$\end{huge}}}
\psline[linestyle=dotted,linewidth=\mylinewidth,linecolor=black](2,-0.7)(2,-2)
\rput(1.7,-2.2){\textbf{\begin{huge}$S_{A60}$\end{huge}}}

\rput(26.6,2.2){\textbf{\begin{huge}$D_{A1}$\end{huge}}}
\psline[linestyle=dotted,linewidth=\mylinewidth,linecolor=black](26.3,2)(26.3,0.6)
\rput(26.6,0.4){\textbf{\begin{huge}$D_{A30}$\end{huge}}}

\rput(26.6,-0.4){\textbf{\begin{huge}$D_{B1}$\end{huge}}}
\psline[linestyle=dotted,linewidth=\mylinewidth,linecolor=black](26.3,-0.7)(26.3,-2)
\rput(26.6,-2.2){\textbf{\begin{huge}$D_{B30}$\end{huge}}}

% \rput(8.25,-3.25){\begin{huge} { \color{blue} Round-trip time, $\tau$}\end{huge}}
% \psarc[linewidth=\mylinewidth](0.8,-1.25){1.25}{90.0}{-90.0} %left bottom semi circle
% \psline[linewidth=\mylinewidth,arrowsize=\myarrowsize,arrowlength=\myarrowlength,arrowinset=0.4]{->}(0.8,0)(2.2,0) % bottom left small horizontal line
\psline[linewidth=\mylinewidth](8.6,0.75)(8.6,-0.75) %queue box - 2nd line
\psline[linewidth=\mylinewidth](9.6,0.75)(9.6,-0.75) %queue box - 3rd line
\psline[linewidth=\mylinewidth](10.6,0.75)(10.6,-0.75) %queue box - 4th line
\rput(7.60,0.0){\huge{$\cdots$}} %dots inside queue box
\psline[linewidth=\mylinewidth](6.4,0.75)(6.4,-0.75) %queue box - 1th line
\rput(6.00,-1.54){\textbf{\begin{huge}${B}$\end{huge}}}

\psframe[linewidth=\mylinewidth,dimen=outer](15.5,0.75)(21.56,-0.75) 
\rput(22.14,0.0){\textbf{\begin{huge}$C$\end{huge}}}
\pscircle[linewidth=\mylinewidth,dimen=outer](22.16,0.0){0.6} %% 2nd queue circle
\psline[linewidth=\mylinewidth](18.54,0.75)(18.54,-0.75) %queue box - 2nd line
\psline[linewidth=\mylinewidth](19.54,0.75)(19.54,-0.75) %queue box - 3rd line
\psline[linewidth=\mylinewidth](20.54,0.75)(20.54,-0.75) %queue box - 4th line
\rput(17.44,0.0){\huge{$\cdots$}} %dots inside queue box
\psline[linewidth=\mylinewidth](16.34,0.75)(16.34,-0.75)
\rput(15.94,-1.54){\textbf{\begin{huge}${B}$\end{huge}}}

%%%%%%%%%%%%%%%%%%%%%%%%%%%%intermediate sources %%%%%%%%%%%%%%%%%%%%%%%%%%%%%%%%%%%%%%%%%%%%%%%%%%%%%%%%%%%%%%%
\psframe[linewidth=\mylinewidth,dimen=outer](12.23,2.4)(12.73,1.9)
\psframe[linewidth=\mylinewidth,dimen=outer](12.83,2.4)(13.33,1.9)
\psline[linestyle=dashed,linecolor=black](13.33,2.15)(13.83,2.15)
\psframe[linewidth=\mylinewidth,dimen=outer](13.83,2.4)(14.33,1.9)
\psframe[linewidth=\mylinewidth,dimen=outer](14.43,2.4)(14.93,1.9)

\rput(12.23,3){\textbf{\begin{huge}$S_{B1}$\end{huge}}}
\psline[linestyle=dotted,linewidth=\mylinewidth,linecolor=black](12.8,3)(14.3,3)
\rput(14.9,3){\textbf{\begin{huge}$S_{B30}$\end{huge}}}

%%%%%%%%%%%%%%%%%%%%%%%%%%%%intermediate sinks %%%%%%%%%%%%%%%%%%%%%%%%%%%%%%%%%%%%%%%%%%%%%%%%%%%%%%%%%%%%%%%
\psframe[linewidth=\mylinewidth,dimen=outer](12.23,-2.4)(12.73,-1.9)
\psframe[linewidth=\mylinewidth,dimen=outer](12.83,-2.4)(13.33,-1.9)
\psline[linestyle=dashed,linecolor=black](13.33,-2.15)(13.83,-2.15)
\psframe[linewidth=\mylinewidth,dimen=outer](13.83,-2.4)(14.33,-1.9)
\psframe[linewidth=\mylinewidth,dimen=outer](14.43,-2.4)(14.93,-1.9)

\rput(12.23,-3){\textbf{\begin{huge}$D_{A31}$\end{huge}}}
\psline[linestyle=dotted,linewidth=\mylinewidth,linecolor=black](12.8,-3)(14.3,-3)
\rput(14.9,-3){\textbf{\begin{huge}$D_{A60}$\end{huge}}}

%%%%%%%%%%%%%%%%%%%%%%%%%%%%%%%%%%arrows from intermediate sources %%%%%%%%%%%%%%%%%%%%%%%%%%%%%%%
\psline[linestyle=solid,linecolor=black,linewidth=\mylinewidth](12.48,1.9)(12.48,1.3)
\psarc[linewidth=\mylinewidth](12.68,1.3){0.2}{180.0}{-100.0} 
\psline[linewidth=\mylinewidth,arrowsize=\myarrowsize,arrowlength=\myarrowlength,arrowinset=0.4]{->}(12.61,1.115)(15.5,0)
\psline[linestyle=solid,linecolor=black,linewidth=\mylinewidth](13.08,1.9)(13.08,1.3)
\psarc[linewidth=\mylinewidth](13.28,1.3){0.2}{180.0}{-105.0} 
\psline[linewidth=\mylinewidth,arrowsize=\myarrowsize,arrowlength=\myarrowlength,arrowinset=0.4]{->}(13.2,1.115)(15.5,0)
\psline[linestyle=solid,linecolor=black,linewidth=\mylinewidth](14.08,1.9)(14.08,1.3)
\psarc[linewidth=\mylinewidth](14.28,1.3){0.2}{180.0}{-112.0} 
\psline[linewidth=\mylinewidth,arrowsize=\myarrowsize,arrowlength=\myarrowlength,arrowinset=0.4]{->}(14.15,1.16)(15.5,0)
\psline[linestyle=solid,linecolor=black,linewidth=\mylinewidth](14.68,1.9)(14.68,1.3)
\psarc[linewidth=\mylinewidth](14.88,1.3){0.2}{180.0}{-142.0} 
\psline[linewidth=\mylinewidth,arrowsize=\myarrowsize,arrowlength=\myarrowlength,arrowinset=0.4]{->}(14.72,1.175)(15.5,0)

%%%%%%%%%%%%%%%%%%%%%%%%%%%%%%%%%%arrows to intermediate sinks %%%%%%%%%%%%%%%%%%%%%%%%%%%%%%%
\psline[linestyle=solid,linecolor=black,linewidth=\mylinewidth](12.83,0)(12.9,-0.25)
\psarc[linewidth=\mylinewidth](12.71,-0.3){0.2}{-25.0}{30.0}
\psline[linewidth=\mylinewidth,arrowsize=\myarrowsize,arrowlength=\myarrowlength,arrowinset=0.4]{->}(12.9,-0.35)(12.48,-1.9)
% % 
\psline[linestyle=solid,linecolor=black,linewidth=\mylinewidth](12.83,0)(13.2,-0.25)
\psarc[linewidth=\mylinewidth](13.075,-0.4){0.2}{-10.0}{60.0}
\psline[linewidth=\mylinewidth,arrowsize=\myarrowsize,arrowlength=\myarrowlength,arrowinset=0.4]{->}(13.275,-0.42)(13.08,-1.9)
% % 
\psline[linestyle=solid,linecolor=black,linewidth=\mylinewidth](12.83,0)(14.2,-0.25)
\psarc[linewidth=\mylinewidth](14.15,-0.44){0.2}{-25.0}{90.0} 
\psline[linewidth=\mylinewidth,arrowsize=\myarrowsize,arrowlength=\myarrowlength,arrowinset=0.4]{->}(14.35,-0.48)(14.05,-1.9)
% %
\psline[linestyle=solid,linecolor=black,linewidth=\mylinewidth](12.83,0)(14.9,-0.25)
\psarc[linewidth=\mylinewidth](14.8,-0.435){0.2}{-25.0}{90.0} 
\psline[linewidth=\mylinewidth,arrowsize=\myarrowsize,arrowlength=\myarrowlength,arrowinset=0.4]{->}(15,-0.475)(14.68,-1.9)

% %%% sinks %%%%%
\psframe[linewidth=\mylinewidth,dimen=outer](25.13,0.4)(25.63,0.9) % top box 3
\psline[linewidth=\mylinewidth,arrowsize=\myarrowsize,arrowlength=\myarrowlength,arrowinset=0.4]{->}(22.76,0)(25.13,0.65)
\psframe[linewidth=\mylinewidth,dimen=outer](25.13,1.0)(25.63,1.5) % top box 2
\psline[linewidth=\mylinewidth,arrowsize=\myarrowsize,arrowlength=\myarrowlength,arrowinset=0.4]{->}(22.76,0)(25.13,1.25)
\psframe[linewidth=\mylinewidth,dimen=outer](25.13,1.6)(25.63,2.1) % top box 1
\psline[linewidth=\mylinewidth,arrowsize=\myarrowsize,arrowlength=\myarrowlength,arrowinset=0.4]{->}(22.76,0)(25.13,1.85)
\psline[linestyle=dashed,linecolor=black](25.38,0.4)(25.38,-0.4)
 \psframe[linewidth=\mylinewidth,dimen=outer](25.13,-0.4)(25.63,-0.9) % bottom box 3
\psline[linewidth=\mylinewidth,arrowsize=\myarrowsize,arrowlength=\myarrowlength,arrowinset=0.4]{->}(22.76,0)(25.13,-0.65)
\psframe[linewidth=\mylinewidth,dimen=outer](25.13,-1.0)(25.63,-1.5) % bottom box 2
\psline[linewidth=\mylinewidth,arrowsize=\myarrowsize,arrowlength=\myarrowlength,arrowinset=0.4]{->}(22.76,0)(25.13,-1.25)
\psframe[linewidth=\mylinewidth,dimen=outer](25.13,-1.6)(25.63,-2.1) % bottom box 1
\psline[linewidth=\mylinewidth,arrowsize=\myarrowsize,arrowlength=\myarrowlength,arrowinset=0.4]{->}(22.76,0)(25.13,-1.85)
\end{pspicture}
}\caption{Parking-lot topology for the performance evaluation of queue policies. There are $60$ end systems, connected to the first a link. A half of them, labelled as $[S_{A1},S_{A30}]$, are connected to $[D_{A1},D_{A30}]$ on the other side of the second link. The rest half of the sources $[S_{A31},S_{A60}]$ are connected to destinations $[D_{A31},D_{A60}]$, which are connected to the first link. The traffic between these pairs of sources and destinations would have to traverse the first link alone. Meanwhile, the sources $[S_{B1},S_{B30}]$ are connected to destinations $[D_{B1}, D_{B30}]$ through the second link. Both links have service capacity $C$ and buffer size $B$.}
\label{fig:parking_lot}
\end{center}
\end{figure}
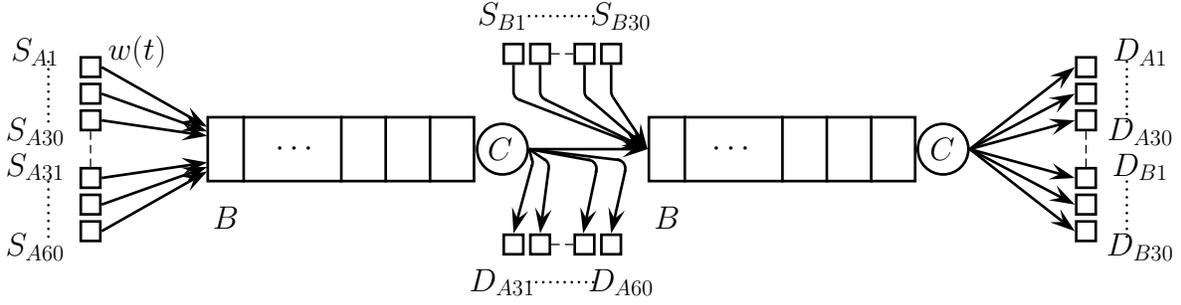

    \begin{figure}[tbh]
  \captionsetup[subfigure]{labelformat=empty}
      \psfrag{0}[b][b]{\scriptsize{$0$}}
      \psfrag{15}[b][b]{\scriptsize{$15$}}
      \psfrag{40}[b][b]{\scriptsize{$40$}}
      \psfrag{25}[b][b]{\scriptsize{$25$}}
      \psfrag{100}[b][b]{\scriptsize{$100$}}
      \psfrag{475}[c][t]{\scriptsize{$475$}}
      \psfrag{500}[c][t]{\scriptsize{$500$}}
  \psfrag{q}{\small\hspace{2mm}{Queue size (pkts)}}
  \psfrag{t}[b][b]{\small{Time (sec)}}
  \begin{center}
     \begin{tabular}{ c c c c}
      \toprule  \multicolumn{2}{c}{RED policy with $\overline{b} = 15,\underline{b}=8$ pkts} &  \multicolumn{2}{c}{Threshold policy with $q_{th} = 15$ pkts}\\ 
    \cmidrule(lr){1-2}\cmidrule(lr){3-4}
    Queue 1 & Queue 2 & Queue 1 & Queue 2\\
    \cmidrule(lr){1-1}\cmidrule(lr){2-2}\cmidrule(lr){3-3}\cmidrule(lr){4-4}
   \subfloat[]{
%   \begin{flushleft}
 \psfrag{T1}{ }
 \psfrag{T2}{ }
  \includegraphics[width=4.26in,height=1.23in,angle=270]{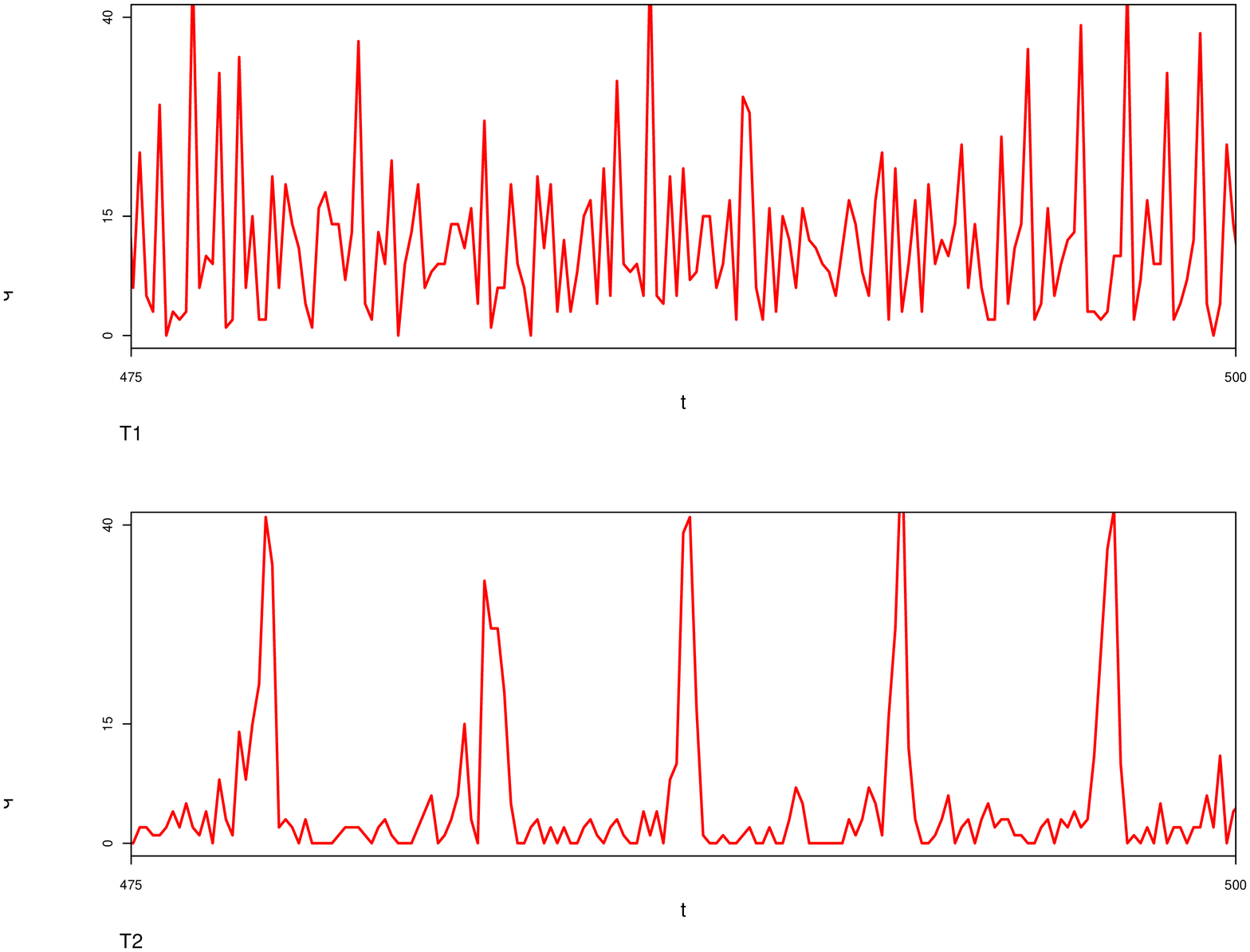}
%    \caption{(a) Queue 1}
%    \end{flushleft}
 }
  &
 \subfloat[]{
%   \begin{flushleft}
\psfrag{T1}{\hspace{8mm}Round-trip time = $10$ ms}
 \psfrag{T2}{\hspace{7.5mm}Round-trip time = $200$ ms}
  \includegraphics[width=4.26in,height=1.23in,angle=270]{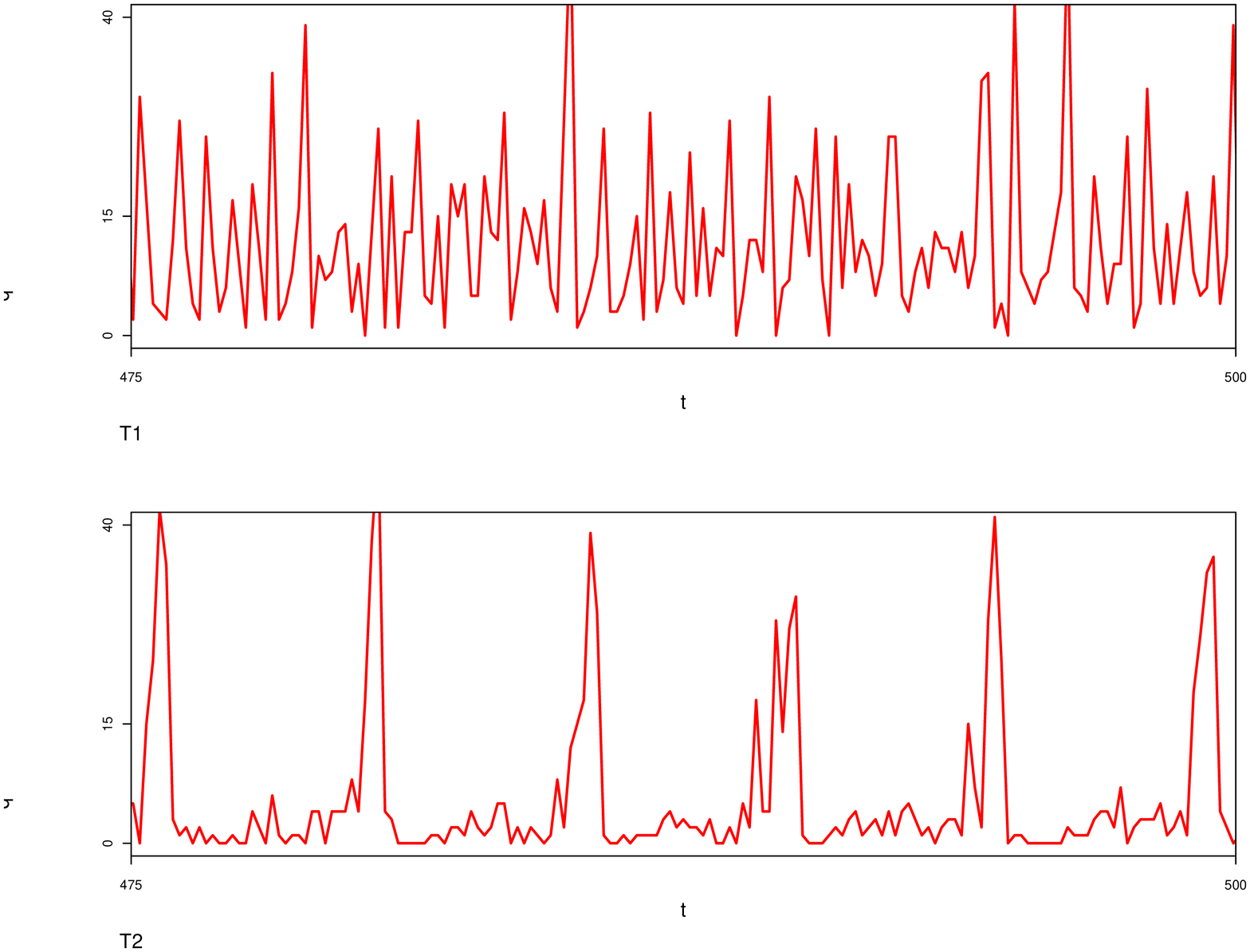}
%   \caption{(b) Queue 2}
%   \end{flushleft}
  }
 &
 \subfloat[]{
%  \begin{flushright}
\psfrag{T1}{ }
 \psfrag{T2}{ }
 \includegraphics[width=4.26in,height=1.23in,angle=270]{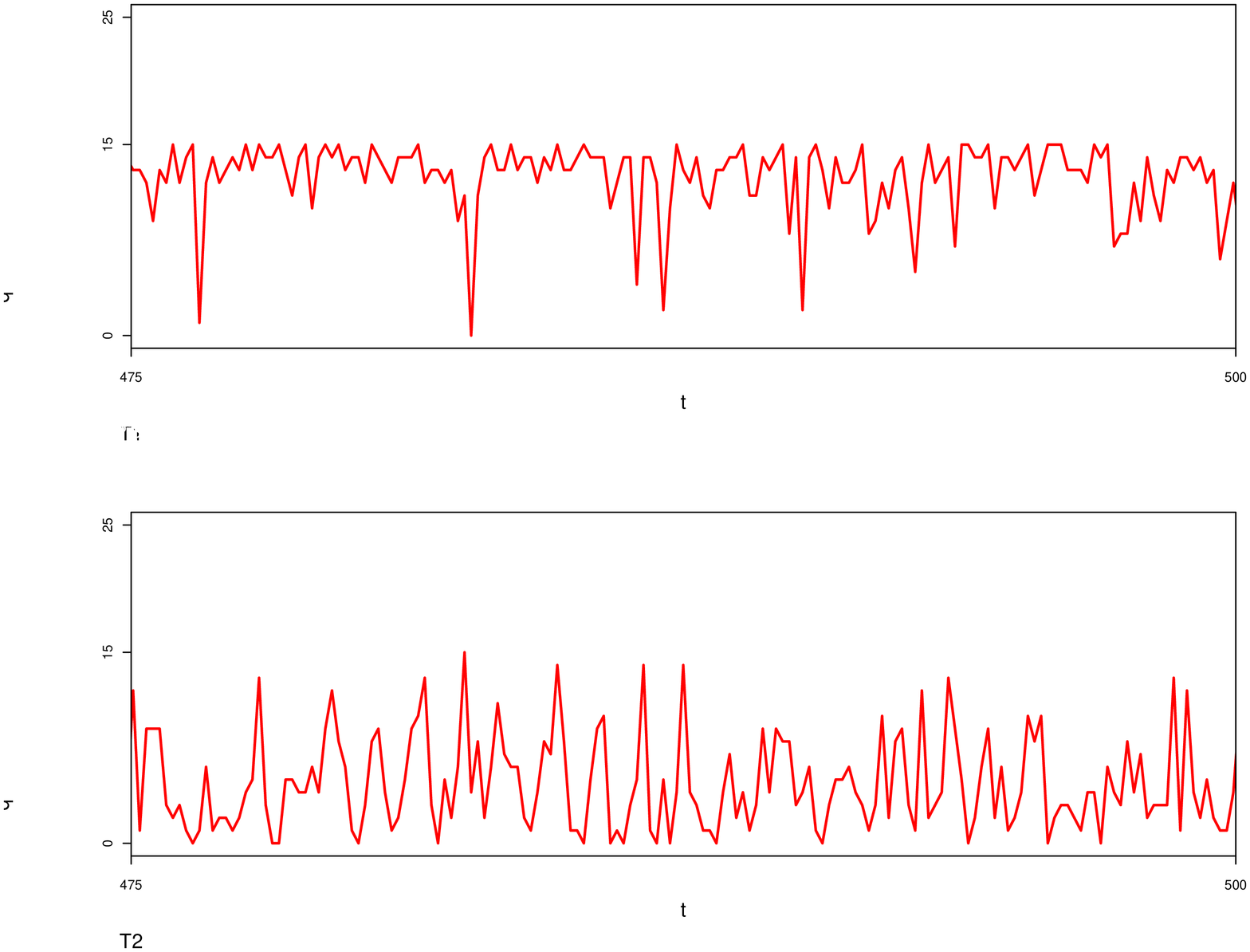}
%  \caption{(a) Queue 1}
%  \end{flushright}
}
 &
 \subfloat[]{
%  \begin{flushright}
\psfrag{T1}{ }
 \psfrag{T2}{ }
 \includegraphics[width=4.26in,height=1.23in,angle=270]{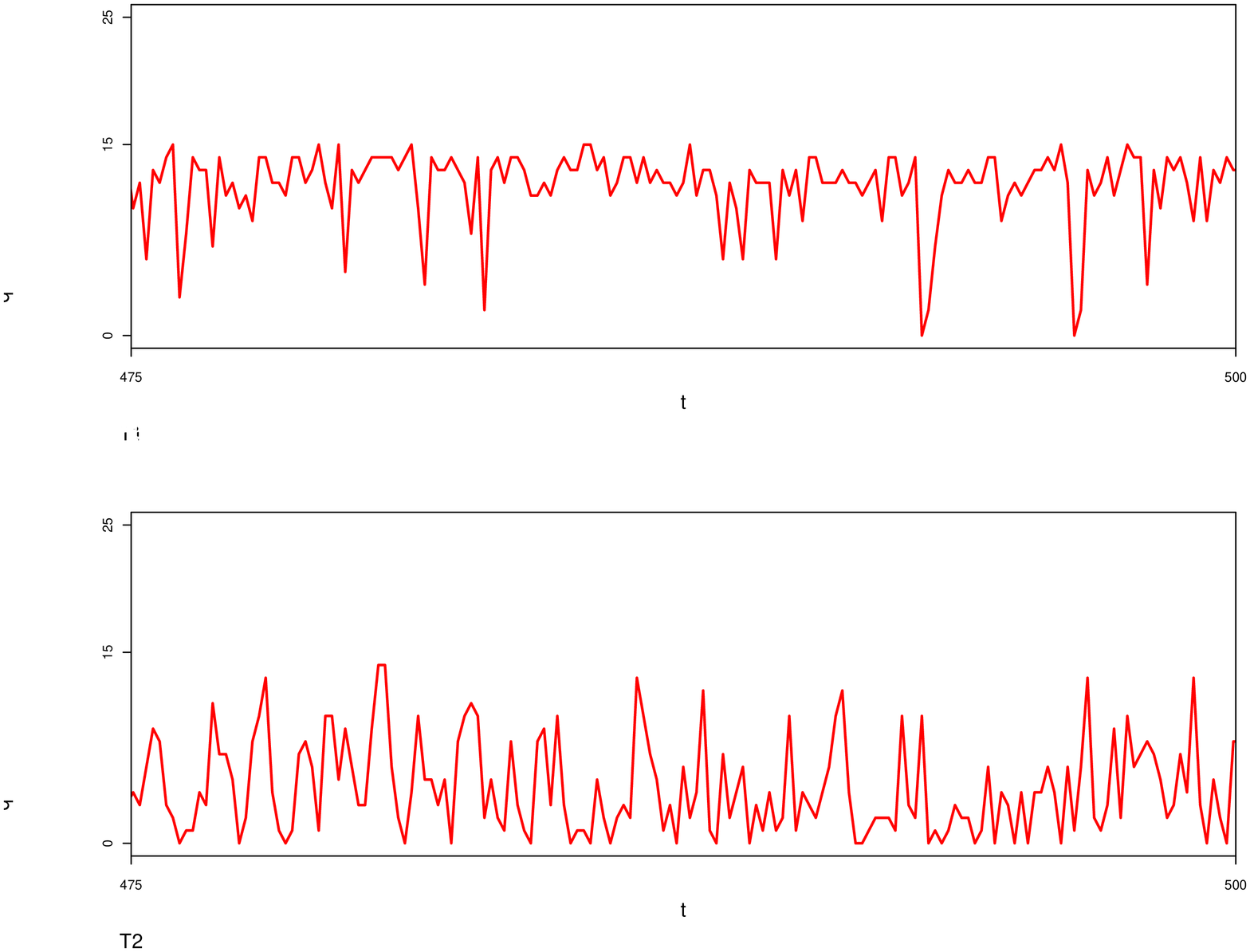}
%  \caption{(b) Queue 2}
%  \end{flushright}
 }
     \\ 
      \bottomrule
     \end{tabular}
  \caption{Comparison of RED and threshold-based queue policy in parking-lot topology with heterogeneous traffic. RED thresholds: $15$ and $8$ pkts. For threshold policy: $q_{th} = 15$ pkts. The threshold policy successfully maintains the queue size below $15$ pkts, for both RTTs.}\label{fig:comparison_RED_Threshold_PL}
\end{center}
 \end{figure}

  \begin{figure}[tbh]
  \psfrag{1}{\tiny{$1$}}
  \psfrag{2}{\tiny{$2$}}
    \psfrag{3}{\tiny{$3$}}
     \psfrag{4}{\tiny{$4$}}
  \psfrag{2.5}{\tiny{$2.5$}}
    \psfrag{0.5}{\tiny{$0.5$}}
  \psfrag{1.5}{\tiny{$1.5$}}
  \psfrag{100}[b][b]{\tiny{$100$}}
  \psfrag{200}[b][b]{\tiny{$200$}}
  \psfrag{300}{\tiny{$300$}}
  \psfrag{50}{\tiny{$50$}}
  \psfrag{150}{\tiny{$150$}}
  \psfrag{250}{\tiny{$250$}}
  \psfrag{0}{\tiny{$0$}}
  \psfrag{7}{\tiny{$7$}}
  \psfrag{9}{\tiny{$9$}}
      \psfrag{75}{\tiny{$75$}}
    \psfrag{25}{\tiny{$25$}}
    \psfrag{5}{\tiny{$5$}}
    \psfrag{8}{\tiny{$8$}}
  \psfrag{15}{\tiny{$15$}}
  \psfrag{40}{\tiny{$40$}}
  \psfrag{Af}{\hspace{1.5mm}\scriptsize{AFCT (mins)}}
  \psfrag{10}[b][b]{\tiny{$10$}}
  \psfrag{20}{\tiny{$20$}}
  \psfrag{30}{\tiny{$30$}}
  \psfrag{T}{\scriptsize{RTT (ms)}}
  \psfrag{Mq}{\hspace{4mm}\scriptsize{Mean QD (ms)}}
     \psfrag{Mp}{\scriptsize{Mean packet loss (\%)}}
        \psfrag{Mt}{\hspace{-2mm}\scriptsize{Mean throughput (Mbps)}}
  \psfrag{RED}[l][c][1][0]{\small{RED}}
  \psfrag{Th}[l][c][1][0]{\small{Threshold}}
 \begin{center}
  \begin{tabular}{*{5}{l  >{\centering\arraybackslash} m{0.21\textwidth} >{\centering\arraybackslash} m{0.21\textwidth} >{\centering\arraybackslash} m{0.21\textwidth} >{\centering\arraybackslash} m{0.21\textwidth}}}
  \toprule
   & \multicolumn{2}{c}{Single bottleneck} & \multicolumn{2}{c}{Parking-lot}\\
   \cmidrule(lr){2-3}\cmidrule(lr){4-5}
   &\multicolumn{1}{c}{ Homogeneous} & \multicolumn{1}{c}{Heterogeneous} & \multicolumn{1}{c}{Homogeneous} & \multicolumn{1}{c}{Heterogeneous}\\
 \cmidrule(lr){2-2} \cmidrule(lr){3-3}  \cmidrule(lr){4-4}  \cmidrule(lr){5-5}
   \rotatebox[origin=c]{90}{\scriptsize{Queueing delay}}
   & 
   \subfloat{
     \includegraphics[height=0.21\textwidth,width=0.21\textwidth,angle=270]{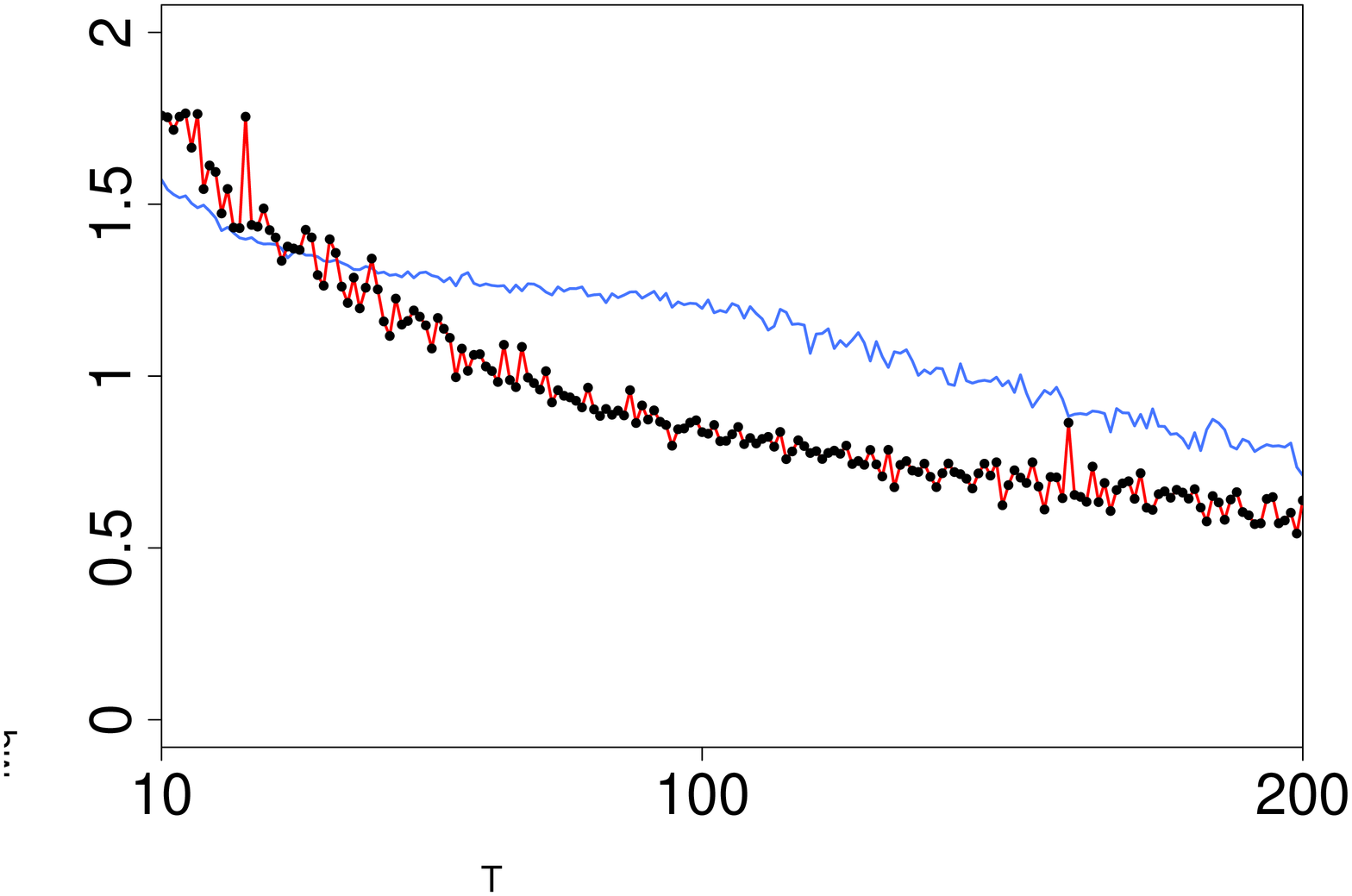}
   }
      &
      \subfloat{
      \includegraphics[height=0.21\textwidth,width=0.21\textwidth,angle=270]{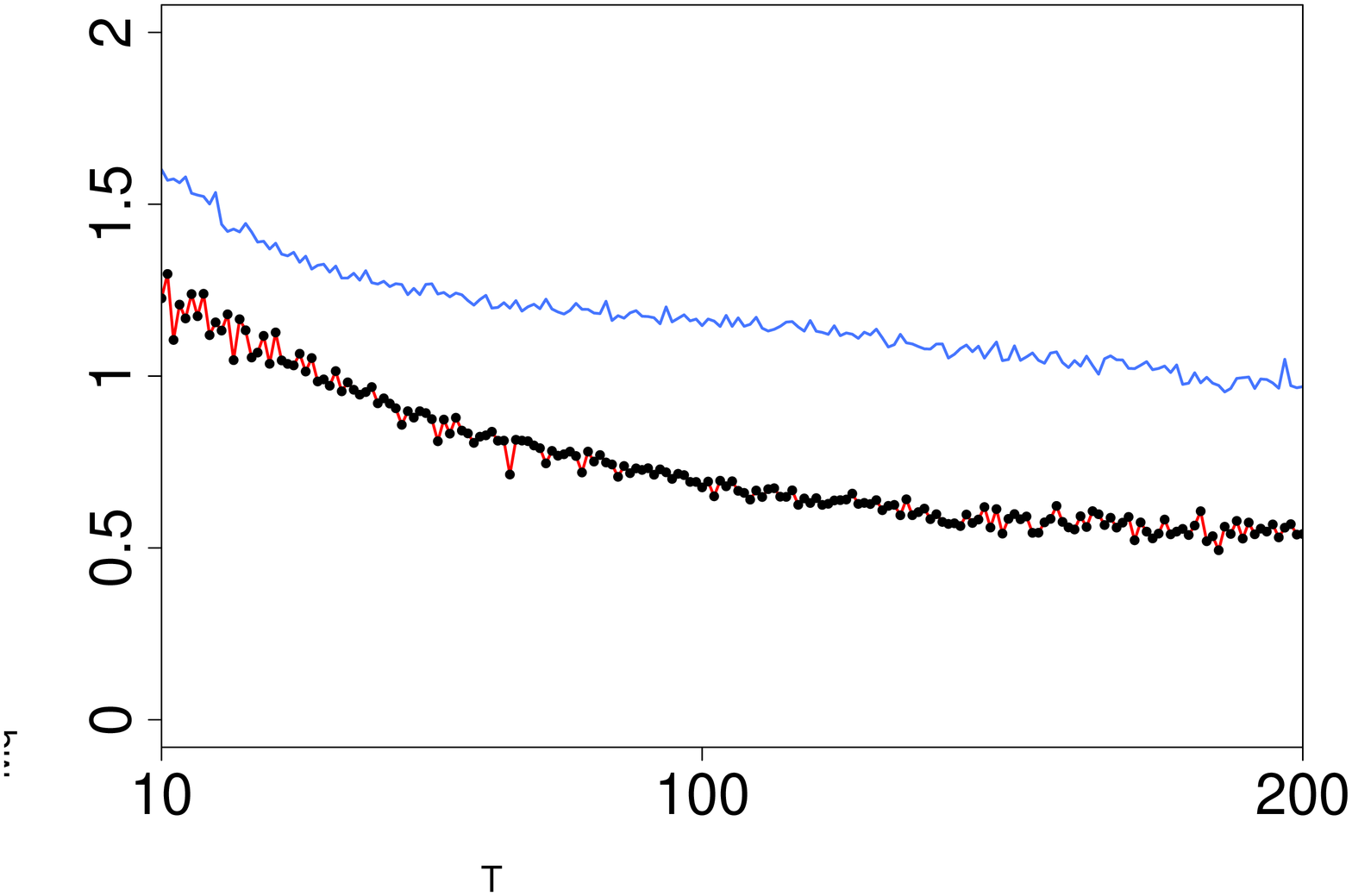}
      }
      &
      \subfloat{
      \includegraphics[height=0.21\textwidth,width=0.21\textwidth,angle=270]{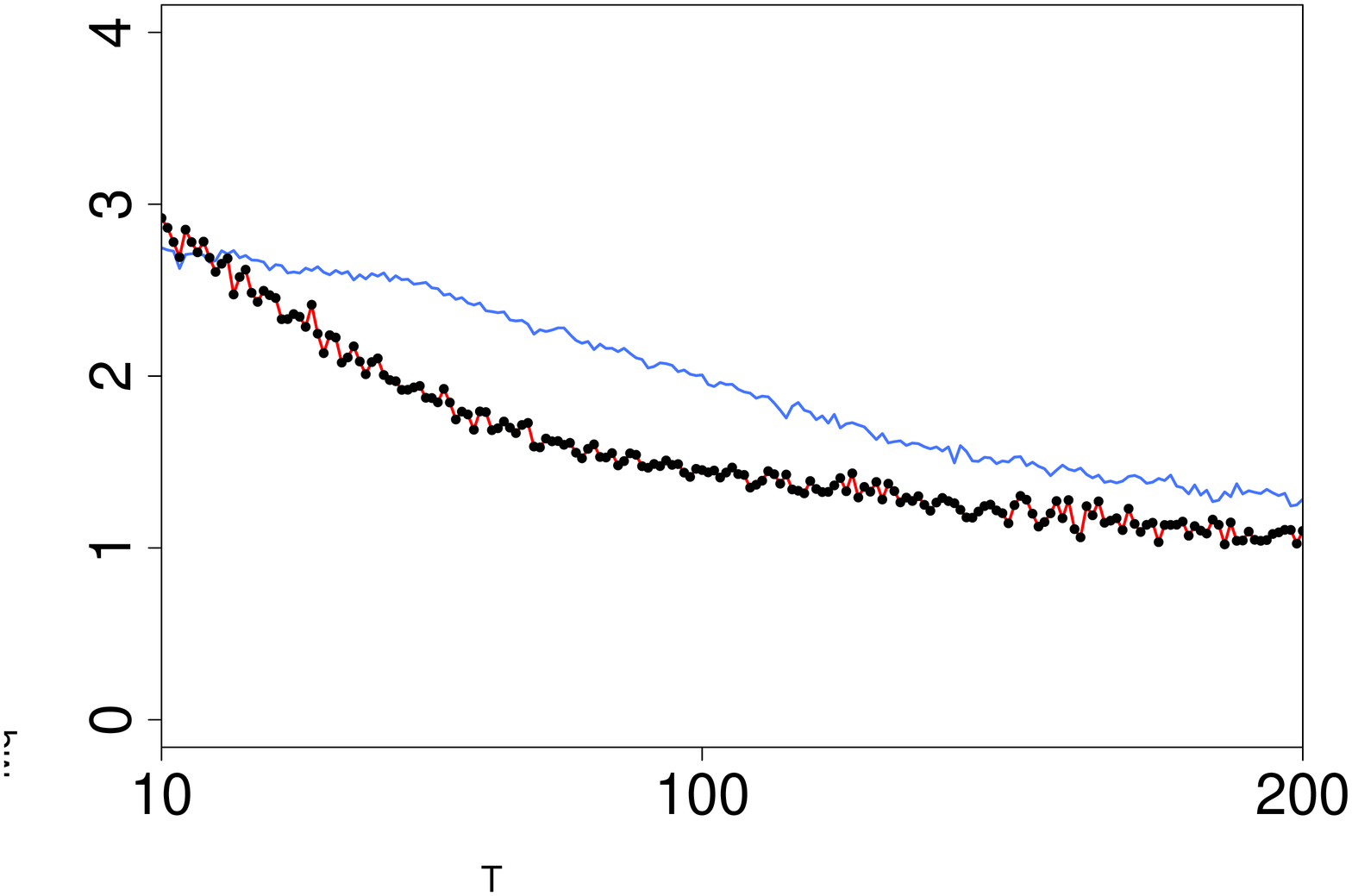}
      }
      &
     \subfloat{
     \includegraphics[height=0.21\textwidth,width=0.21\textwidth,angle=270]{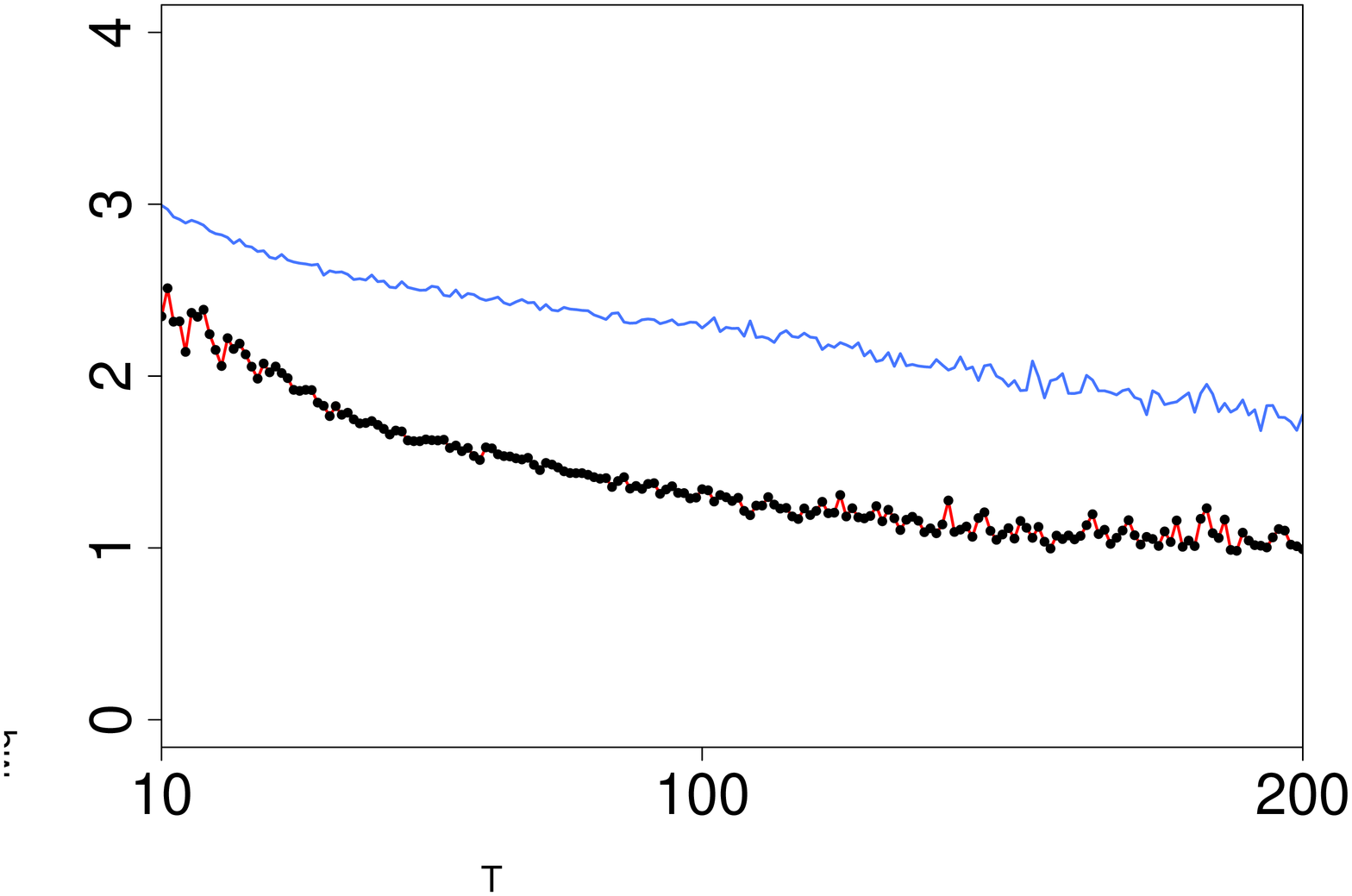}
     }
       \\
      \hline
      \rotatebox[origin=c]{90}{\scriptsize{Flow completion time}}  & 
      \subfloat{
      \includegraphics[height=0.21\textwidth,width=0.21\textwidth,angle=270]{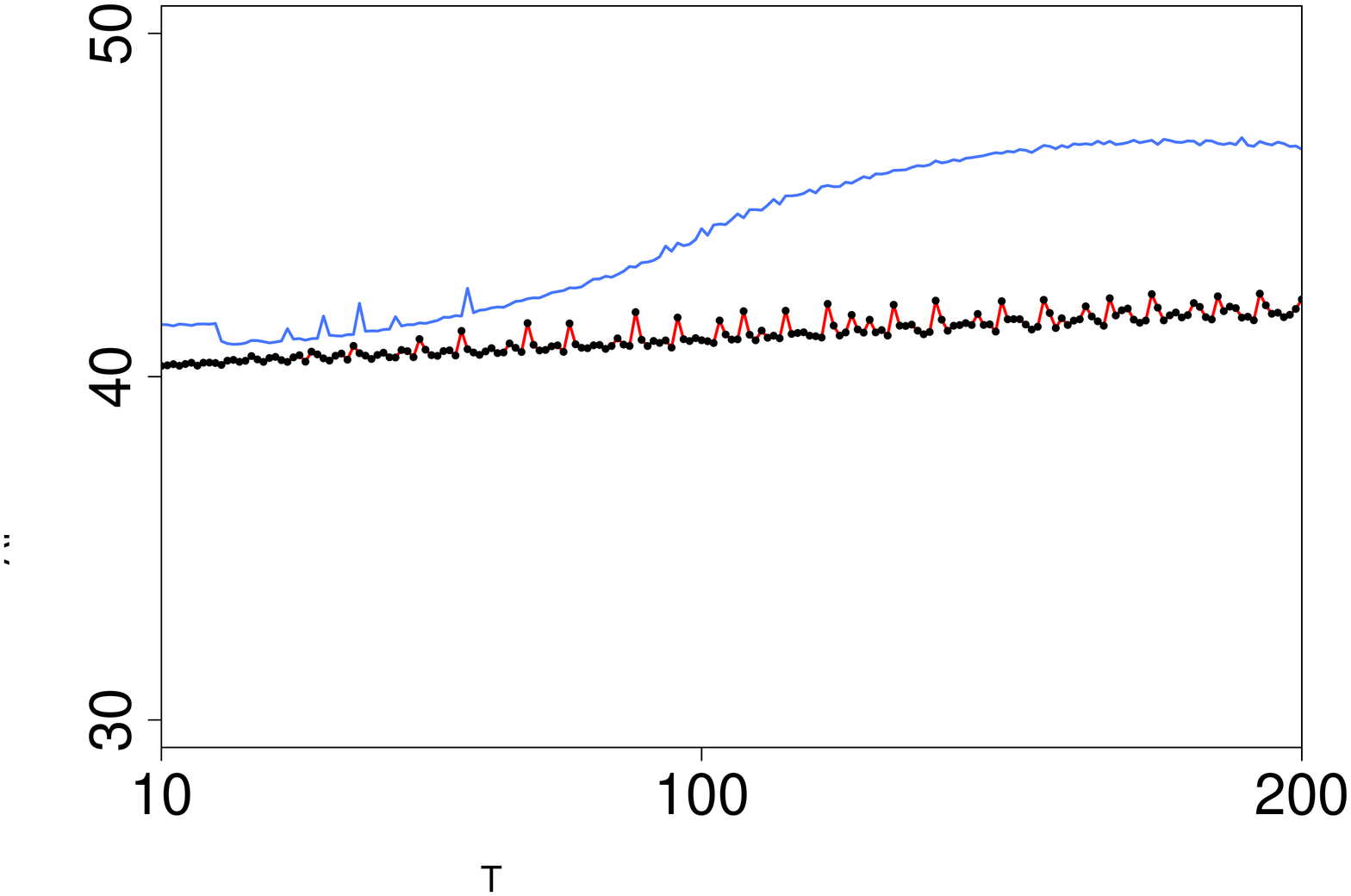}
      }
      &
      \subfloat{
      \includegraphics[height=0.21\textwidth,width=0.21\textwidth,angle=270]{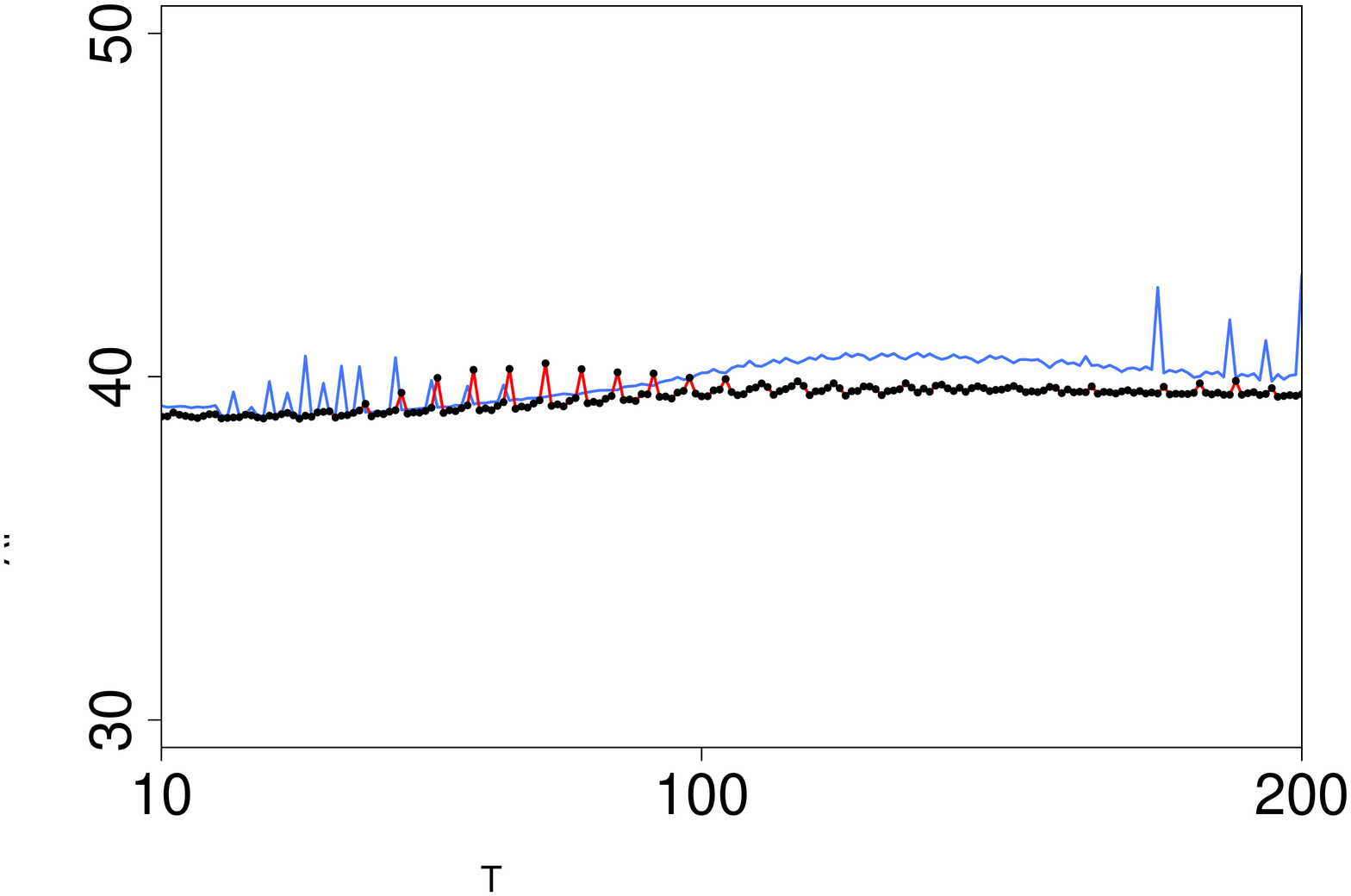}
      }
      &
     \subfloat{
      \includegraphics[height=0.21\textwidth,width=0.21\textwidth,angle=270]{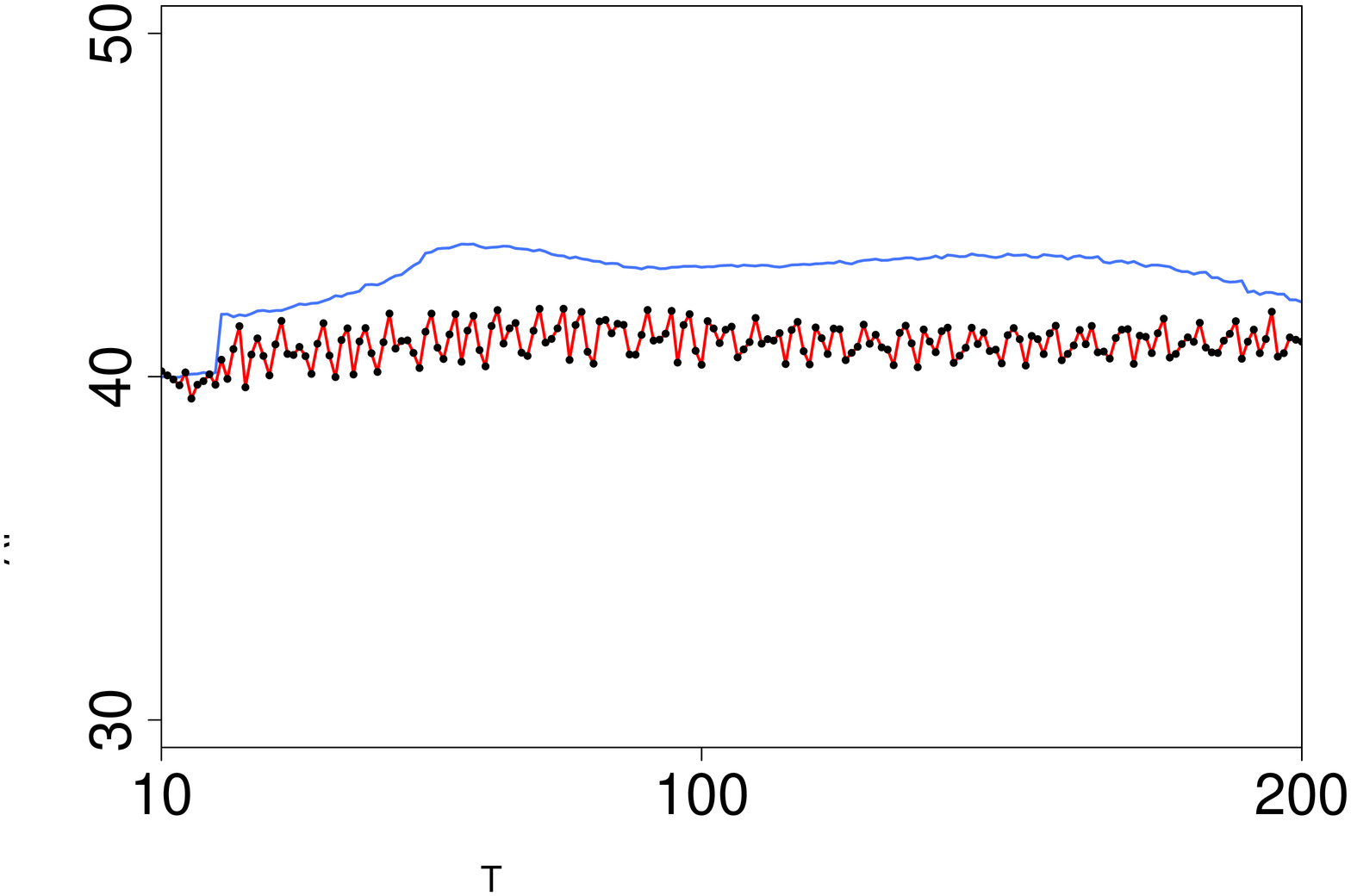}
      }
      &
      \subfloat{
      \includegraphics[height=0.21\textwidth,width=0.21\textwidth,angle=270]{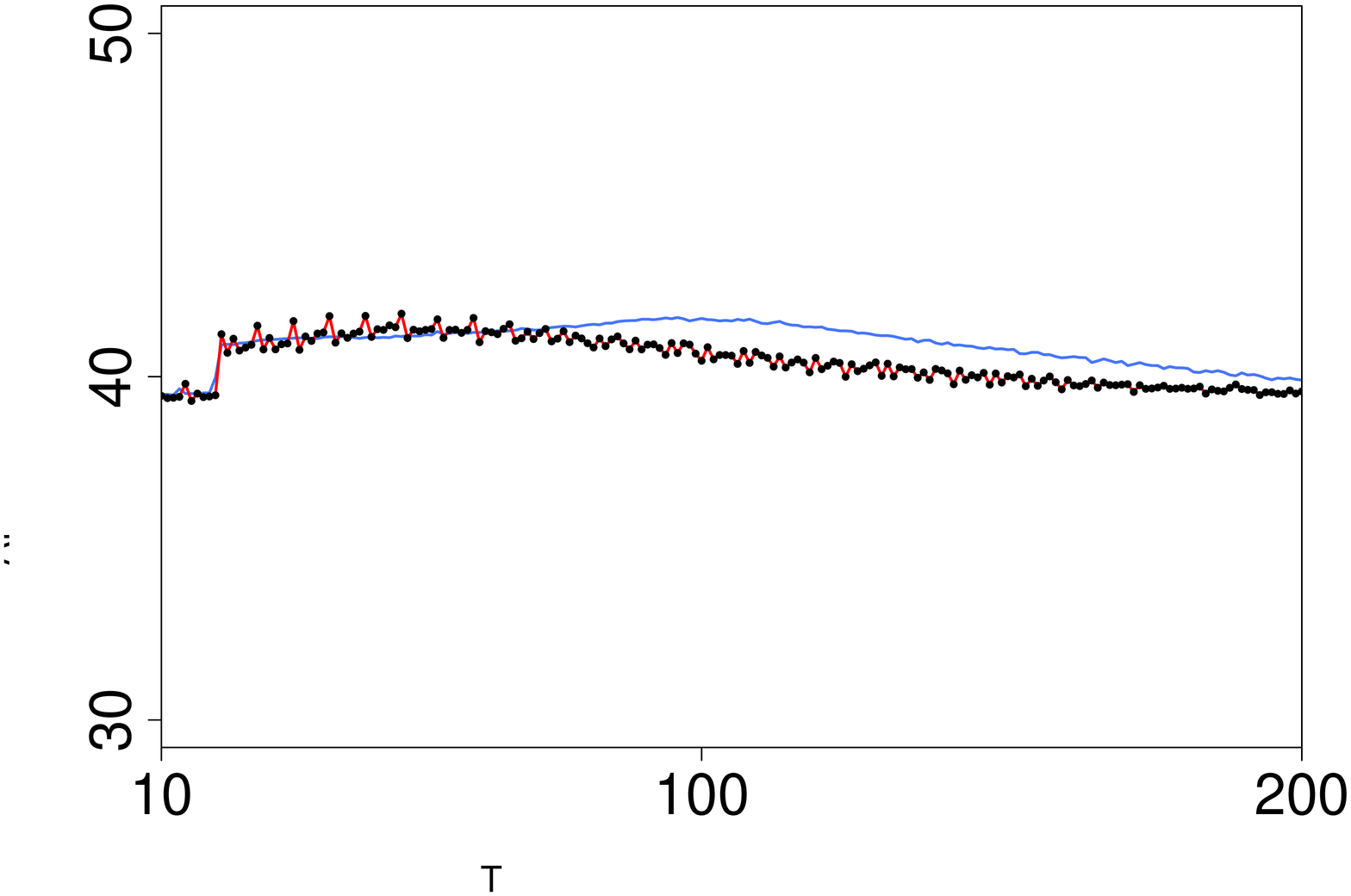}
      }\\
%       \hline
%            \rotatebox[origin=c]{90}{\scriptsize{Packet loss}}  & 
%    \subfloat{
%       \includegraphics[height=0.21\textwidth,width=0.19\textwidth,angle=270]{packet_loss_mean_RED_Th_SL_CTCP.eps}
%      }
%       &
%       \subfloat{
%       \includegraphics[height=0.21\textwidth,width=0.19\textwidth,angle=270]{packet_loss_mean_RED_Th_SL_MT.eps}
%      }
%       &
%      \subfloat{
%       \includegraphics[height=0.21\textwidth,width=0.19\textwidth,angle=270]{packet_loss_mean_RED_Th_PL_CTCP.eps}
%       }
%       &
%       \subfloat{
%       \includegraphics[height=0.21\textwidth,width=0.19\textwidth,angle=270]{packet_loss_mean_RED_Th_PL_MT.eps}
%       }\\
%       \hline
%        \rotatebox[origin=c]{90}{\scriptsize{Throughput}}  & 
%       \subfloat{
%       \includegraphics[height=0.21\textwidth,width=0.19\textwidth,angle=270]{throughput_mean_RED_Th_SL_CTCP.eps}
%       }
%       &
%       \subfloat{
%       \includegraphics[height=0.21\textwidth,width=0.19\textwidth,angle=270]{throughput_mean_RED_Th_SL_MT.eps}
%       }
%       &
%      \subfloat{
%       \includegraphics[height=0.21\textwidth,width=0.19\textwidth,angle=270]{throughput_mean_RED_Th_PL_CTCP.eps}
%       }
%       &
%       \subfloat{
%       \includegraphics[height=0.21\textwidth,width=0.19\textwidth,angle=270]{throughput_mean_RED_Th_PL_MT.eps}
%       }\\
      \hline
            &
      \subfloat{
      \includegraphics[width=0.25in,height=1in,angle=270]{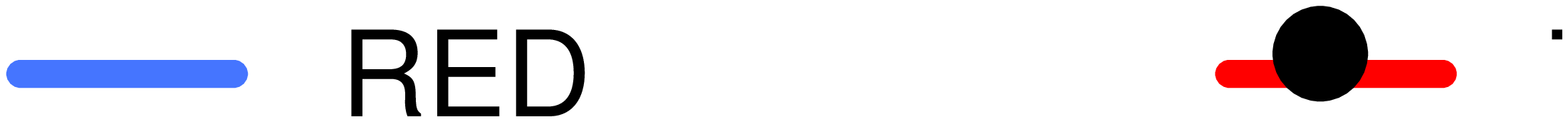}
     } & & &\\
      \bottomrule
  \end{tabular}
 \end{center}
 \caption{Performance evaluation of RED and threshold-based queue policy for single-bottleneck (Figure~\ref{fig:single_bottleneck_dumbbell}) and parking-lot (Figure~\ref{fig:parking_lot}) topologies, with homogeneous (refer Section~\ref{sec:RED_sims_homo}) and heterogeneous (refer Section~\ref{sec:RED_sims_hetro}) traffic setting. It can be observed that the threshold-based queue policy ensures comparatively lower queueing delay and flow completion time.}
 \label{fig:PE_RED_Th_1}
\end{figure}
  \begin{figure}[tbh]
\psfrag{1}{\tiny{$1$}}
  \psfrag{2}{\tiny{$2$}}
    \psfrag{3}{\tiny{$3$}}
     \psfrag{4}{\tiny{$4$}}
  \psfrag{2.5}{\tiny{$2.5$}}
    \psfrag{0.5}{\tiny{$0.5$}}
  \psfrag{1.5}{\tiny{$1.5$}}
  \psfrag{100}[b][b]{\tiny{$100$}}
  \psfrag{200}[b][b]{\tiny{$200$}}
  \psfrag{300}{\tiny{$300$}}
  \psfrag{50}{\tiny{$50$}}
  \psfrag{150}{\tiny{$150$}}
  \psfrag{250}{\tiny{$250$}}
  \psfrag{0}{\tiny{$0$}}
  \psfrag{7}{\tiny{$7$}}
  \psfrag{9}{\tiny{$9$}}
      \psfrag{75}{\tiny{$75$}}
    \psfrag{25}{\tiny{$25$}}
    \psfrag{5}{\tiny{$5$}}
    \psfrag{8}{\tiny{$8$}}
  \psfrag{15}{\tiny{$15$}}
  \psfrag{40}{\tiny{$40$}}
  \psfrag{Af}{\hspace{1.5mm}\scriptsize{AFCT (mins)}}
  \psfrag{10}[b][b]{\tiny{$10$}}
  \psfrag{20}{\tiny{$20$}}
  \psfrag{30}{\tiny{$30$}}
  \psfrag{T}{\scriptsize{RTT (ms)}}
  \psfrag{Mq}{\hspace{4mm}\scriptsize{Mean QD (ms)}}
     \psfrag{Mp}{\scriptsize{Mean packet loss (\%)}}
        \psfrag{Mt}{\hspace{-2mm}\scriptsize{Mean throughput (Mbps)}}
  \psfrag{RED}[l][c][1][0]{\small{RED}}
  \psfrag{Th}[l][c][1][0]{\small{Threshold}}
 \begin{center}
  \begin{tabular}{*{5}{l  >{\centering\arraybackslash} m{0.21\textwidth} >{\centering\arraybackslash} m{0.21\textwidth} >{\centering\arraybackslash} m{0.21\textwidth} >{\centering\arraybackslash} m{0.21\textwidth}}}
  \toprule
   & \multicolumn{2}{c}{Single bottleneck} & \multicolumn{2}{c}{Parking-lot}\\
   \cmidrule(lr){2-3}\cmidrule(lr){4-5}
   &\multicolumn{1}{c}{ Homogeneous} & \multicolumn{1}{c}{Heterogeneous} & \multicolumn{1}{c}{Homogeneous} & \multicolumn{1}{c}{Heterogeneous}\\
 \cmidrule(lr){2-2} \cmidrule(lr){3-3}  \cmidrule(lr){4-4}  \cmidrule(lr){5-5}
%    \rotatebox[origin=c]{90}{\scriptsize{Queueing delay}}
%    & 
%    \subfloat{
%      \includegraphics[height=0.21\textwidth,width=0.19\textwidth,angle=270]{que_delay_mean_RED_Th_SL_CTCP.eps}
%    }
%       &
%       \subfloat{
%       \includegraphics[height=0.21\textwidth,width=0.19\textwidth,angle=270]{que_delay_mean_RED_Th_SL_MT.eps}
%       }
%       &
%       \subfloat{
%       \includegraphics[height=0.21\textwidth,width=0.19\textwidth,angle=270]{que_delay_mean_RED_Th_PL_CTCP.eps}
%       }
%       &
%      \subfloat{
%       \includegraphics[height=0.21\textwidth,width=0.19\textwidth,angle=270]{que_delay_mean_RED_Th_PL_MT.eps}
%      }
%        \\
%       \hline
%       \rotatebox[origin=c]{90}{\scriptsize{Flow completion time}}  & 
%       \subfloat{
%       \includegraphics[height=0.21\textwidth,width=0.19\textwidth,angle=270]{afct_RED_Th_SL_CTCP.eps}
%       }
%       &
%       \subfloat{
%       \includegraphics[height=0.21\textwidth,width=0.19\textwidth,angle=270]{afct_RED_Th_SL_MT.eps}
%       }
%       &
%      \subfloat{
%       \includegraphics[height=0.21\textwidth,width=0.19\textwidth,angle=270]{afct_RED_Th_PL_CTCP.eps}
%       }
%       &
%       \subfloat{
%       \includegraphics[height=0.21\textwidth,width=0.19\textwidth,angle=270]{afct_RED_Th_PL_MT.eps}
%       }\\
           \rotatebox[origin=c]{90}{\scriptsize{Packet loss}}  & 
   \subfloat{
      \includegraphics[height=0.21\textwidth,width=0.21\textwidth,angle=270]{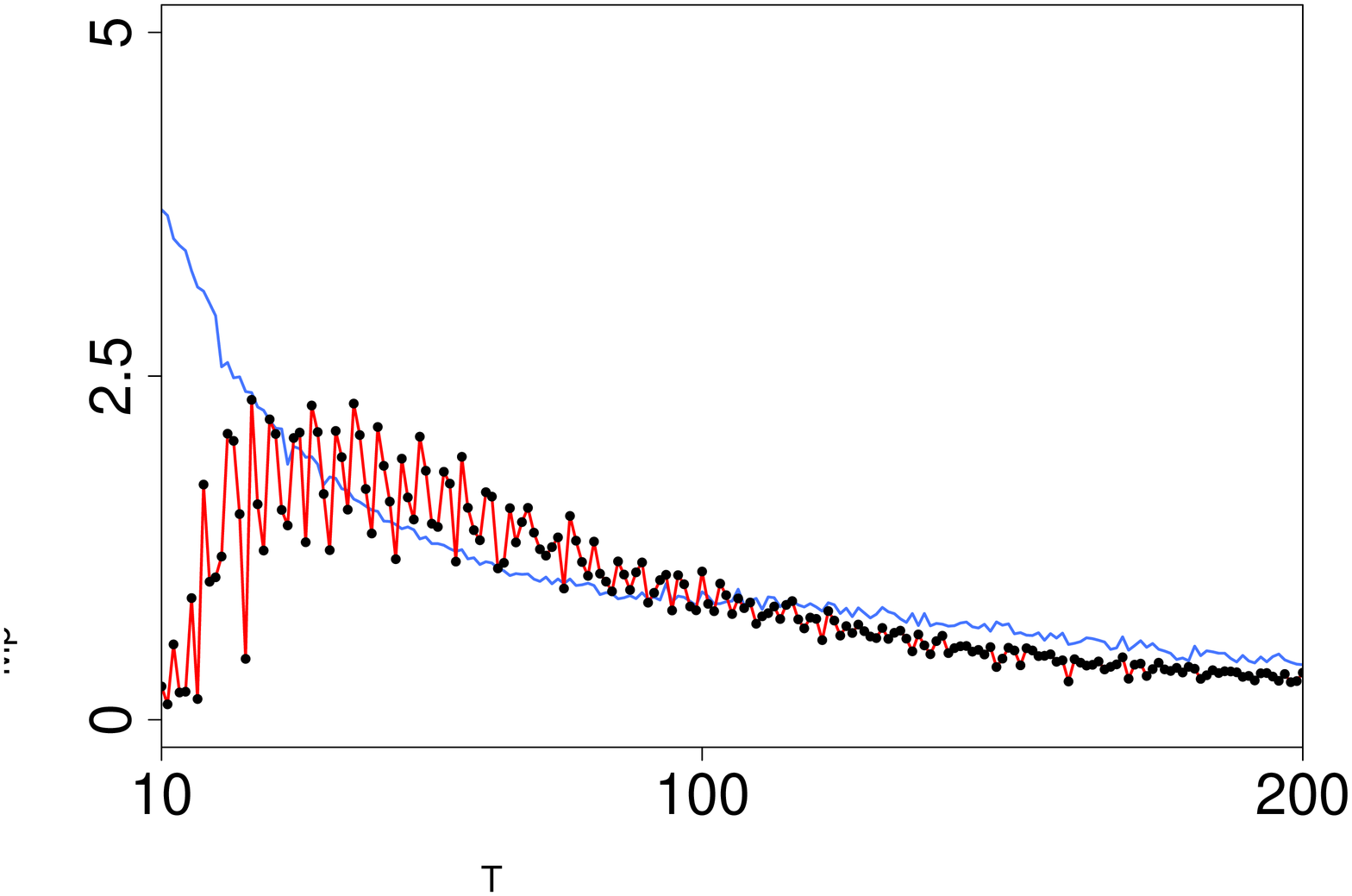}
     }
      &
      \subfloat{
      \includegraphics[height=0.21\textwidth,width=0.21\textwidth,angle=270]{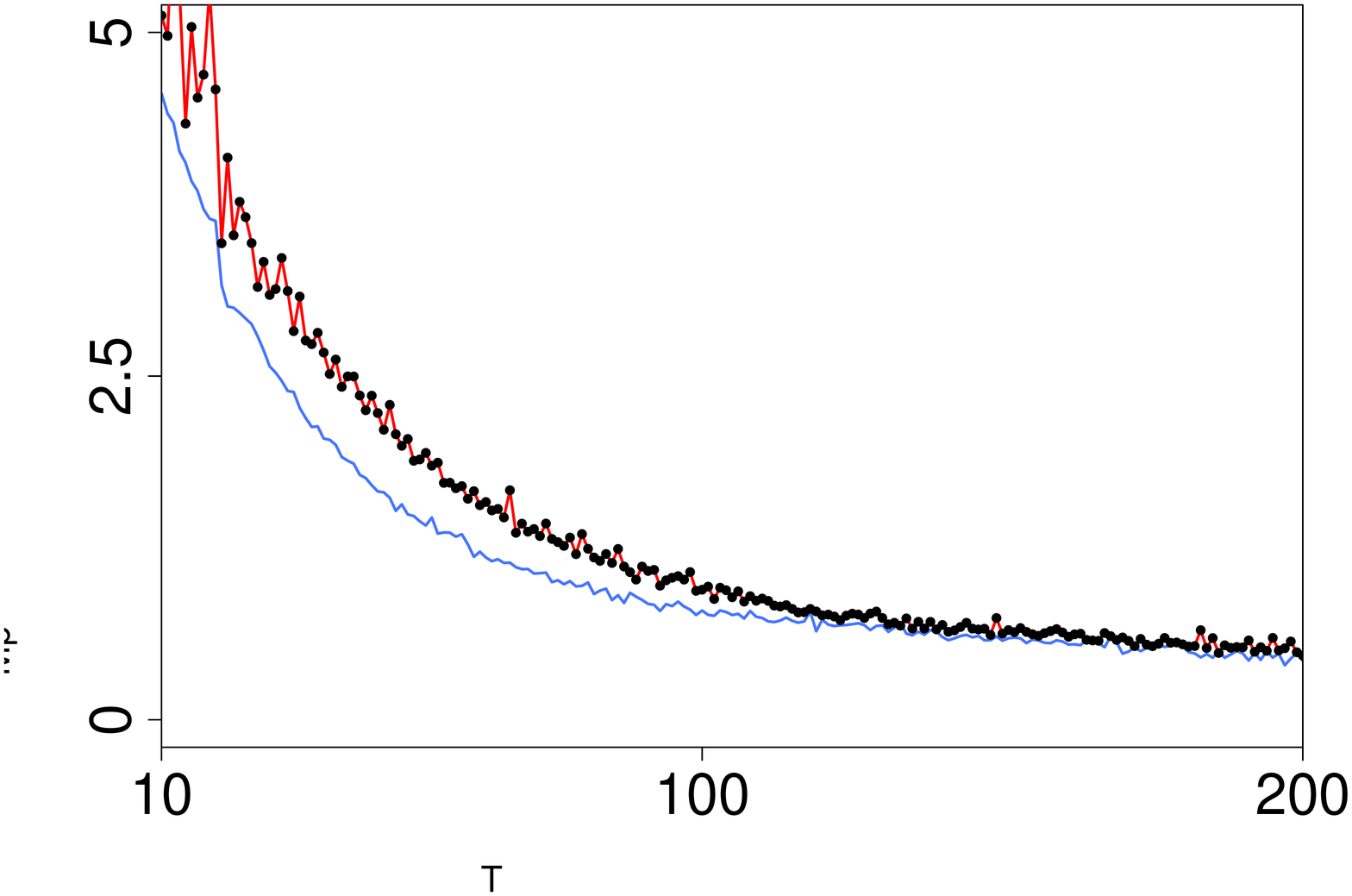}
     }
      &
     \subfloat{
      \includegraphics[height=0.21\textwidth,width=0.21\textwidth,angle=270]{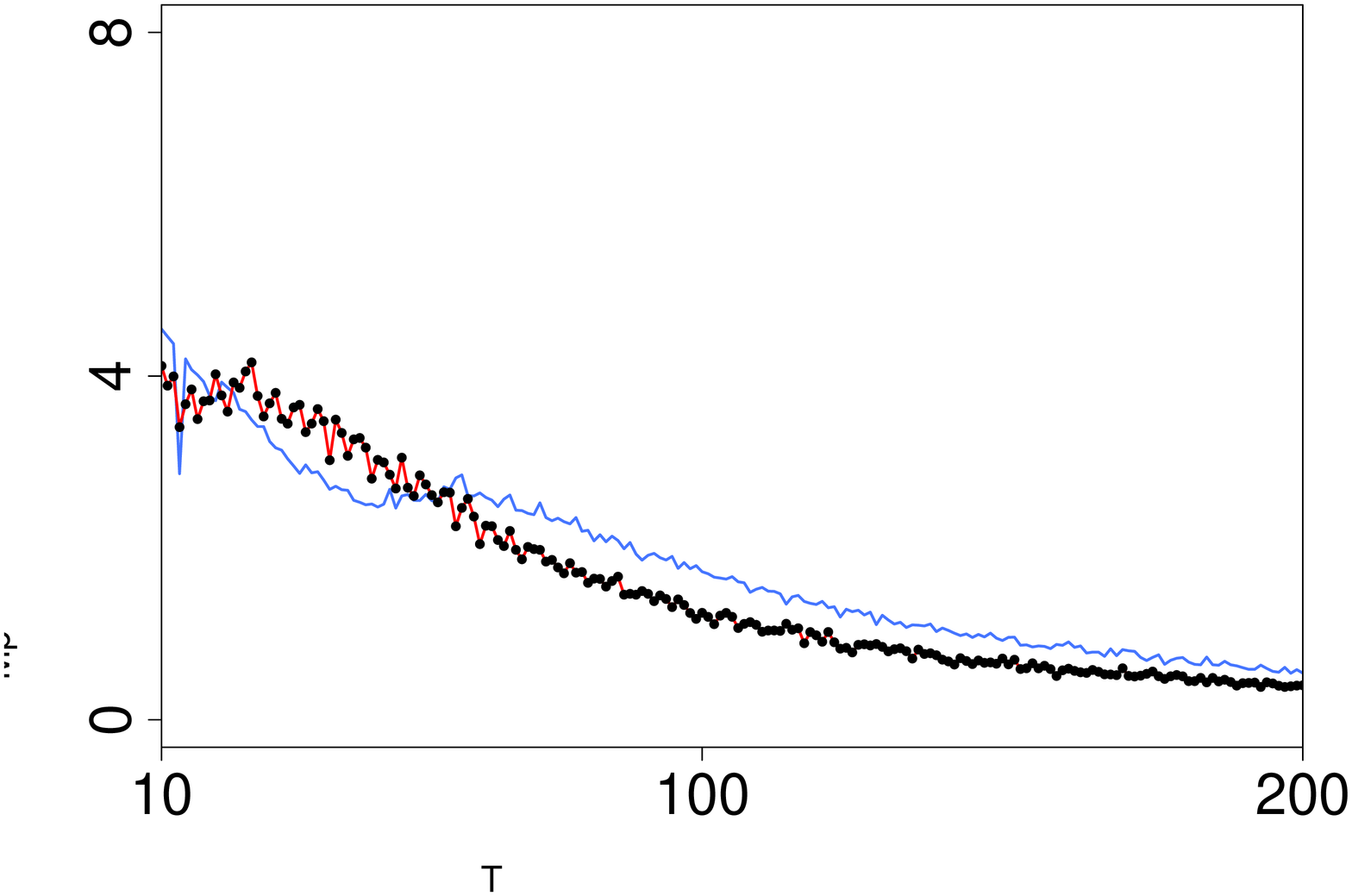}
      }
      &
      \subfloat{
      \includegraphics[height=0.21\textwidth,width=0.21\textwidth,angle=270]{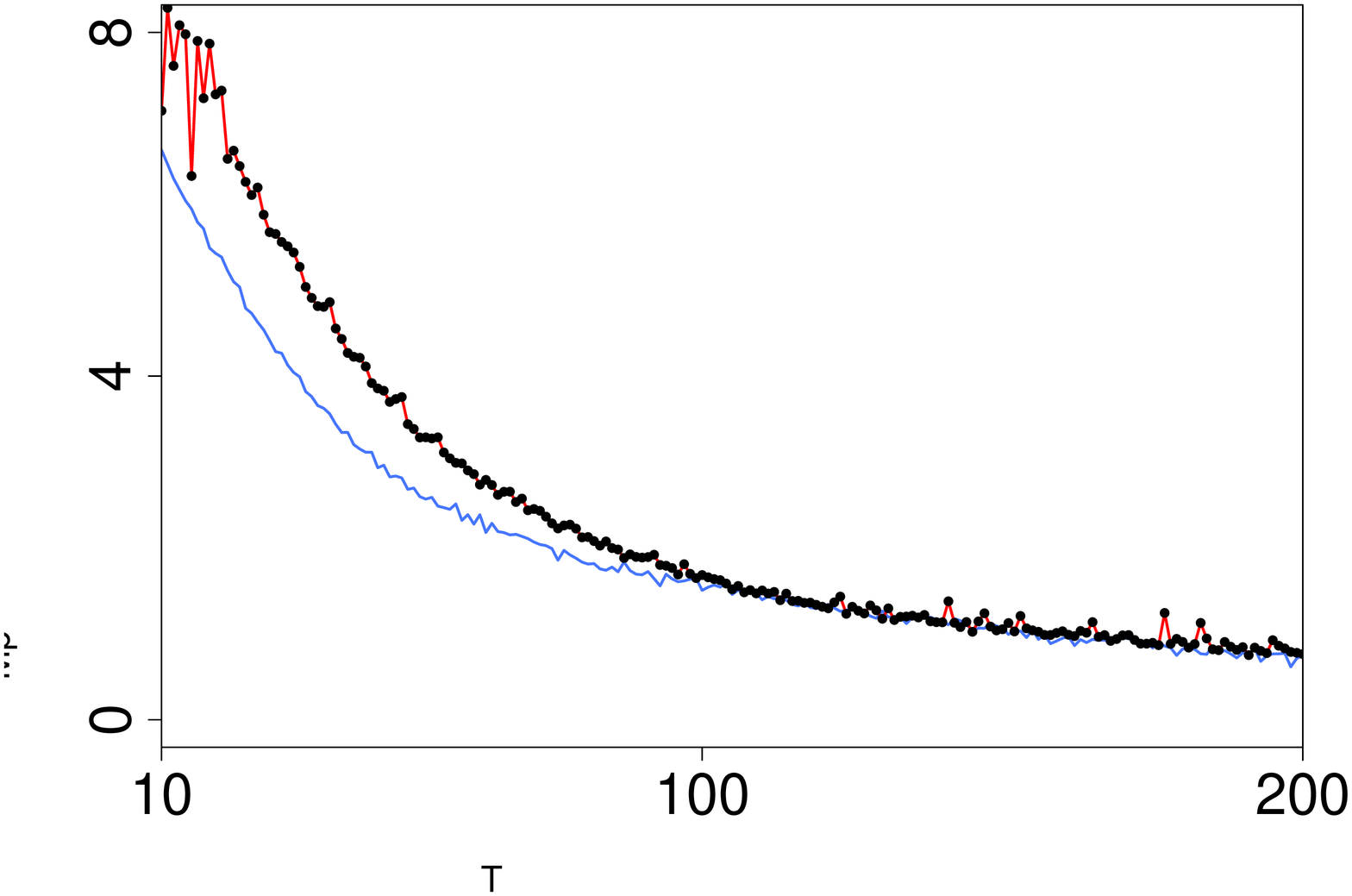}
      }\\
      \hline
       \rotatebox[origin=c]{90}{\scriptsize{Throughput}}  & 
      \subfloat{
      \includegraphics[height=0.21\textwidth,width=0.21\textwidth,angle=270]{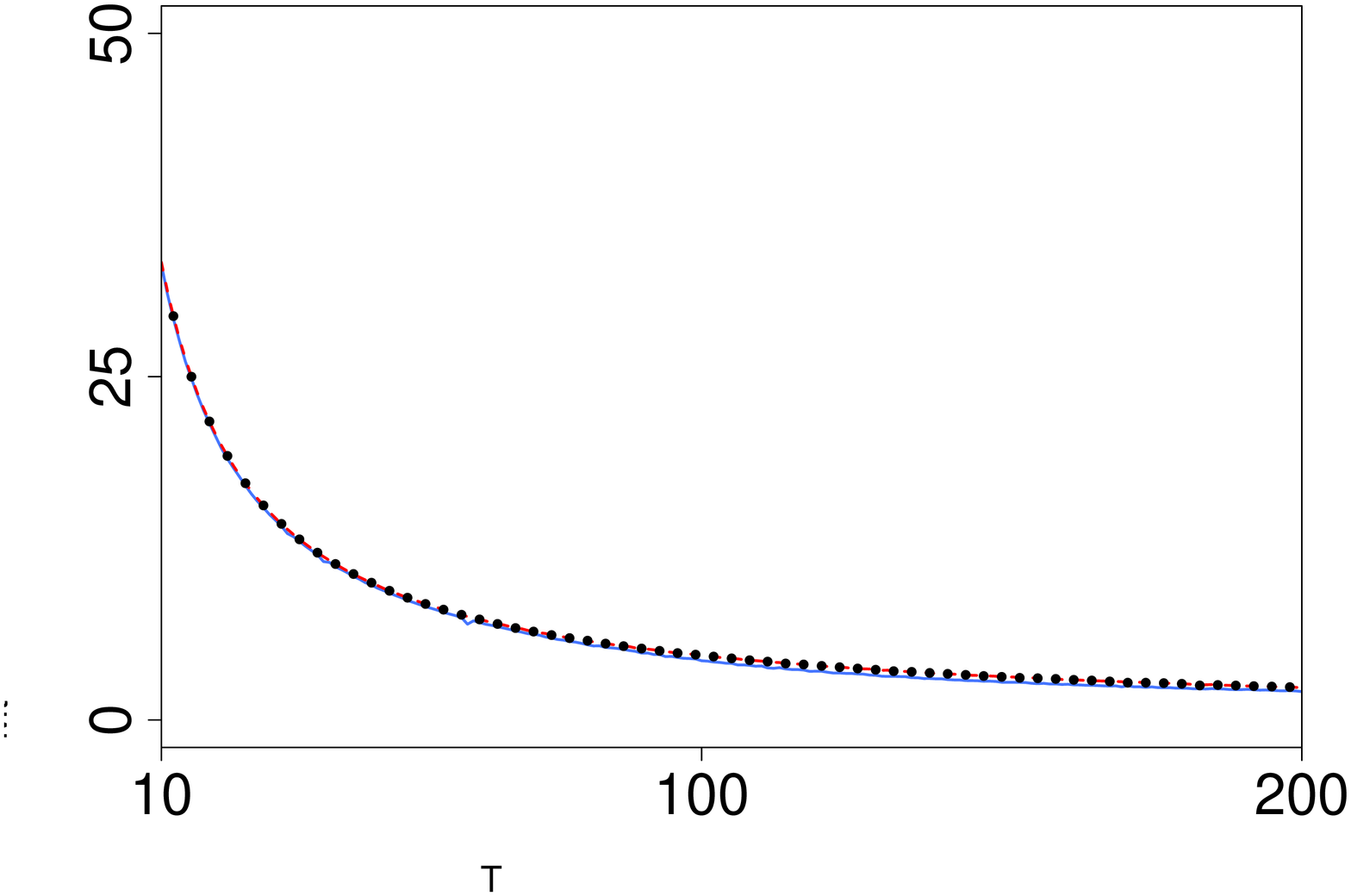}
      }
      &
      \subfloat{
      \includegraphics[height=0.21\textwidth,width=0.21\textwidth,angle=270]{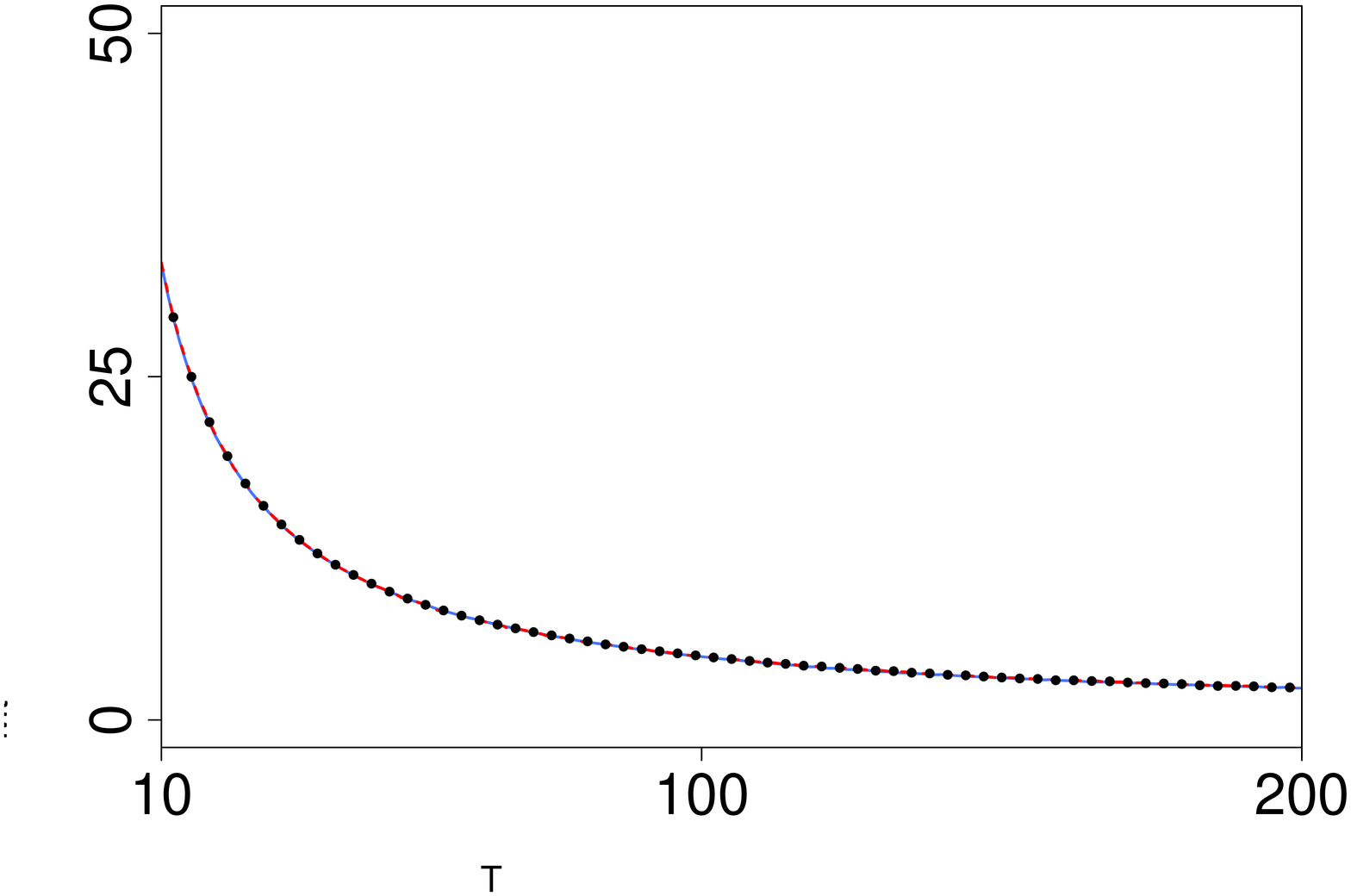}
      }
      &
     \subfloat{
      \includegraphics[height=0.21\textwidth,width=0.21\textwidth,angle=270]{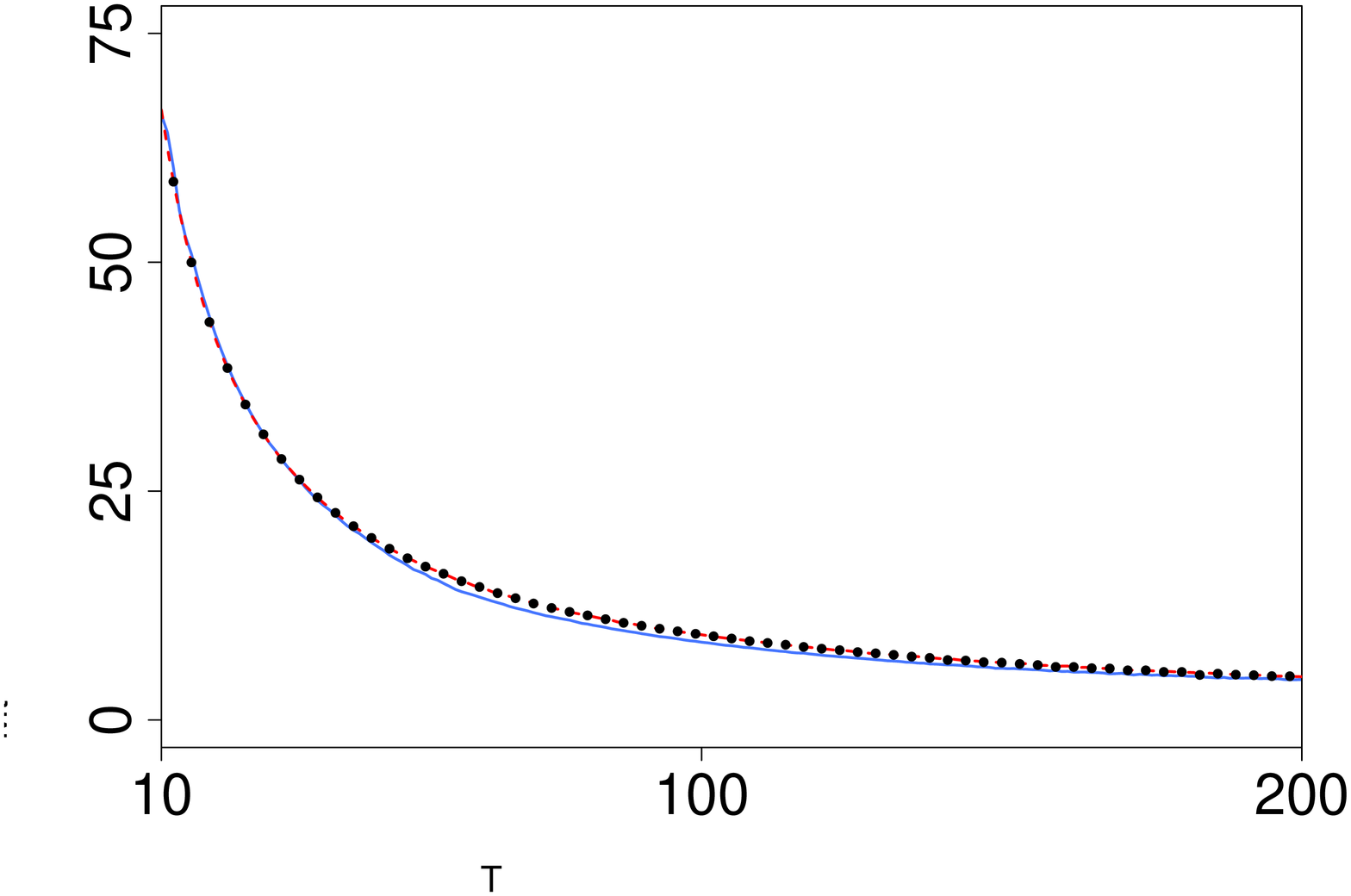}
      }
      &
      \subfloat{
      \includegraphics[height=0.21\textwidth,width=0.21\textwidth,angle=270]{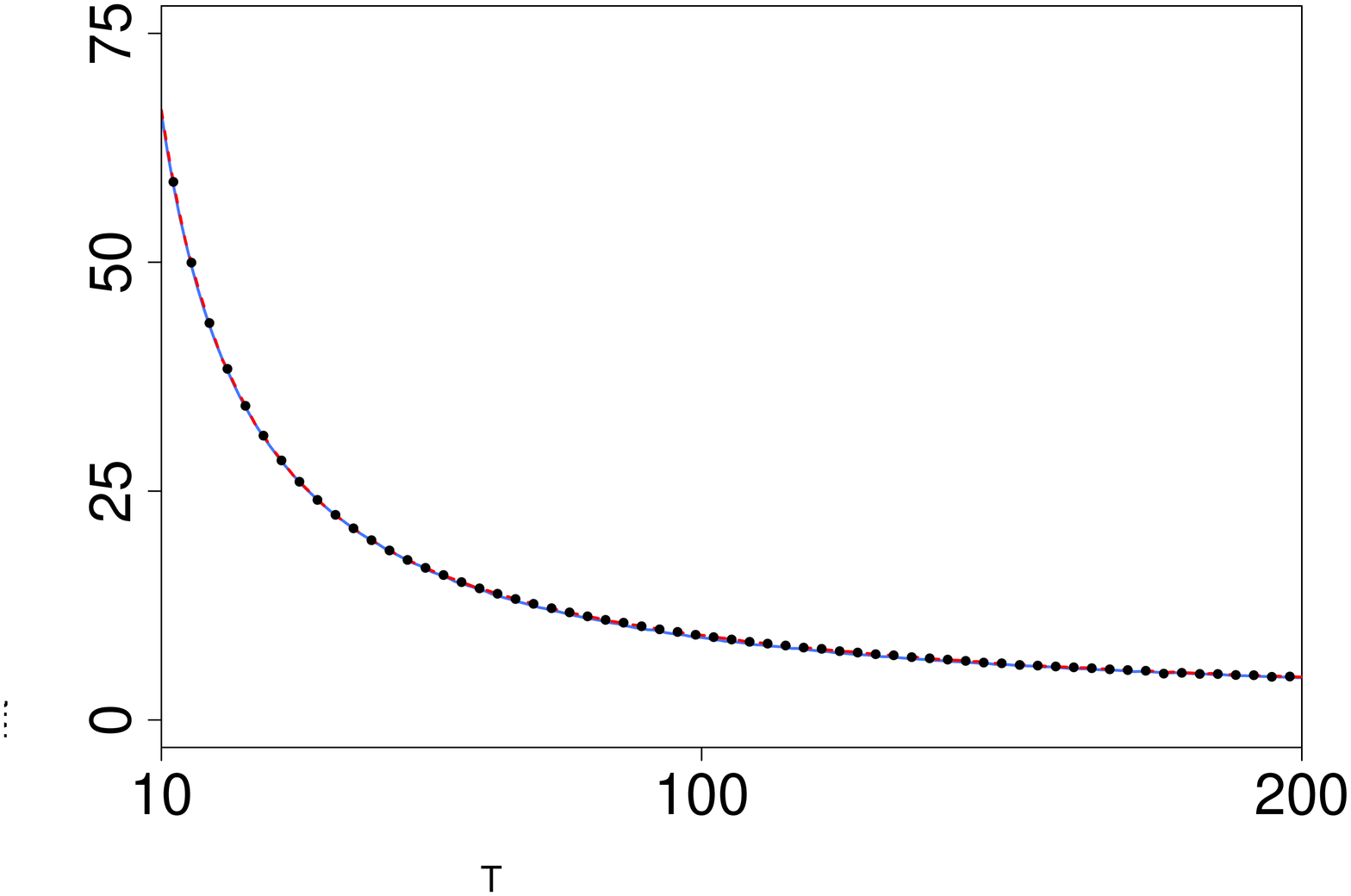}
      }\\
      \hline
            &
      \subfloat{
      \includegraphics[width=0.25in,height=1in,angle=270]{legend.eps}
     } & & &\\
      \bottomrule
  \end{tabular}
 \end{center}
 \caption{Performance evaluation of RED and threshold-based queue policy for single-bottleneck (Figure~\ref{fig:single_bottleneck_dumbbell}) and parking-lot (Figure~\ref{fig:parking_lot}) topologies, with homogeneous (refer Section~\ref{sec:RED_sims_homo}) and heterogeneous (refer Section~\ref{sec:RED_sims_hetro}) traffic setting. It can be observed that the threshold-based queue policy ensures comparatively lower queueing delay and flow completion time. Both policies appear to drop approximately same fraction of the incoming packets, and offer the same throughput.}
 \label{fig:PE_RED_Th_2}
\end{figure}
 
\subsubsection{Parking-lot topology}
 This topology comprises of two bottleneck links connected in series such that the output of the first feeds into the second. A schematic of the topology is presented in Figure~\ref{fig:parking_lot}. There are 60 end systems, that act as traffic sources, connected to the first a link. A half of them, labelled as $[S_{A1},S_{A30}]$ in the diagram, are connected to destinations $[D_{A1},D_{A30}]$ which are on the other side of the second link. The packets exchanged between these end systems would have to be served at both the links. The rest half of the sources $[S_{A31},S_{A60}]$ are connected to destinations $[D_{A31},D_{A60}]$, which are connected to the first link. The traffic between these pairs of sources and destinations would have to traverse the first link alone. Meanwhile, a bunch of sources $[S_{B1},S_{B30}]$ are connected to destinations $[D_{B1}, D_{B30}]$ through the second link. The service capacity of both the links is fixed as $100$ Mbps. Both links have buffers of size of $2084$ pkts. 
With this network setting, we consider both homogeneous and heterogeneous traffic scenarios. For the homogeneous traffic setting, sources $[S_{A1},S_{A30}]$ generate $30$ Compound TCP long-lived flows, and sources $[S_{A31},S_{A60}]$ and $[S_{B1},S_{B30}]$ generate $30$ Compound TCP long-lived flows. The packet size is $1500$ bytes.
 
 For the heterogeneous traffic mix of Compound TCP, CUBIC, UDP and HTTP flows, the sources $[S_{A1},S_{A30}]$ generate $13$ Compound TCP and $14$ CUBIC long-lived flows, $8$ UDP and short-lived HTTP flows generated at the rate of $50$ flows per second. Sources $[S_{A31},S_{A60}]$ and $[S_{B1},S_{B30}]$ generate $14$ Compound TCP and $14$ CUBIC long-lived flows. 
 This combination is obtained by dividing the total of $55$ long-lived flows among sets $[S_{A1},S_{A30}]$ and $[S_{A31},S_{A60}]$. Among the flows in each of these sets, half are set as Compound TCP and the other half as CUBIC. The sources $[S_{B1},S_{B30}]$ generate the same traffic combination as $[S_{A31},S_{A60}]$. The Compound TCP and CUBIC parameters are fixed at their default values. The packet size is fixed as $1500$ bytes. 
 
 With this setup, we conduct simulations with the RED and threshold policies. For the RED policy we set the thresholds as $\overline{b}=15, \underline{b}=8$ pkts. The rest of the RED parameters are retained at their default values. For the threshold policy, we fix the queue threshold as $q_{th}=15$ pkts. We first fix RTT as $10$ and $200$ ms and observe the queue sizes for the two policies. The traces of the queue sizes observed at both the links, for RED and threshold policies are plotted in Figure~\ref{fig:comparison_RED_Threshold_PL}. These results are qualitatively similar to what was observed in the case of the single bottleneck topology. While the threshold policy successfully maintains the queue size at the defined threshold, the RED policy allows the queue size to increase up to about $40$ pkts. 
 
We then vary the RTT continuously in the range of $[10,200]$ ms, and conduct simulations to compare the two policies in terms queueing delay, packet loss percentage, throughput and average flow completion time. These plots for both homogeneous and heterogeneous traffic scenarios are tabulated in Figures~\ref{fig:PE_RED_Th_1} and~\ref{fig:PE_RED_Th_2}. The queueing delay plotted is the sum of the mean queueing delay observed in both the queues. It can be observed from these plots that the threshold policy ensures lower queueing delay compared to RED, in this network setting as well. The packet loss percentage represents the probability that a packet is dropped at either of the two links. Here, we see that for lower RTTs the packet loss in case of the threshold policy is marginally higher than that for RED. As RTT increases, the two curves converge, indicating that both policies drop packets equally for larger RTTs. The total throughput observed at both links is seen to be equal for both queue policies. For comparing the flow completion time of both the policies, we plotted the average flow completion time observed across all sources, \emph{i.e.}, $[S_{A1},S_{A60}]$ and $[S_{B1},S_{B30}]$. It is seen that flows are completed in a shorter time 
with the threshold policy across the entire range of RTT considered. To that end, one may argue that the threshold offers better performance compared to RED, and this observation is seen to be consistent across the network settings that we 
consider.
 
% \subsection{Threshold policy with other flavours of TCP}
\subsection{Threshold policy with TCP Reno}
Transport protocols primarily vary in the metric they use to infer incipient network congestion. Compound TCP, studied in this paper, uses a combination of queueing delay and packet loss as congestion feedback. There are other varieties of TCP that use either packet loss or queueing delay, and not both. Before proposing a queue policy for implementation at routers, it would be desirable to verify if it can ensure stable operation with these other varieties of TCP as well. To start with, one may study the threshold policy with the basic TCP Reno, which is one of the earliest proposals for loss-based TCP~\cite{jacobson1990modified}.
% We now discuss the stability of a network of loss-based and delay-based TCP flows feeding into a router with the threshold-based queue policy.
% \subsubsection{TCP Reno: loss-based protocol}

Loss-based protocols, such as TCP Reno, use packet loss to infer congestion, \emph{i.e.} when a sent packet is not acknowledged TCP infers network congestion and reduces its sending window. 
% TCP Reno is one of the earliest proposals for loss-based TCP~\cite{jacobson1990modified}. 
TCP Reno uses an additive increase multiplicative decrease rule for window update, wherein the sending window is increased by one packet over a round-trip time when all packets and acknowledged, and halved if a packet loss is detected. This mechanism can be captured by the following window increase and decrease functions
\begin{align*}
 i\big(w(t)\big) = 1/w(t) && d\big(w(t)\big) = w(t)/2.
\end{align*}
Observe that this can be considered as a special case of the window update functions defined for Compound TCP~\ref{sec:CompoundTCP}, where the protocol parameters are set as: $\alpha = 1, k = 0$ and $\beta = 1/2$. Therefore, the stability conditions for the system of TCP Reno with the threshold-based policy can be derived by substituting these values in equations~\eqref{eq:Compound_Threshold_necc_buffer} and~\eqref{eq:Compound_Threshold_suff_buffer}, which are the conditions for the stability of Compound TCP with the threshold policy. Upon doing so, we obtain the necessary and sufficient condition for local stability as
\begin{align*}
 \frac{1}{w^\ast}\sqrt{q_{th}^2 - 4(1-p^\ast)^2} < \cos^{-1}\bigg(\frac{-2(1-p^\ast)}{q_{th}}\bigg).
\end{align*}
A sufficient condition for local stability is given by
\begin{align}
 q_{th}/w^\ast < \pi/2.\label{eq:Reno_Threshold_suff}
\end{align}
It can be argued that $1/w^\ast$ is a decreasing function in $\tau$, as shown in the case of Compound TCP. Therefore, the threshold-based queue policy ensures stability of TCP Reno irrespective of the round-trip time of the TCP flows. The packet-dropping threshold $q_{th}$ can be tuned to a few tens of packets, as per the sufficient condition~\eqref{eq:Reno_Threshold_suff}, to ensure stable and low-latency operation of TCP Reno.
Thus, the threshold-based queue policy is shown to ensure stable operation of TCP Reno (a loss-based protocol), and Compound TCP (a delay-and-loss-based protocol). 

%  \subsection{Pseudo code}
%  The threshold-based queue policy drops an arriving packet only when the queue size $q$ reaches the threshold $q_{th}$. This requires the packet-drop probability to be decided as follows:
% % \begin{center}
% %  \begin{tabular}{c|c} 
% %  Condition & Probability\\
% %  \hline
% %  $q < q_{th}$ & 0\\
% %  $q \geq q_{th}$ & 1
% % %  $q > q_{th}$ & 1
% % \end{tabular}
% % \end{center}
% \begin{verbatim}
%  if q < q_th, then p = 0,  else  p = 1.
% \end{verbatim}
% Upon arrival of each packet, the queue size must be compared with the fixed threshold, and the packet-drop probability must be decided. 
% This decision-making process of the AQM can be realised by the following pseudo code:
% \begin{verbatim}
%   p := !max{sgn(q_th-q),0},
% \end{verbatim}
% where $!$ represents the binary NOT operator, and the function \texttt{max} returns the maximum of the two elements within the braces. The function \texttt{sgn(x)} is the signum function that returns $1$ for positive argument, $-1$ for negative argument and $0$ when the argument is $0$.
 
\section{Conclusions}
\label{sec:conclude}
The problem of \emph{bufferbloat} remains relevant even after many queue management proposals, and calls for a detailed study of these queue policies with current transport protocols. 
% 
% One such proposal is the Random Early Detection (RED) policy. The RED policy computes an exponentially weighted moving average of the queue size, and uses this average to compute the packet-drop probability.
% % Such an approach could tackle bursty traffic and penalise users that take up more than their fair share of available bandwidth. 
% RED is implemented in routers, however owing to the extensive parameter tuning required, it is not deployed. 
% The performance of RED with currently deployed TCP flavours needs to be studied before its deployment. 
% In particular, we analysed the stability of the Compound TCP-RED system and explored the impact of queue size averaging.
We studied the currently deployed Compound TCP in conjunction with the RED queue policy. 

We first studied a non-linear fluid model for Compound TCP-RED, and derived a sufficient condition for its local stability. We explicitly established that the system transits into instability via a Hopf bifurcation as system parameters are varied. In non-linear systems, a Hopf bifurcation indicates the emergence of limit cycle oscillations in system dynamics. Stability charts indicate that large round-trip times of the TCP flows or large values of queue thresholds can destabilise the system, and that stability is sensitive to the queue averaging parameter. We then studied a regime where queue size averaging is not performed, and the packet-drop probability is a function of the instantaneous queue size itself. We derived the necessary and sufficient condition for local stability of Compound TCP-RED in this regime. We also showed that the system transits into instability via a Hopf bifurcation in this regime as well. However, it is seen that Compound TCP-RED may remain stable for comparatively large round-trip 
times when queue size 
averaging is not performed. Local stability results also suggest that low thresholds for dropping packets could aid system stability. Packet-level simulations are presented to corroborate the analytical insight. It is observed that (i) large round-trip times are detrimental to system stability, (ii) averaging over queue size may not be beneficial to system performance, (iii) smaller thresholds for dropping packets indeed aid stability.

We then proposed a simple threshold-based queue policy that can be tuned to maintain small queues and hence reduce queueing delays. The threshold-based policy is observed to ensure stable operation of Compound TCP regardless of the round-trip time of the TCP flows. We then conducted a simulation-based performance evaluation of RED and the threshold policy. The threshold policy appears to outperform RED, in terms of queueing delay, flow completion time and packet loss, consistently across the considered network settings. Finally, we outlined some stability results for the threshold-based policy with a couple of other transport protocols. It is observed that the threshold policy could guarantee stable operation of networks while ensuring reduced queueing delays.

\subsection{Avenues for future work}
In our analysis we only consider the average RTT of all the TCP flows, thus incorporating only one time delay in the fluid model. It would be interesting to analyse the system dynamics in the presence of heterogeneous delays. As a natural extension to our study, one may also explore the case of a multiple bottleneck topology. It would also be interesting to study the system dynamics and network performance when packets are not dropped, but marked using Explicit Congestion Notification (ECN).
% % % % % % % % % % % % % % % % % % % % % % % % % % % % % % % remove for full version% % % % % % % % % % % % % % % % % % % % % % % % % % %
 \appendix
 \section{Hopf bifurcation}
 \label{sec:appendix}
 A bifurcation is a phenomenon in which the system dynamics undergoes a change\textemdash such as creation or destruction of equilibrium points, or change in their stability\textemdash as system parameters are varied~\cite{strogatz2018nonlinear}. Such a change in dynamics occurs when the characteristic roots of the linearised system cross over the imaginary axis to the right half of the Argand plane. The Hopf bifurcation is a type of bifurcation wherein a complex conjugate pair of characteristic roots crosses over the imaginary axis through a pair of purely imaginary roots $\lambda = \pm j\omega$. 
 % 
 % Hopf bifurcation is a phenomenon in which a non-linear system loses local stability; and the phase portrait of the system exhibits a topological change from a stable equilibrium to a limit cycle~\cite{Kuznetsov_13}.
 In other words, a system 
 \begin{align}
  \dot{x}=f(x,\tilde{p}),\,\,\,\,\,\,\,\, x = [x_1 \,\,x_2]^T \in \mathbb{R}^2, \label{eq:gen_2d_system}
 \end{align}
 is said to undergo a Hopf bifurcation, when the characteristic roots of the linearised system satisfy the following: (a) $\lambda = -\sigma\pm j\omega$ ($\sigma > 0$) for $\tilde{p} < \tilde{p}_c$, (b) $\lambda = \pm j \omega$ for $\tilde{p} = \tilde{p}_c$, and (c)  $\lambda = \sigma\pm j\omega$ for $\tilde{p} > \tilde{p}_c$. The parameter $\tilde{p}$ is referred to as the bifurcation parameter, and the Hopf bifurcation occurs at the critical value $\tilde{p}_c$. 
 % 
 % The loss of system stability via a Hopf bifurcation can be studied by observing the roots of the characteristic equation corresponding to the linear approximation of the system about its equilibrium $x^\ast$. When the system is stable, the characteristic roots are of the form $\lambda = -\sigma \pm j\omega, \sigma > 0$. As $\tilde{p}$ is varied, at least one pair of complex roots crosses over the imaginary axis and takes the form $\lambda = \sigma \pm j\omega, \sigma > 0$. At the critical value of the bifurcation parameter $\tilde{p}=\tilde{p}_c$, one pair of characteristic roots is of the form $\lambda = j\omega$, which causes the emergence of limit cycle oscillations in the state variables.
 The Hopf bifurcation is classified as:
 \begin{enumerate}
  \item [(i)] Super-critical Hopf: The trajectories of the system converge to a stable equilibrium asymptotically, when $\tilde{p} < \tilde{p}_c$. As $\tilde{p}$ is increased beyond $\tilde{p}_c$, the stable equilibrium gives rise to an asymptotically orbitally stable limit cycle. 
  \item [(ii)] Sub-critical Hopf: The trajectories converge to a stable equilibrium, when $\tilde{p} < \tilde{p}_c$. As $\tilde{p}$ is increased beyond $\tilde{p}_c$, either the trajectories could blow up to infinity in finite time, or they could converge to a limit cycle of large amplitude.  
 \end{enumerate}
 For a detailed discussion of Hopf bifurcation in non-linear systems, the reader is referred to ~\cite[Section 8.2]{strogatz2018nonlinear},\cite{hassard1981theory}.
 
 In engineered systems, it would be preferable to have a stable equilibrium. However, if the system does lose stability due to variation in system parameters or feedback delay, it would be desirable to have an asymptotically orbitally stable limit cycle of small amplitude. To that end, a super-critical Hopf may be preferable over a sub-critical Hopf bifurcation.
 % The rest of this section describes some results pertaining to local stability and Hopf bifurcation of system~\eqref{eq:2nd_order_model}.}
 
 Recall that, we use the exogenous parameter $\kappa$ as the bifurcation parameter. Introducing this parameter, the general non-linear system~\eqref{eq:gen_2d_system} becomes
 \begin{align}
  \dot{x}=\kappa f(x,\tilde{p}),\,\,\,\,\,\,\,\, x = [x_1 \,\,x_2]^T \in \mathbb{R}^2, \label{eq:gen_2d_system_kappa}
 \end{align}
 A close look at the equations in~\eqref{eq:gen_2d_system_kappa} reveals that this system boils down to the original system~\eqref{eq:gen_2d_system} when $\kappa = 1$. Consider system~\eqref{eq:gen_2d_system}, let the system parameters be tuned such that this system is at the edge of stability. For the same parameter setting, system~\eqref{eq:gen_2d_system_kappa} would also be at the edge of stability at $\kappa = 1$. We may then increase $\kappa$ marginally from this point, to push the system into the unstable regime. We then analyse the system dynamics in the unstable regime, to derive the quantities required to characterise the type of the Hopf bifurcation and determine the stability of the limit cycles.
 
 % In order to establish the existence of a Hopf bifurcation, as discussed above, one needs to choose one of the model parameters as the bifurcation parameter. It is known that time-delayed systems could become unstable as the feedback delay increases~\cite{hassard1981theory}. Further, variations in other system parameters, like the Compound TCP parameters such as $\alpha$ and $\beta$, are also seen to affect system stability (Section~\ref{sec:CompoundTCP-RED}). Therefore, any of these parameters or the round-trip time may induce the bifurcation phenomenon, and can thus be chosen as the bifurcation parameter. However, this may not be desirable owing to the following reasons:
 % \begin{enumerate}
 %  \item [(i)]The equilibrium of the Compound TCP-RED system, is seen to be dependent on the system parameters. Consequently, varying any of these parameters may shift the equilibrium, making it difficult to observe the system dynamics with respect to change in the particular parameter.
 %  \item [(ii)] Varying any of these parameters may also affect the other system parameters, thus making it cumbersome to study the system dynamics.
 % \end{enumerate}
 % Therefore, instead of choosing one of the system parameters as the bifurcation parameter, we introduce an exogenous non-dimensional parameter 
 % ($\kappa > 0$) that can act as the bifurcation parameter is introduced~\cite{raina2005local}. 
 \subsection{Local bifurcation analysis}
 This appendix presents the Hopf bifurcation analysis for the Compound TCP-RED system, in the regime where queue size averaging is not performed~\eqref{eq:Compound_RED_nav_eta}. However, instead of using the specific function form for the packet-drop probability given by equation~\eqref{eq:prob_RED_nav}, we use a general functional form $p(q(\cdot))$. This enables us to provide an framework for the Hopf bifurcation analysis of not only the Compound TCP-RED system under study, but also for a system where RED is replaced by any other queue policy that computes the packet-drop probability as a linear function of the queue size. Following is the requisite analytical framework to characterise the type of the Hopf bifurcation and determine the stability of the limit cycles in such a system. This is followed by a numerical example to illustrate the use of the analytical framework.
 
 Taking a Taylor series expansion of system~\eqref{eq:Compound_RED_nav_eta} about $(w^\ast,q^\ast)$, using similar perturbations as before, yields
 % \begin{small}
 \begin{align}
  \dot{u}_{1}(t) =&\, \kappa\Big(\xi_{x}u_1(t) + \xi_s u_2(t-\tau) + \xi_{xx}u^2_1(t)+ \xi_{xr}u_1(t)u_1(t-\tau) + \xi_{xs}u_1(t)u_2(t-\tau) \notag\\
  &+ \xi_{rs}u_1(t-\tau)u_2(t-\tau)+ \xi_{xxx}u_1^3(t)+\xi_{xxr}u_1^2(t)u_1(t-\tau)+\xi_{xxs}u_1^2(t)u_2(t-\tau)\notag\\
  & +\xi_{xrs}u_1(t)u_1(t-\tau)u_2(t-\tau)\Big),\notag\\
  \dot{u}_2(t) =&\,\kappa \Big(\chi_x u_1(t) + \chi_y u_2(t) + \chi_{xy} u_1(t)u_2(t)\Big),\label{eq:lin_expand}
 \end{align}
 % \end{small}
 where the coefficients are as outlined in Table~\ref{tab:Taylor_series_coefficients}. 
 % Considering only the linear terms and looking for exponential solutions gives
 % \begin{align}
 %  \lambda^2 + \kappa a\lambda + \kappa^2 b + \kappa^2 c e^{-\lambda\tau} = 0,\label{eq:char_eta}
 % \end{align}
 % where $a,b,c$ are as given in~\eqref{eq:abc}. We now show that the system undergoes a Hopf bifurcation with variation in $\kappa$.
 % To verify the transversality condition of the Hopf spectrum~\cite{Hassard_81}, we need to show that $\text{Re}(\mathrm{d}\lambda/\mathrm{d}\kappa)_{\lambda=i\omega_0} \neq 0$. 
 Using notation $\mathbf{u}=[u_1\hspace{1ex}u_2]^{T}$, can be rewritten \eqref{eq:lin_expand} as
 \begin{align}
  \mathbf{\dot{u}}(t) = \mathcal{L}_{\mu}\mathbf{u}_t + \mathcal{F}(\mathbf{u}_t,\mu),\label{eq:geneq}
 \end{align}
 $t>0, \mu \in \mathbb{R}$, where for $\tau > 0$
 \begin{equation*}
  \mathbf{u}_{t}(\theta) = \mathbf{u}(t+\theta)\quad \mathbf{u}:[-\tau,0]\rightarrow\mathbb{R} \quad \theta\in [-\tau,0].
 \end{equation*}
 $\mathcal{L}_{\mu}$ is a one parameter family of continuous (bounded) linear operators, $\mathcal{L}_\mu : C[-\tau,0]\rightarrow \mathbb{R}^2,$ given by
 \begin{align*}
  \mathcal{L}_{\mu}\mathbf{u}_t = \int_{-\tau}^0 \mathrm{d}\Gamma(\theta,\mu)\mathbf{u}(t+\theta),\,\,\,\,\,\,\,\,\,\,\,\text{where} &&
 % \end{align*}
 % where
 %  \begin{align*}
  \mathrm{d}\Gamma(\theta,\mu) = \,\kappa\begin{bmatrix} \xi_x \delta(\theta) & \xi_s\delta(\theta+\tau) \\
   \chi_x\delta(\theta) &\chi_y\delta(\theta)\end{bmatrix}\mathrm{d}\theta,
  \end{align*}
 where $\delta(\theta)$ is the Dirac delta function. The operator $\mathcal{F}(\mathbf{u}_t,\mu)$ contains the non-linear terms, and is given by
 % \begin{small}
 \begin{align}
  &\,\,\,\,\,\,\,\,\,\,\,\,\,\,\,\,\mathcal{F}(\mathbf{u}_t,\mu) = \begin{bmatrix}
                                                    \mathcal{F}_1,
                                                    \mathcal{F}_2
                                   \end{bmatrix}^T,
                                   %\,\,\,\,\,\,\,\,\,\, \text{where}\notag\\
 \end{align}
 where $\mathcal{F}_1$ and $\mathcal{F}_2$ are the non-linear terms present in $\dot{u}_1(t)$ and $\dot{u}_2(t)$, given by~\eqref{eq:lin_expand}, respectively.
 Further, assume that both $\mathcal{L}_\mu$ and $\mathcal{F}$ depend analytically on the bifurcation parameter $\mu$.
 % \begin{align}
 %  \mathcal{F}_1 =&\, \kappa\big(\xi_{xx}u_1^2(t)+\xi_{xr}u_1(t)u_1(t-\tau)+\xi_{xs}u_1(t)u_2(t-\tau)\notag\\
 %  &+\xi_{rs}u_1(t-\tau)u_2(t-\tau) +\xi_{xxx}u^3_1(t)+\xi_{xxr}u_1^2(t)u_1(t-\tau)\notag\\
 %  &+\xi_{xxs}u_1^2(t)u_2(t-\tau)+\xi_{xrs}u_1(t)u_1(t-\tau)u_2(t-\tau),\notag\\
 %  \mathcal{F}_2 =&\, \kappa\big(\chi_{xy}u_1(t)u_2(t)+\chi_{yy}u_2^2(t)+\chi_{xyy}u_1(t)u_2^2(t)\notag\\
 %  &+\chi_{yyy}u_2^3(t)\big).\label{eq:non-linear_terms}
 % \end{align}
 % \end{small}
 %\hspace{-1.2mm}
 Equation~\eqref{eq:geneq} is to be cast into the following form
   \begin{align}
    \mathbf{\dot{u}}_{t} = \mathcal{A}(\mu)\mathbf{u}_{t} + \mathcal{R}\mathbf{u}_{t}.\label{eq:operatorform}
   \end{align}
 For $\mathbf{\phi} \in C^{1}[-\tau,0]$, the following operators can be defined,
 % \begin{small}
  \begin{align}
   \mathcal{A}(\mu)\mathbf{\phi}(\theta) &= \begin{cases}\begin{array}{ll}\frac{\mathrm{d}\mathbf{\phi}(\theta)}{d\theta},& \theta \in [-\tau,0),\\ \small \int_{-\tau}^{0}\,\mathrm{d}\Gamma(s,\mu)\mathbf{\phi}(s)\equiv \mathcal{L}_{\mu}\mathbf{\phi},&\theta=0,\end{array}\end{cases}\notag\\
 %   \end{align}
 %   \begin{align}
    \mathcal{R}\mathbf{\phi}(\theta)&= \begin{cases} \begin{array}{ll} 0,& \theta \in [-\tau,0),\\ \mathcal{F}(\mathbf{\phi},\mu),&\theta = 0.\end{array}\end{cases}\label{eq:defineA}
   \end{align}
 %   \end{small}
 Then, as $\mathrm{d}\mathbf{u}_t/\mathrm{d}\theta \equiv \mathrm{d}\mathbf{u}_t/\mathrm{d}t$, equation \eqref{eq:geneq} becomes \eqref{eq:operatorform}. Let $\kappa=\kappa_c+\mu$ be the bifurcation parameter, then the Hopf bifurcation occurs at $\mu = 0$. Hence, set $\mu =0$. Let $\mathbf{q}(\theta)$ be the eigenvector of $\mathcal{A}(0)$ corresponding to $\lambda(0)$, namely
 $\mathcal{A}(0) \mathbf{q}(\theta) = i\omega_0 \mathbf{q}(\theta).$ The eigenvector is derived as
 % \begin{small}
 \begin{align*}
  \mathbf{q}(\theta) &= \begin{bmatrix}1 & \phi_1\end{bmatrix}^{T}e^{i\omega\theta},\quad\text{where}\quad
  \phi_1 =\kappa\chi_x/(i\omega_0-\kappa\chi_y).
 \end{align*}
 % \end{small}
 Define the adjoint operator $\mathcal{A}^{\ast}(0)$ as 
 $$\mathcal{A}^{\ast}(0)\rho(s) = \begin{cases} \begin{array}{ll}-\frac{d\rho(s)}{ds},&s\in (0,\tau]\\\int_{-\tau}^{0}d\Gamma^{T}(t,0)\rho(-t)&s=0.\end{array}\end{cases}$$
  The eigenvector $\mathbf{q}^{\ast}(\theta)$ of the adjoint operator corresponding to eigenvalue $\bar{\lambda}(0)$ may be defined as $$\mathcal{A}^{\ast}\mathbf{q}^{\ast}~=~-i\omega_0\mathbf{q}^{\ast}.$$
 % \begin{align*}
 %   \mathcal{A}^{\ast}\mathbf{q}^{\ast} = -i\omega_0\mathbf{q}^{\ast}.
 % \end{align*}
 This gives $\mathbf{q}^{\ast}$ as
 % \begin{small}
 \begin{align*}
  \mathbf{q}^{\ast}(\theta) &= B \begin{bmatrix}\phi_2& 1\end{bmatrix}^{T}e^{i\omega_0\theta},\,\,\,\text{where}\,\,\,
  \phi_2 = -\kappa\chi_x/(\kappa\xi_x+i\omega_0).
 \end{align*}
 % \end{small}
 Now, define the inner-product of the functions $\boldsymbol{\psi} \in C[0,\tau]$ and $\boldsymbol{\phi} \in C[-\tau,0]$ as 
 \begin{small}
 \begin{align*}\left\langle \boldsymbol{\psi}, \boldsymbol{\phi} \right\rangle =&\, \boldsymbol{\overline{\psi}}(0)^{T}\boldsymbol{\phi}(0)- \int_{\theta = -\tau}^{0} \int_{\zeta=0}^{\theta} \boldsymbol{\overline{\psi}}^{T}(\zeta - \theta) \mathrm{d}\Gamma(\theta, \mu) \boldsymbol{\phi}(\zeta)\mathrm{d}\zeta\mathrm{d}\theta.
 \end{align*}
 \end{small}
 The scalar $B$ is to be found such that the orthogonality condition $\langle \mathbf{q}^{\ast},\mathbf{q}\rangle = 1$ is satisfied. Using the dot product defined above along with the orthogonality condition yields
 \begin{align*}
  B = \big(\phi_2(1+\kappa\overline{\phi}_1\tau  \xi_s e^{i\omega_0\tau})+\overline{\phi}_1\big)^{-1}.
 \end{align*}
 For the above $B$, one may also verify that $\langle \mathbf{q}^{\ast},\mathbf{\bar{q}}\rangle = 0$.
 For $\mathbf{u}_t$, a solution of \eqref{eq:operatorform} at $\mu=0$, define
 \begin{align*}
  z(t) &= \langle \mathbf{q}^{\ast},\mathbf{u}_t\rangle, \hspace{1.75ex}\text{and}\hspace{1.75ex}
  \mathbf{w}(t,\theta) = \mathbf{u_t}(\theta) - 2\text{Re}\big(z(t)\mathbf{q}(\theta)\big).
 \end{align*}
 Then on the manifold, $C_0$, using $\mathbf{w}_{ij}(\theta)=[\mathsf{w}_{ij1}(\theta)\hspace{1ex} \mathsf{w}_{ij2}(\theta)]^{T},$
 \begin{align}\label{eq:expandw}
 % \mathbf{w}(t,\theta) &= \mathbf{w}\big(z(t),\bar{z}(t),\theta\big),\hspace{1.2ex}\text{where}\notag\\
 \mathbf{w}(z,\bar{z},\theta) &= \mathbf{w}_{20}(\theta)\frac{z^{2}}{2}+ \mathbf{w}_{11}(\theta)z\bar{z} + \mathbf{w}_{02}(\theta)\frac{\bar{z}^{2}}{2}.
 \end{align}
 Here, $z$ and $\bar{z}$ are coordinates for $C_{0}$ in $C$ in the directions of $\mathbf{q}^{\ast}$ and $\mathbf{\bar{q}}^{\ast}$, respectively. The next step is to reduce~\eqref{eq:operatorform} to a differential equation for a complex variable on $C_{0}$. At $\mu = 0$,
 \begin{align}
 z'(t) &= \langle \mathbf{q}^{\ast},\mathcal{A}\mathbf{u_{t}} + \mathcal{R}\mathbf{u_{t}}\rangle
 % &= i\omega_{0}z(t) + \mathbf{\bar{q}}^{\ast}(0)\cdot \mathcal{F}\big(\mathbf{w}(z,\bar{z},\theta)+ 2\text{Re}\left(z(t)\mathbf{q}(\theta)\right)\big)\notag\\
 = i\omega_{0}z(t) + \mathbf{\bar{q}}^{\ast}(0)\cdot \mathcal{F}(z,\bar{z}).\label{eq:zbar}
 \end{align}
 which is written in abbreviated form as 
 $z'(t) = i\omega_{0}z(t) + g(z,\bar{z}).$
 Define
 \begin{align}
 g(z,\bar{z}) &=\mathbf{\bar{q}}^{\ast}(0)\cdot \mathcal{F}(z,\bar{z})\,=\, g_{20}\frac{z^{2}}{2} + g_{11}z \bar{z} + g_{02}\frac{\bar{z}^{2}}{2}+ g_{21}\frac{z^{2}\bar{z}}{2}+\cdot\cdot,\label{eq:gdeff}
 \end{align}
 and
 % \begin{small}
 \begin{align}
 \mathbf{u}_{t}(\theta) =&\, \mathbf{w}_{20}(\theta)\frac{z^{2}}{2} + \mathbf{w}_{11}(\theta)z\bar{z} + \mathbf{w}_{02}(\theta)\frac{\bar{z}^{2}}{2}+ z\mathbf{q}(0)e^{i\omega_{0}\theta}+ \bar{z}\mathbf{\bar{q}}(0)e^{-i\omega_{0}\theta}+ \cdots,\label{eq:u} 
 \end{align}
 % \end{small}
 from which $u_{1t}(0),u_{1t}(-\tau),u_{2t}(0)$ and $u_{2t}(-\tau)$ are obtained. One may then use these expressions and expand the  non-linear terms in \eqref{eq:lin_expand}.
 Non-linear terms in~\eqref{eq:lin_expand} in terms of $z,\bar{z}$ are
 
 \vspace{-3mm}
 \begin{small}
 \begin{align}
  \mathcal{F}(\mathbf{u}_t,\mu) =& \begin{bmatrix}\mathcal{F}_{201}\\ \mathcal{F}_{202}\end{bmatrix}\frac{z^2}{2} +\begin{bmatrix}\mathcal{F}_{111}\\\mathcal{F}_{112}\end{bmatrix}z\bar{z}+\begin{bmatrix}\mathcal{F}_{021}\\\mathcal{F}_{022}\end{bmatrix}\frac{\bar{z}^2}{2}+ \begin{bmatrix}\mathcal{F}_{211}\\\mathcal{F}_{212}\end{bmatrix}\frac{z^2\bar{z}}{2},\label{eq:fexpand}
 \end{align}
 \end{small}
 \vspace{-3mm}
 
 \noindent where the coefficients $\mathcal{F}_{ij1},\mathcal{F}_{ij2}$ can be obtained by substituting the non-linear terms in~\eqref{eq:lin_expand} by their expansion obtained using~\eqref{eq:u}.
 Substituting these in equation~\eqref{eq:gdeff}, and collecting the coefficients of $z^2,z\bar{z},\bar{z}^2$ and $z^2\bar{z}$, yields
 \begin{table}[t]
 \centering
 \caption{Coefficients in the Taylor series expansion of the Compound TCP-RED system in the absence of queue size averaging, evaluated at equilibrium $(w^\ast,q^\ast)$.}
 \begin{tabular}{l | l}
 \hline\label{tab:Taylor_series_coefficients}
 \\ 
 $\displaystyle\xi_x = \displaystyle\Big(i'(w^\ast)\big(1-p(q^\ast)\big)-d'(w^\ast)p(q^\ast)\Big)\frac{w^\ast}{\tau}$ & $\displaystyle\xi_s = \displaystyle -p'(q^\ast)\big(i(w^\ast)+d(w^\ast)\big)\frac{w^\ast}{\tau}$\\[2ex]
 $\displaystyle\xi_{xr} =\displaystyle\Big(i'(w^\ast)\big(1-p(q^\ast)\big)-d'(w^\ast) p(q^\ast)\Big)\frac{1}{\tau}$ &$\displaystyle\xi_{xx} = \displaystyle\frac{1}{2} i''(w^\ast)\big(1-p(q^\ast)\big) \frac{w^\ast}{\tau}$
 \\[2ex]
 $\displaystyle\xi_{xs} = \displaystyle-p'(q^\ast)\big(i'(w^\ast)+ d'(w^\ast)\big)\frac{w^\ast}{\tau}$ &
 $\displaystyle\xi_{rs} = \displaystyle-p'(q^\ast)\big(i(w^\ast)+d(w^\ast)\big)\frac{1}{\tau}$\\[2ex]
 $\displaystyle\xi_{xxx} = \displaystyle\frac{1}{6}i'''(w^\ast)\big(1-p(q^\ast)\big)\frac{w^\ast}{\tau}$ &
 $\displaystyle\xi_{xxr} =\displaystyle \frac{1}{2} i''(w^\ast)\big(1-p(q^\ast)\big)\frac{1}{\tau}$\\[2ex] 
 $\displaystyle\xi_{xrs} =\displaystyle-\frac{1}{2}p'(q^\ast)\big(i'(w^\ast)+ d'(w^\ast)\big)\frac{1}{\tau}$ & $\displaystyle\xi_{xxs} =\displaystyle-\frac{1}{2}p'(q^\ast)i''(w^\ast)\frac{w^\ast}{\tau}$
 \\[2ex]
 $\displaystyle\chi_{x} =\displaystyle\big(1-p(q^\ast)\big)\frac{1}{\tau}$ &
 $\displaystyle\chi_{y} = \displaystyle-p'(q^\ast)\frac{w^\ast}{\tau}$\\[2ex]
 $\displaystyle\chi_{xy} =\displaystyle-p'(q^\ast)\frac{1}{\tau}$ & \\[2ex]
 \hline \end{tabular}
 \end{table} 
 \begin{align}
  g_{20}=&\, \overline{B}\left(\bar{\phi}_2\, \mathcal{F}_{201} + \mathcal{F}_{202}\right), & g_{11}=&\, \overline{B}\left(\bar{\phi}_2\, \mathcal{F}_{111} + \mathcal{F}_{112}\right),\notag\\
  g_{02}=&\, \overline{B}\left(\bar{\phi}_2\, \mathcal{F}_{021} + \mathcal{F}_{022}\right), &  g_{21}=&\, \overline{B}\left(\bar{\phi}_2\, \mathcal{F}_{211} + \mathcal{F}_{212}\right).\label{eq:gcoeff}
 \end{align}
 % \end{small}
 To evaluate $\mathbf{w}_{11}(0),\mathbf{w}_{11}(-\tau),\mathbf{w}_{20}(0)$ and $\mathbf{w}_{20}(-\tau)$ seen in the expression for $g_{21}$, 
 one may write $\mathbf{w}' = \mathbf{u'_{t}} - z'\mathbf{q} - \bar{z}'\mathbf{\bar{q}}$, as done in \cite{hassard1981theory}.
 % \begin{align*}
 % 
 % \end{align*}
 Using \eqref{eq:operatorform} and \eqref{eq:zbar}, one obtains 
 \begin{equation*}
   \mathbf{w}' = \begin{cases} \begin{array}{l} \mathcal{A}\mathbf{w} - 2\text{Re}\big(\mathbf{\bar{q}}^{\ast}(0)\cdot\mathcal{F}\mathbf{q}(\theta)\big),\\ \mathcal{A}\mathbf{w} - 2\text{Re}\big(\mathbf{\bar{q}}^{\ast}(0)\cdot\mathcal{F}\mathbf{q}(0)\big) + \mathcal{F},\end{array}
   \begin{array}{l}\theta \in [-\tau,0)\\\theta = 0,\end{array}\end{cases}
 \end{equation*}
 which can be rewritten, using equation \eqref{eq:expandw}, as 
 \begin{align}
  \mathbf{w}' &= \mathcal{A}\mathbf{w} + \mathbf{H}(z,\bar{z},\theta),\hspace{0.85ex}\text{where}\label{eq:wderivative}\\
 % \end{align} where
 % \begin{equation}
    \mathbf{H}(z,\bar{z},\theta) &= \mathbf{H}_{20}(\theta)\frac{z^{2}}{2} + \mathbf{H}_{11}(\theta)z\bar{z} + \mathbf{H}_{02}(\theta)\frac{\bar{z}^{2}}{2} \cdots.
    \label{eq:expandH}
 \end{align}
 Now, on $C_0$, near the origin $\mathbf{w}' = \mathbf{w}_{z}z' + \mathbf{w}_{\bar{z}}\bar{z}'.$
 % \begin{equation*}
 %   \end{equation*} 
 Replacing $\mathbf{w}_z,z'$ from \eqref{eq:expandw} and \eqref{eq:zbar}, and equating with \eqref{eq:wderivative}, gives 
 \begin{align}\label{eq:Hoperator}
    (2i\omega_{0}- \mathcal{A})\mathbf{w}_{20}(\theta) &= \mathbf{H}_{20}(\theta), &  -\mathcal{A}\mathbf{w}_{11}(\theta) &= \mathbf{H}_{11}(\theta),
 %   -(2i\omega_{0}+ \mathcal{A})\mathbf{w}_{02}(\theta) &= \mathbf{H}_{02}(\theta).
  \end{align}
  For $\theta\in[-\tau,0)$
   \begin{align*}
    \mathbf{H}(z,\bar{z},\theta) =& -2\text{Re}\big(\mathbf{\bar{q}}^{\ast}(0)\cdot \mathcal{F}_{0} \mathbf{q}(\theta)\big) 
   %  =& -2 Re \{g(z,\bar{z})q(\theta) \}\\
     = -g(z,\bar{z})\mathbf{q}(\theta) - \bar{g}(z,\bar{z})\mathbf{\bar{q}}(\theta),
 %    =& -\left(g_{20}\frac{z^{2}}{2} + g_{11}z\bar{z} + g_{02}\frac{\bar{z}^{2}}{2}+\cdots \right)\mathbf{q}(\theta)\\ &- \left(\bar{g}_{20}\frac{\bar{z}^{2}}{2} + \bar{g}_{11}z\bar{z} + \bar{g}_{02}\frac{z^{2}}{2}+\cdots \right)\mathbf{\bar{q}}(\theta),
   \end{align*}
 which when compared with \eqref{eq:expandH} gives
 % \begin{small}
  \begin{align}
  \label{eq:Hterms}
    \mathbf{H}_{20}(\theta) &= -g_{20}\mathbf{q}(\theta)-\bar{g}_{02}\mathbf{\bar{q}}(\theta), &   \mathbf{H}_{11}(\theta) &= -g_{11}\mathbf{q}(\theta)-\bar{g}_{11}\mathbf{\bar{q}}(\theta).
    \end{align}
 %    \end{small}
 % From \eqref{eq:defineA} and \eqref{eq:Hoperator}, we have
 %  \begin{align}\label{eq:wdashes}
 %    \mathbf{w}'_{20}(\theta) &= 2i\omega_{0}\mathbf{w}_{20}(\theta) + g_{20}\mathbf{q}(\theta)+ \bar{g}_{02}\mathbf{\bar{q}}(\theta),\notag\\
 %    \mathbf{w}'_{11}(\theta) &= g_{11}\mathbf{q}(\theta) + \bar{g}_{11}\mathbf{\bar{q}}(\theta).
 %   \end{align}
   Using~\eqref{eq:defineA},~\eqref{eq:Hoperator} yields
  \begin{align}\label{eq:ws}
    \mathbf{w}_{20}(\theta) &= \frac{-g_{20}}{i\omega_{0}}\mathbf{q}(0)e^{i\omega_{0}\theta}-\frac{\bar{g}_{02}}{3i\omega_{0}}\mathbf{\bar{q}}(0)e^{-i\omega_{0}\theta} + \mathbf{E}e^{2i\omega_{0}\theta},\notag\\
    \mathbf{w}_{11}(\theta) &= \frac{g_{11}}{i\omega_{0}}\mathbf{q}(0)e^{i\omega_{0}\theta} - \frac{\bar{g}_{11}}{i\omega_{0}}\mathbf{\bar{q}}(0)e^{-i\omega_{0}\theta} + \mathbf{F}.
   \end{align}
 The next step is to determine $\mathbf{E}$ and $\mathbf{F}$. It is known that
 \begin{align}
  \mathbf{H}(z,\bar{z},0) &= -2\text{Re}\big(\mathbf{\bar{q}}^{\ast}(0)\cdot\mathbf{\mathcal{F}} \mathbf{q}(0)\big)+\mathbf{\mathcal{F}}.\label{eq:Hdeff}
  \end{align}
  Substituting the expansion of $\mathcal{F}$ from \eqref{eq:fexpand} gives
  \begin{align}
  \mathbf{H}_{20}(0) &= -g_{20}\mathbf{q}(0)-\bar{g}_{02}\mathbf{\bar{q}}(0)+\begin{bmatrix}\mathcal{F}_{201}&\hspace{-2.5mm}\mathcal{F}_{202}\end{bmatrix}^{T},\notag\\
  \mathbf{H}_{11}(0) &= -g_{11}\mathbf{q}(0)-\bar{g}_{11}\mathbf{\bar{q}}(0)+\begin{bmatrix}\mathcal{F}_{111}&\hspace{-2.5mm}\mathcal{F}_{112}\end{bmatrix}^{T}.\label{eq:Hzeros}
 \end{align}
 Then, the definition of the linear operator in~\eqref{eq:defineA} is used to find $H_{20}(0)$ and $H_{11}(0)$ from~\eqref{eq:Hoperator}, and then this is equated with~\eqref{eq:Hzeros} to obtain
 % From \eqref{eq:Hoperator} and \eqref{eq:defineA}, we obtain
 \begin{align}
  g_{20}\mathbf{q}(0)+\bar{g}_{02}\mathbf{\bar{q}}(0) =& \begin{bmatrix}\mathcal{F}_{201}\\\mathcal{F}_{202}\end{bmatrix}- \begin{bmatrix} (2i\omega_0-\kappa\xi_x)\mathsf{w}_{201}(0) - \kappa \xi_s\mathsf{w}_{202}(-\tau)\notag\\
  -\kappa \chi_x \mathsf{w}_{201}(0)+(2i\omega_0-\kappa\chi_y)\mathsf{w}_{202}(0)\end{bmatrix},\notag\\
 %  \end{align}
 %  \begin{align}
 g_{11}\mathbf{q}(0)+\bar{g}_{11}\mathbf{\bar{q}}(0) =& \begin{bmatrix}\mathcal{F}_{111}\\\mathcal{F}_{112}\end{bmatrix}- \begin{bmatrix}-\kappa \xi_x\mathsf{w}_{111}(0)-\kappa \xi_s\mathsf{w}_{112}(-\tau)\\ -\kappa \chi_x\mathsf{w}_{111}(0)-\kappa \chi_y\mathsf{w}_{112}(0)\end{bmatrix}.\label{eq:getE&F}
 \end{align}
 % Substituting  in \eqref{eq:getE&F}, we get 
 Upon substituting $\mathbf{w_{20}}(0),\mathbf{w_{20}(-\tau)},\mathbf{w_{11}}(0)$ and $\mathbf{w_{11}(-\tau)}$ from \eqref{eq:ws} to get 
 $\mathbf{E}$ and $\mathbf{F}$ of the form
 \begin{align}
  \mathbf{E} = \begin{bmatrix}E_{1}& E_{2}\end{bmatrix}^{T} \quad\text{and}\quad \mathbf{F} = \begin{bmatrix}F_{1}& F_{2}\end{bmatrix}^{T},
 \end{align}
 where $E_1,E_2,F_1$ and $F_2$ are given by
 \begin{align*}
  \begin{bmatrix}E_1 \\ E_2\end{bmatrix} = \frac{1}{A_1B_2 - A_2B_1}\begin{bmatrix}C_1B_2 - C_2B_1 \\ C_2A_1 - C_1A_2 \end{bmatrix}, && \begin{bmatrix}F_1 \\ F_2\end{bmatrix} = \frac{1}{K_1L_2 - K_2L_1}\begin{bmatrix}J_1L_2 - J_2L_1\\ J_2K_1  -J_1K_2\end{bmatrix},
 \end{align*}
 % 
 % \begin{align*}
 %  E_1 &= \frac{C_1B_2 - C_2B_1}{A_1B_2 - A_2B_1}, & E_2 &= \frac{C_2A_1 - C_1A_2}{A_1B_2 - A_2B_1}, &
 %  F_1 &= \frac{J_1L_2 - J_2L_1}{K_1L_2 - K_2L_1}, & F_2 &= \frac{J_2K_1  -J_1K_2}{K_1L_2 - K_2L_1},
 %  \end{align*}
   where
   \begin{align*}
  A_1 &=\,  \kappa\xi_x-2i\omega_0,& A_2 &=\, \kappa\chi_x, &  B_1 &=\, \kappa\xi_se^{-2i\omega_0\tau}, &   B_2 &=\, \kappa\chi_y - 2i\omega_0, \\
  C_1 &=\,-\mathcal{F}_{201},&  C_2 &=\, -\mathcal{F}_{202}, & J_1 &=\, -\mathcal{F}_{111},& J_2 &=\, -\mathcal{F}_{112},\\
  K_1 &=\, \kappa\xi_x, &  K_2 &=\, \kappa \chi_x, & L_1 &=\, \kappa \xi_s,& L_2 &=\, \kappa \chi_y.\notag
 \end{align*}
 %  \begin{align}
 %    C_1 =&\, \frac{g_{20}}{i\omega_0}\left(-i\omega_0+\kappa\xi_x+\kappa\xi_s\phi_1e^{-i\omega_0\tau}\right)-\mathcal{F}_{201}\notag\\
 %  &+\frac{\bar{g}_{02}}{3i\omega_0}\left(i\omega_0+\kappa \xi_x +\kappa\xi_s\bar{\phi}_1e^{i\omega_0\tau}\right),\notag\\
 %   C_2 =&\, \frac{g_{20}}{i\omega_0}\left(-i\omega_0\phi_1+\kappa\chi_x+\kappa\chi_y\phi_1\right)-\mathcal{F}_{202}\notag\\
 %  &+\frac{\bar{g}_{02}}{3i\omega_0}\big(i\omega_0\bar{\phi}_1 + \kappa \chi_x+\kappa\chi_y\bar{\phi_1}),\notag\\ 
 %  J_1 =&\, \frac{g_{11}}{i\omega_0}\left(i\omega_0-\kappa \xi_x-\kappa\xi_s\phi_1 e^{-i\omega_0\tau})\right)-\mathcal{F}_{111}\notag\\
 %  &+ \frac{\bar{g}_{11}}{i\omega}\left(i\omega_0 + \kappa\xi_x+\kappa\xi_s\bar{\phi}_1e^{i\omega_0\tau})\right),\notag\\
 %  \end{align}
 %  \begin{align}
 %  J_2 =&\, \frac{g_{11}}{i\omega_0}\big(i\omega_0\phi_1-\kappa \chi_x-\kappa \chi_y\phi_1\big)-\mathcal{F}_{112}\notag\\
 %  &+\frac{\bar{g}_{11}}{i\omega_0}\big(i\omega_0\bar{\phi}_1 + \kappa\chi_x+\kappa \chi_y\bar{\phi}_1\big).
 % \end{align}
 Using $\mathbf{E}$ and $\mathbf{F}$ one can compute $\mathbf{w}_{20}$ and $\mathbf{w}_{11}$, which in turn enables one to evaluate $g_{21}$. All the terms required for the 
 Hopf bifurcation analysis have now been derived. Using these, the following quantities are computed~\cite{hassard1981theory}:
 \begin{align}
  \hspace{-3mm}c_1(0) &= \frac{i}{2\omega_0}\left(g_{20}g_{11}-2|g_{11}|^2-\frac{1}{3}|g_{02}|^2\right)+\frac{g_{21}}{2},\label{eq:cterm}\\
  \mu_2 &= -\frac{\text{Re}\big(c_1(0)\big)}{\alpha'(0)},\quad\quad\beta_2 = 2\text{Re}\big(c_1(0)\big),\label{eq:muterm&betaterm}
 \end{align}
 where $c_1(0)$ is the lyapunov coefficient, $\beta_2$ is the Floquet exponent, and $\alpha'(0) =~\text{Re}\left(\mathrm{d}\lambda/\mathrm{d}\kappa\right)|_{\kappa=1}$.
 % \begin{align}
 % 
 % \end{align}
 Substituting \eqref{eq:gcoeff} in \eqref{eq:cterm} gives the expression for $c_1(0)$, which is the lyapunov coefficient. The values of $\mu_2$ and $\beta_2$ are computed using \eqref{eq:muterm&betaterm}. 
 \begin{itemize}
  \item The Hopf bifurcation is \emph{super-critical} if $\mu_2 > 0$, and \emph{sub-critical} if $\mu_2 <0$.
  \item The limit cycles are \emph{asymptotically orbitally stable} if $\beta_2< 0$, and \emph{unstable} if $\beta_2>0$.
 \end{itemize}
 We now exemplify the above analysis using a numerical example. 
 
 \subsection{Numerical example}
\begin{figure}
  \psfrag{n}{$\kappa$}
   \psfrag{m}{\hspace{-1.5mm}$\mu_2$}
   \psfrag{b}{\hspace{-1.5mm}$\beta_2$}
    \psfrag{0}{\scriptsize{$0$}}
   \psfrag{2.5}{\scriptsize{$2.5$}}
   \psfrag{0.5}{\scriptsize{$0.5$}}
   \psfrag{1.5}{\scriptsize{$1.5$}}
   \psfrag{1}{\scriptsize{$1$}}
   \psfrag{-2}{\scriptsize{$-2$}}
   \psfrag{1.025}{\scriptsize{$1.025$}}
   \psfrag{1.05}{\scriptsize{$1.05$}}
   \psfrag{-1}{\scriptsize{$-1$}}
   \psfrag{-5}{\scriptsize{$-5$}}
   \psfrag{-2.5}{\scriptsize{$-2.5$}}
   \psfrag{9}{\scriptsize{$9$}}
   \psfrag{p1}[b1][b1][0.8][-90]{\scriptsize{$\times 10^{4}$}}
   \begin{center}
   \subfloat[Type of the Hopf bifurcation]{
   \includegraphics[width=1.4in,height=2in,angle=270]{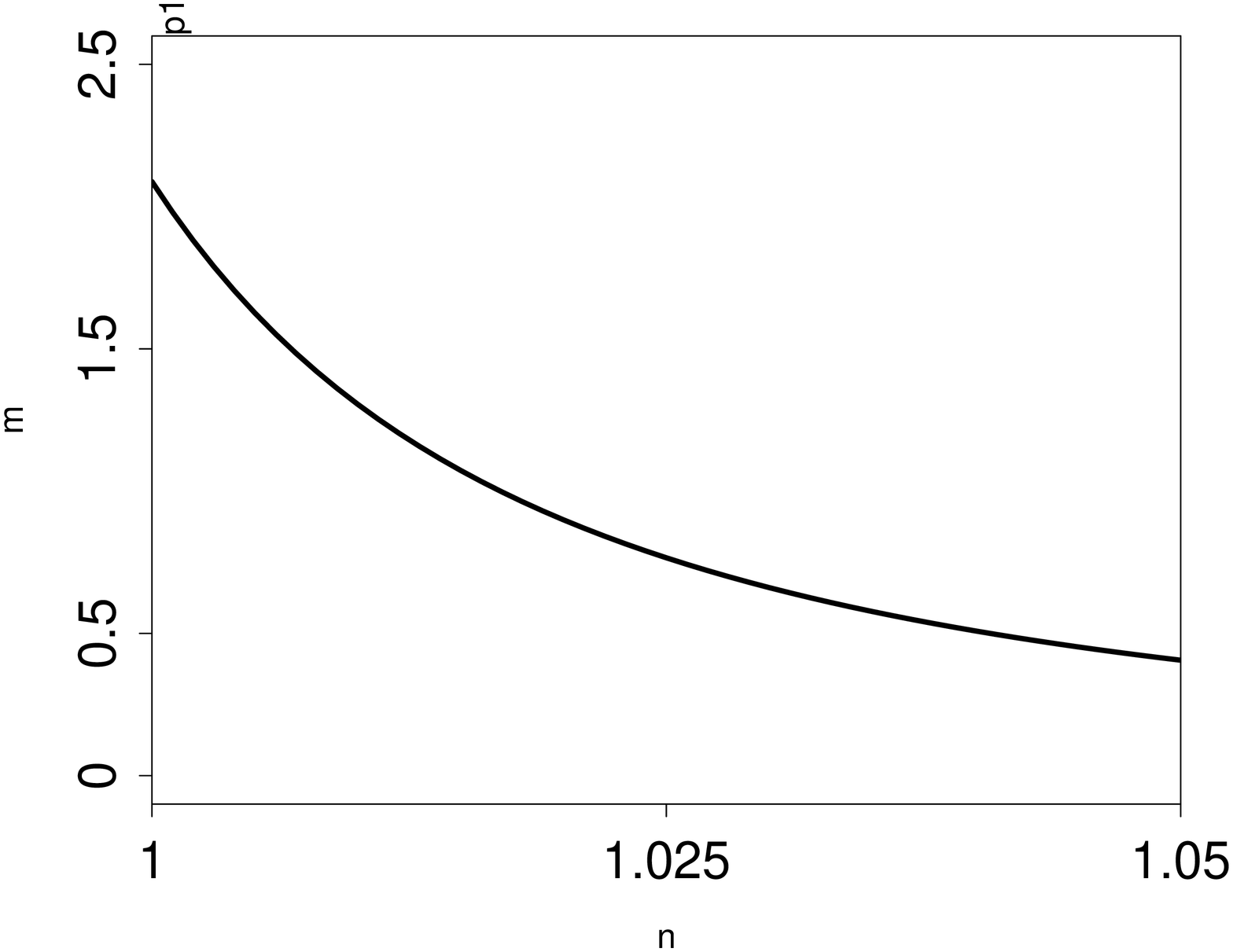}
   \label{fig:numerical_example_mu2}
   }
   \qquad
   \subfloat[Orbital stability of limit cycle]{
   \includegraphics[width=1.4in,height=2in,angle=270]{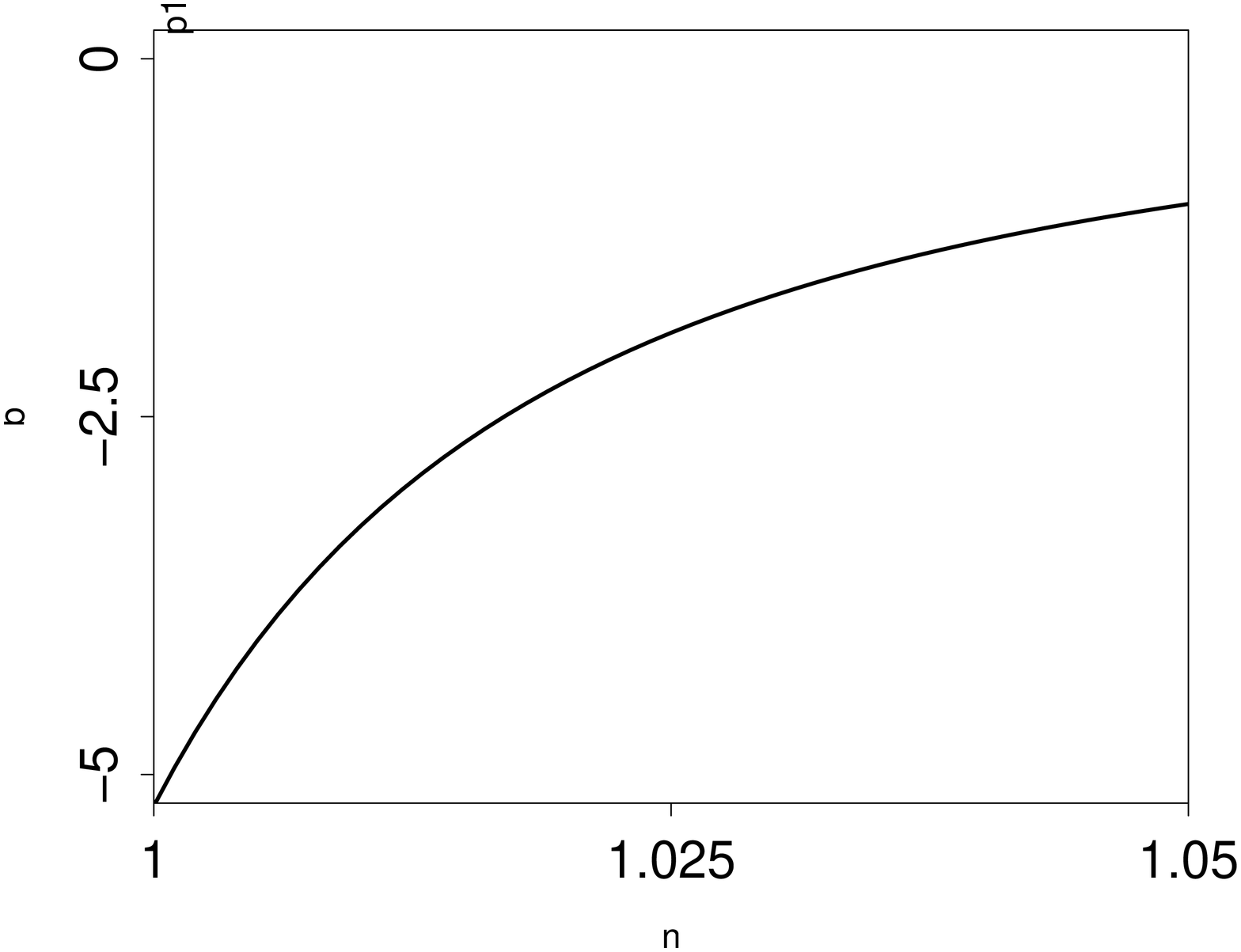}
   \label{fig:numerical_example_beta2}
   }
   \caption{Plots of $\mu_2$ and $\beta_2$ versus the non-dimensional bifurcation parameter $\kappa$. Observe that $\mu_2 > 0$ and $\beta_2 < 0$. This implies that, 
 for the chosen parameter values, the Hopf bifurcation is super-critical, and the emergent limit cycles are asymptotically orbitally stable.}
   \label{fig:numerical_example}
 \end{center}
  \end{figure}

 The Compound TCP and RED parameters are set at their default values. Then for $C = 100$, at the resulting equilibrium, the system undergoes a Hopf bifurcation at $\tau = 0.7751$. At this point, by construction, $\kappa = 1$. Now, the bifurcation parameter $\kappa$ is marginally increased, by about $5\%$, to push the system into a locally unstable regime. Using the Hopf bifurcation analysis, the values of $\mu_2$ and $\beta_2$ can be computed. The plots of $\mu_2$ and $\beta_2$, with respect to the bifurcation parameter $\kappa$, are presented in Figure~\ref{fig:numerical_example}. Observe from the plots that $\mu_2 > 0$ and $\beta_2 < 0$. Therefore, for these parameter values, the Hopf bifurcation is super-critical and the limit cycles are asymptotically orbitally stable.
 
\bibliographystyle{IEEEtran}
\bibliography{Arxiv_version}

\end{document}